\documentclass[onecollarge,final]{svjour2}                  
\smartqed 
\usepackage{graphicx}
\usepackage[francais,english]{babel}
\usepackage[latin1]{inputenc}
\usepackage[T1]{fontenc}
\usepackage{amsmath}
\usepackage{amssymb}
\usepackage{color}
\usepackage{psfrag}
\usepackage{graphics}
\usepackage{mathrsfs}
\usepackage{verbatim}
\usepackage{textcomp}
\usepackage{xspace}

\newcommand{\nn}{\nonumber}

\newcommand{\rmd}{{\mathrm{d}}}

\newcommand{\rmext}{{\mathrm{ext}}}

\begin{document}

\title{On certain functionals of the maximum of Brownian motion and their applications}

\author{Anthony Perret \and Alain Comtet \and \\Satya N.~Majumdar \and Gr\'egory Schehr}

\institute{A. Perret \at Laboratoire de Physique Th\'eorique et Mod\`eles
  Statistiques, Universit\'e Paris-Sud, B\^at. 100, 91405 Orsay Cedex, France
   \\ \and A. Comtet \at  Laboratoire de Physique Th\'eorique et Mod\`eles
  Statistiques, Universit\'e Paris-Sud, B\^at. 100, 91405 Orsay Cedex, France, Universit\'e Pierre et Marie Curie, Paris 6, 75252 Paris Cedex 05, France
   \\ \and S. N. Majumdar \at  Laboratoire de Physique Th\'eorique et Mod\`eles
  Statistiques, Universit\'e Paris-Sud, B\^at. 100, 91405 Orsay Cedex, France
  \\ \and G. Schehr \at  Laboratoire de Physique Th\'eorique et Mod\`eles
  Statistiques, Universit\'e Paris-Sud, B\^at. 100, 91405 Orsay Cedex, France}

\date{\today}

\maketitle

\begin{abstract}
We consider a Brownian motion (BM) $x(\tau)$ and its maximal value $x_{\max} = \max_{0 \leq \tau \leq t} x(\tau)$ on a fixed time interval $[0,t]$.
We study functionals of the maximum of the BM, of the form ${\cal O}_{\max}(t)=\int_0^t\, V(x_{\max} - x(\tau))  \rmd \tau$  where $V(x)$ can be any arbitrary function and develop various analytical tools to compute their statistical properties. These tools rely in particular on (i) a ``counting paths'' method and (ii) a path-integral approach. 
In particular, we focus on the case where $V(x) = \delta(x-r)$, with $r$ a real parameter, which is relevant to study the density of near-extreme values of the BM (the so called density of states), $\rho(r,t)$, which is the local time of the BM spent at given distance $r$ from the maximum. We also provide a thorough analysis of the family of functionals ${T}_{\alpha}(t)=\int_0^t (x_{\max} - x(\tau))^\alpha \, {\rmd}\tau$, corresponding to $V(x) = x^\alpha
$, with $\alpha$ real. As $\alpha$ is varied, $T_\alpha(t)$ interpolates between different interesting observables. For instance, for $\alpha =1$, 
$T_{\alpha = 1}(t)$ is a random variable of the ``area'', or ``Airy'', type while for $\alpha=-1/2$ it corresponds to the maximum time spent by a ballistic particle through a Brownian random potential. On the other hand, for $\alpha = -1$, it corresponds to the cost of the optimal algorithm to find the maximum of a discrete random walk, proposed by Odlyzko. We revisit here, using tools of theoretical physics, the statistical properties of this algorithm which had been studied before using probabilistic methods. Finally, we extend our methods to constrained BM, including in particular the Brownian bridge, i.e., the Brownian motion starting and ending at the origin.


\end{abstract}

\newpage

\section{Introduction}

\begin{figure}[ht]
\begin{center}
\rotatebox{-90}{\resizebox{70mm}{!}{\includegraphics{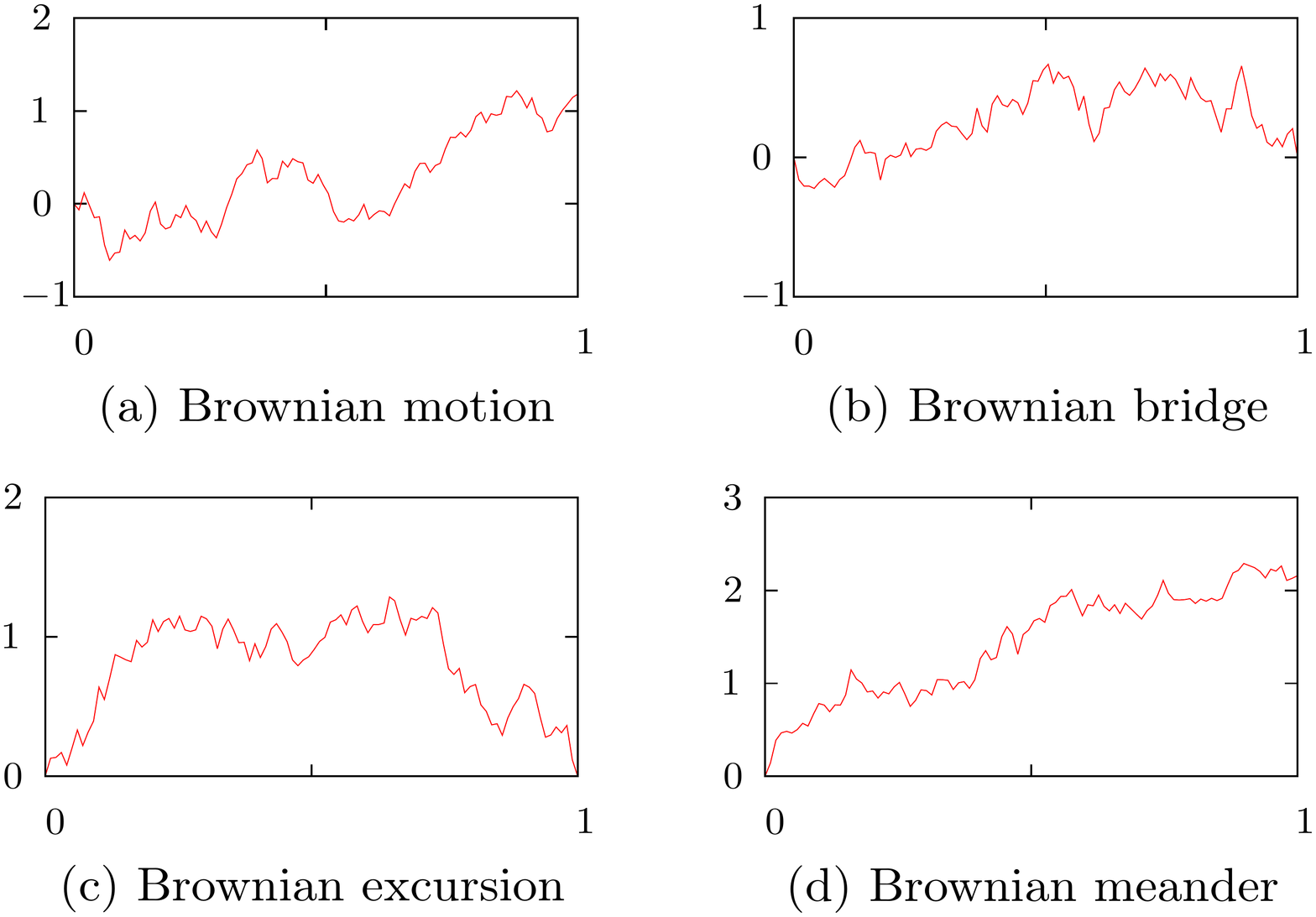}}}
\caption{Brownian motion $x(\tau)$ (a) and its variants: the Brownian bridge $x_B(\tau)$ in (b), the Brownian excursion $x_E(\tau)$ in (c) and the Brownian meander $x_{Me}(\tau)$ in (d).}\label{fig_Brownian}
\end{center}
\end{figure}

Stochastic processes are at the heart of many fundamental problems in statistical physics. In particular, it was realized a long time ago that Brownian motion (BM) is the process underlying many physical systems and corresponding random models. Since then, BM has not only become a cornerstone of statistical physics \cite{Cha43,Fel68,Hug80} but has also found numerous applications in various areas of science, including biology \cite{Kos80}, computer science \cite{Asm03,Kearney04,KM05,Maj05a,Maj05b} or financial mathematics \cite{Wil06,MB08}. These various applications have motivated the study of functionals of Brownian motion \cite{Maj05a,Wil06,Comtet05,Yor00}, which are observables of the form ${\cal O}(t) = \int_0^t V(x(\tau))\, \rmd\tau$ where $V(x)$ can be any function and $x(\tau)$ is a BM (see Fig. \ref{fig_Brownian} (a)). It might also be relevant to consider functionals of variants of BM, like the Brownian bridge (BB), $x_B(\tau)$, which is a BM conditioned to start and end at the origin (see Fig. \ref{fig_Brownian} (b)), the Brownian excursion (BE), $x_E(\tau)$, which is a BB conditioned to stay positive on the whole time interval $[0,t]$ (see Fig. \ref{fig_Brownian} (c)) as well as the Brownian meander, $x_{Me}(\tau)$, which is constrained to stay positive on $[0,t]$ but can end up at any point at time $t$ (see Fig. \ref{fig_Brownian} (d)). For instance, if $V(x) = \delta(x-a)$, the Brownian functional ${\cal O}(t)$ corresponds to the local time at the fixed level $a$, which is an important quantity in probability theory \cite{Pit99}. Another interesting example concerns the case where $V(x) = x$ for a BE, which corresponds to the area under a BE. In this case, the distribution of ${\cal O}_E(t) = \int_0^t x_E(\tau) \,\rmd\tau$ is given by the so called Airy-distribution which appears in computer science \cite{Darling83,Louchard84,FPV99,JL07} as well as in the extreme statistics of elastic interfaces \cite{Maj05a,MC2005,MC2005b}. Extensions of the Airy-distribution to the area under Bessel processes ({\it i.e.}, radius of the $d$-dimensional process) have been recently discussed in Ref.~\cite{Barkai14}. Yet another example which is relevant in finance is the case where $V(x) = \exp{(x)}$, which describes the price of an Asian stock option in the Black-Scholes framework \cite{BS73}. Note that in this exponential case, ${\cal O}(t) = \int_0^t \exp{(x(\tau))} \,\rmd\tau$ also represents the stationary current of a disordered Sinai chain connected to two reservoirs of particles \cite{KKS75,OMM93,MC94,ORS13}. Quite interestingly, these functionals of BM and its variants can be studied using powerful tools of theoretical physics, namely path integrals methods (leading to the so called Feynman-Kac formula). This allows to recast the study of Brownian functionals in a quantum mechanical framework~\cite{Maj05a,Kac49}.

\begin{figure}[h]
\begin{center}
\rotatebox{-90}{\resizebox{60mm}{!}{\includegraphics{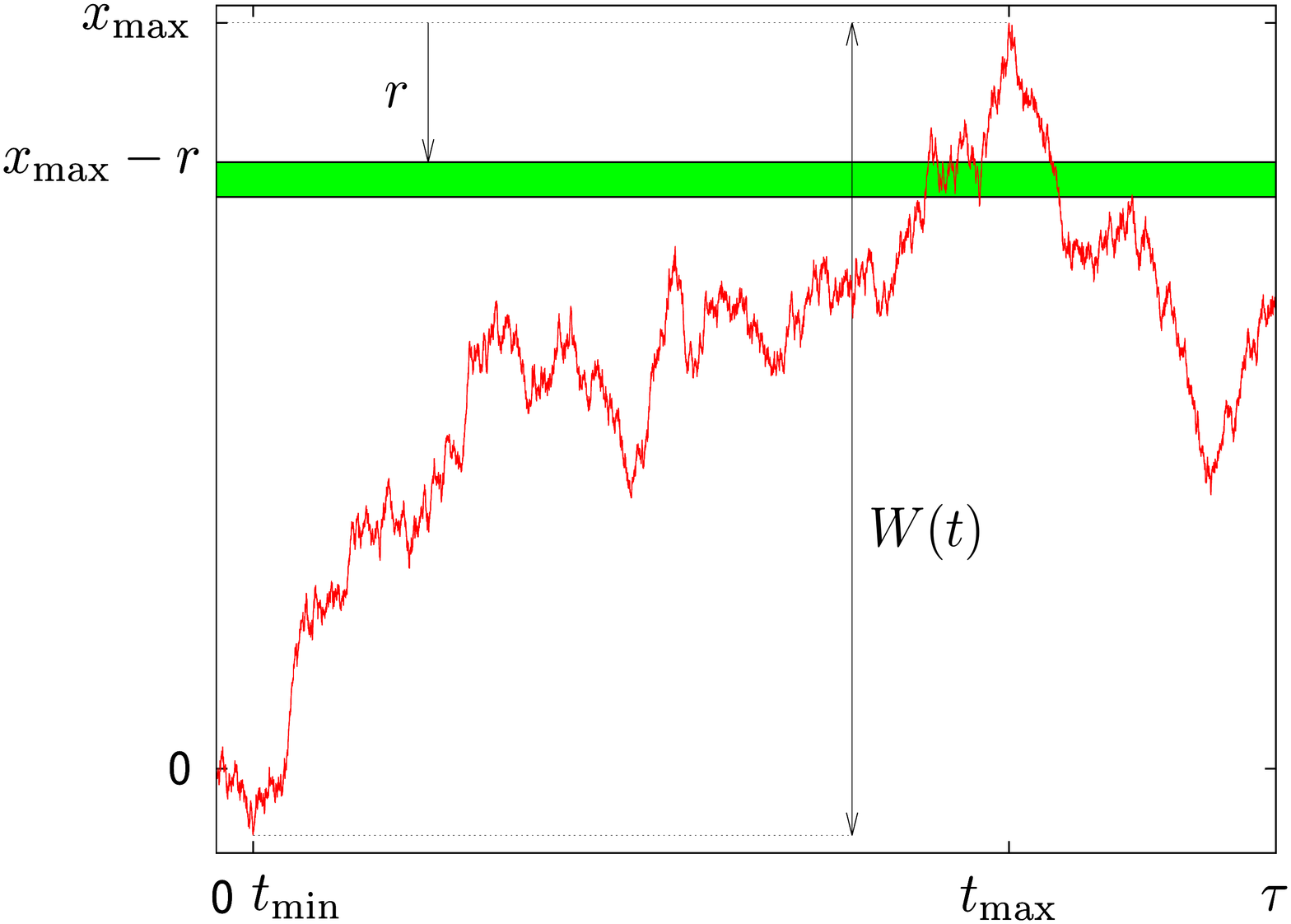}}}
\caption{(Color online) One realization of the process $x(\tau)$ on the time interval $[0,t]$, with a width $W(t) =  \max_{\tau \in [0,t]} x(\tau) - \min_{\tau \in [0,t]} x(\tau)$. $x(\tau)$ spends a time $\rho(r,t) {\rmd r}$ at a distance within $[r, r + {\rmd}r]$ (the green stripe) from the maximum $x_{\max}$, with $\rho(r,t)$ being the DOS (\ref{def_rho}).}
\label{fig_intro}
\end{center}
\end{figure}

While such Brownian functionals are thus well understood, several recent works have pointed out the physical relevance of {\it functionals of the maximum of Brownian motion}, which are much less studied. In this case, one is interested in observables of the form ${\cal O}_{\max}(t) = \int_0^t V(x_{\max} - x(\tau)) \,\rmd\tau$ with $x_{\max} = \max_{0 \leq \tau \leq t} x(\tau)$ where $x(\tau)$ is BM or one of its variants (see Fig. \ref{fig_Brownian}). An important case corresponds to the case where $V(x) = \delta(x-r)$ where 
\begin{eqnarray}\label{def_rho}
\rho(r,t) = \int_0^t \delta(x_{\max} - x(\tau) - r) \,\rmd\tau \;, 
\end{eqnarray}
which is the so called density of states (DOS) near the maximum. This is a natural and useful quantity to characterize the crowding of near-extremes~\cite{sabhapandit2007}. Indeed, $\rho(r,t) {\rmd}r$ denotes the amount of time spent by $x(\tau)$ at a distance within the interval $[r,r+{\rmd}r]$ from $x_{\max}$ (see Fig. \ref{fig_intro}). Hence $\rho(r,t)$ is similar to the local time with the major difference that here the distances are measured from $x_{\max}$, which is itself a random variable. The statistics of the DOS was recently studied by us in the context of near-extreme statistics \cite{Perret_PRL}. Note that, by definition, $\int_0^\infty \rho(r,t) \,\rmd r = t$. Therefore its average value, $\langle \rho(r,t)\rangle/t$, where $\langle \ldots \rangle$ means an average over the trajectories of BM has a natural probabilistic interpretation as it is the probability density function to find the BM at a given distance $r$ from the maximum in the time interval $[0,t]$. In particular, the average value of any functional of the maximum can be expressed as
\begin{eqnarray}\label{property_average}
\langle {\cal O}_{\max}(t) \rangle = \langle \int_0^t V(x_{\max} - x(\tau)) \,\rmd\tau\rangle = \int_0^\infty \langle \rho(r,t) \rangle V(r) \,\rmd r \;,
\end{eqnarray} 
which naturally holds not only for BM but also for its variants, like Brownian bridge or Brownian excursion. 

Another very interesting application of functionals of the maximum of Brownian motion concerns the case where $V(x) = 1/(2 x)$, which enters into the analysis of the optimal algorithm to find the maximum of a discrete random walk of $n$ steps with $n \gg 1$ \cite{odlyzko}. Indeed, let us consider a discrete random walk (RW), starting from $X_0 = 0$ and evolving via the Markov rule: $X_{k} = X_{k-1} + \eta_k$ where $\eta_k = \pm 1$ with equal probability $1/2$. We study the search problem of finding the maximum of the RW, $M_n = \max_{0 \leq i \leq n} X_i$, while minimizing the number of values of $X_k$'s that are probed. The cost of the algorithm is identified by the number of probes used to find $M_n$. The simplest algorithm consists of probing all positions $X_k$ for $1 \leq k \leq n$: its cost is $n$. In Fig. \ref{exemple_algo}, we show an example for $n=14$ where we find $M_{14}=7$ in $4$ probes: this shows that, because of strong correlations between the positions of the RW, $n$ is actually a rough upper bound of the cost of the optimal algorithm. Of course, some RWs need more probes than others in order to find $M_n$. For example, the RW with $n$ jumps $+1$ needs only one probe in $X_n=n$ but the RW with alternating jumps $\pm 1$ ($X_{2k}=0$ and $X_{2k+1}=1$) needs $\lfloor \frac{n}{2} \rfloor +1$ probes to be sure that none $X_{2k}=2$. But of course these cases are rare. Let $A_n$ be the ensemble of algorithms that find $M_n$, according to the above rules, and let us denote by $C(a)$ the cost, as defined above, of the algorithm $a$ that belongs to $A_n$. Of course $C(a)$ is a random variable, which varies from one realization of the RW to another, its average value being denoted by $\langle C(a) \rangle$. In Ref.~\cite{odlyzko}, Odlyzko studied the minimal average cost of such algorithms and he showed that, for large $n$, the minimal average cost is proportional to $\sqrt{n}$, much smaller than the cost of the aforementioned naive algorithm, which necessitates $n$ probes. Indeed, one has \cite{odlyzko}
\begin{eqnarray}\label{average_cost}
\min_{a \in A_n} \langle C(a) \rangle = c_0 \sqrt{n} + o(\sqrt{n}) \;, \;
\end{eqnarray} 
where $c_0$ is a constant given by
\begin{eqnarray}\label{def_I}
c_0 = \langle I \rangle \;, \; I = \frac{1}{2} \int_0^1 \frac{d\tau}{[x_{\max} - x(\tau)]} \;,
\end{eqnarray}
hence the relevance of the functional of the maximum with $V(x) = 1/(2x)$ for this search problem. 
Note that Odlyzko found an expression of $c_0$ in terms of a complicated double integral which was then evaluated independently by Hwang \cite{Hwang1997} and Chassaing \cite{chassaing1999many} to be 
\begin{eqnarray}\label{c0}
c_0 = \sqrt{\frac{8}{\pi}} \log 2 = 1.1061\ldots\;.
\end{eqnarray}
Furthermore, Odlyzko found an algorithm, denoted as Od($n$), belonging to $A_n$, such that
\begin{eqnarray}\label{average_odlyzko}
\langle C({\rm Od}(n)) \rangle  = c_0 \sqrt{n} + o(\sqrt{n}) \;.
\end{eqnarray}  
Hence Eqs. (\ref{average_cost}) and (\ref{average_odlyzko}) show that Odlyzko's algorithm Od($n$) is on average asymptotically optimal. In a subsequent work, Chassaing, Marckert and Yor showed that Od($n$) is not only asymptotically optimal on average (\ref{average_odlyzko}) but also in distribution \cite{chassaing_yor}, which means the following. If one defines $\Phi_n(x)$ as
\begin{eqnarray}
\Phi_n(x) = \min_{a \in A_n} \Pr \left(\frac{C(a)}{\sqrt{n}} \geq x \right) \;,
\end{eqnarray}
then one has for any $x$ \cite{chassaing_yor}:
\begin{eqnarray}
\lim_{n \to \infty} \Phi_n(x) = \Pr(I \geq x) \;,
\end{eqnarray}
where $I$ is the random variable defined above in (\ref{def_I}). In Ref. \cite{chassaing_yor}, the authors studied the distribution of $I$ as well as its moments, using rather involved probabilistic methods. We will show here how these results can be derived simply using path integrals techniques.  

Yet another case of a functional of the maximum, ${\cal O}_{\max}(t)$, corresponds to the case where $V(x) \propto 1/\sqrt{x}$. In this case ${\cal O}_{\max}(t)$ 
describes the largest exit time of a particle, of unit mass, moving ballistically through a random potential on the segment $[0,1]$ (at zero temperature). Consider indeed a random Brownian potential $x(y)$ over a line segment $y\in [0,1]$. Imagine shooting a classical particle
of unit mass with fixed energy $E$ from the left of the segment at $y=0$. The energy conservation leads to
\begin{equation}
\frac{1}{2} \left(\frac{dy}{dt}\right)^2 + x(y) =E \;.
\label{energy_cons.1}
\end{equation}
Clearly, this classical particle can penetrate the region $y\in [0,1]$ if and only if its energy $E$ is
bigger than the maximum value of the potential $x(y)$ over $y\in [0,1]$, i.e., if $E> x_{\rm max}$
where $x_{\rm max}= {\max}_{0\le y\le 1}[x(y)]$. Now, imagine sending a beam of classical particles
with varying energy through this potential barrier. Only those particles with energy larger than $x_{\rm max}$
will go through the barrier. The time taken for such a penetrating particle to exit the region $y\in [0,1]$
through its right can be computed from Eq. (\ref{energy_cons.1}) as
\begin{equation}
{\mathcal T} (E)= \frac{1}{\sqrt{2}}\, \int_0^1 \frac{dy}{\sqrt{E-x(y)}} \;,
\label{crossing_time}
\end{equation}
where $E\ge x_{\rm max}$.
Clearly, the time needed by a penetrating particle to cross the region $y\in [0,1]$ depends on the energy $E$ of the particle. The slowest particle, i.e., the one that takes the longest time to cross, is the one
that has the lowest allowed energy to penetrate, i.e., the one with energy $E= x_{\rm max}$. Hence,
the maximum time needed by a particle to cross the barrier is given by

\begin{equation}
\label{time_ballistic}
{\mathcal T}_{\rm max}= {\max}_{E\ge x_{\rm max}} [{\mathcal T}(E)]= [{\mathcal T}(x_{\rm max})]= \frac{1}{\sqrt{2}}\, \int_0^1 \frac{dy}{\sqrt{x_{\rm max}-x(y)}} \;,
%
%
\end{equation}
which thus corresponds to a functional of the maximum of BM, ${\cal O}_{\max}(t=1)$, with $V(x) = 1/\sqrt{2x}$. In view of these two interesting physical examples in Eqs. (\ref{def_I}) and (\ref{time_ballistic}), it is rather natural to consider the family of functionals of the maximum of BM ${\cal O}_{\max}(t)$ with $V(x) = x^\alpha$ such that
\begin{eqnarray}\label{T_alpha}
T_\alpha(t) = \int_0^t \rmd\tau\, (x_{\max} - x(\tau))^\alpha \;,
\end{eqnarray}
indexed by a real $\alpha \in ]-2,\infty[$. Note that, by using the self-affinity of BM, $T_{\alpha}(t) \overset{\rm law}{=} t^{1+\alpha/2} T_{\alpha}(t=1)$. In particular for $\alpha = -1$ this corresponds to $I$ in (\ref{def_I}) while for $\alpha = -1/2$ this corresponds to ${\cal T}_{\max}$ in (\ref{time_ballistic}). On the other hand, for $\alpha = 1$, $T_{\alpha=1}(t)$ is called the area under a Brownian double meander \cite{JL07,MC2005b}. For other values of $\alpha$, $T_\alpha(t)$ generalizes these three cases.

Of course, one can consider similar observables as in (\ref{T_alpha}) for the Brownian bridge, i.e. when $x(\tau) \to x_B(\tau)$ (see Fig. \ref{fig_Brownian} (b)) and $x_{\max,B}$ denotes its maximum on the time interval $[0,t]$:
\begin{eqnarray}\label{T_alpha_B}
T_{\alpha,B}(t) = \int_0^t \, (x_{\max,B} - x_B(\tau))^\alpha \rmd\tau \;,
\end{eqnarray}
with, as above, $T_{\alpha,B}(t) \overset{\rm law}{=} t^{1+\alpha/2} T_{\alpha,B}(t=1)$. The simplest case is $\alpha = 1$, which corresponds to the area under a Brownian excursion: this can be easily seen by permuting the pre-minimum and the post-minimum part of a Brownian bridge, i.e. by using Vervaat's construction \cite{Ver1979}. Hence $T_{\alpha=1,B}(t)$ is distributed according to the Airy distribution, discussed above \cite{MC2005,MC2005b}. For $\alpha = -1/2$, this yields again the maximal time spent by particles to pass through a disordered periodic potential which is a Brownian bridge. Finally we conjecture, following the lines of reasoning of Refs. \cite{odlyzko,chassaing_yor}, that for $\alpha = -1$, i. e. the equivalent of $I$ in (\ref{def_I}) where $x(\tau)$ is replaced by $x_B(\tau)$ and $x_{\max}$ by $x_{\max,B}$, $T_{\alpha=-1,B}(t)$ yields the cost of the optimal algorithm to find the maximum of a random walk bridge (see Appendix \ref{odl_bridge}). As shown below, this random variable turns out to be related to the maximum of a Brownian excursion (see also Refs. \cite{BY87,CMY00}).  

\section{Summary of main results}

The goal of this work is to present various tools to study the statistics of such functionals ${\cal O}_{\max}(t)$ of the maximum of Brownian motion and its variants. It is useful to summarize the different approaches developed here as well as the main results obtained in the present paper. The first natural observable to compute is the average value of such functionals, $\langle {\cal O}_{\max} (t)\rangle$. According to Eq. (\ref{property_average}), the average can be obtained by computing the average density of states $\langle \rho(r,t)\rangle$. As we show here, this quantity can be calculated using rather elementary computations, based on the propagator of the Brownian motion with appropriate boundary conditions. In particular, this method allows us to recover in a very simple way the result for $c_0$ given in Eq. (\ref{c0}) and previously obtained in Refs. \cite{Hwang1997} and \cite{chassaing1999many} from the analysis of a rather complicated double integral. The expression $\langle \rho(r,t)\rangle$ for the free BM, as well as for the bridge, was recently announced
by us in a short Letter \cite{Perret_PRL}. In this paper, in addition to providing details of these computations, we
extend our techniques in several directions obtaining many new results. In particular, we show that this method, relying on propagators, can    
be easily adapted to a variety of other constrained Brownian motions, including the excursion, the meander, as well as the reflected BM and the reflected Brownian bridge. The main characteristics of $\langle \rho(r,t)\rangle$  for these various constrained BM are summarized in Tables \ref{summary_rho_asympt} and \ref{summary_rho} below. From it, we obtain in particular from (\ref{property_average}) the average of the functional $T_\alpha(t)$ for the free BM, with $\alpha \in ]-2,+\infty[$ as
\begin{eqnarray}\label{intro_Talpha_average}
\langle T_\alpha(t) \rangle =\frac{(2t)^{1+\frac{\alpha}{2}}(2-2^{-\alpha})\Gamma\left(\frac{1+\alpha}{2}\right)}{(2+\alpha)\sqrt{\pi}}\,,
\end{eqnarray}
from which we get in particular the constant $c_0$ in Eq. (\ref{def_I}) as $c_ 0 = T_{\alpha=-1}(t=1)/2 = \sqrt{8/\pi} \log 2 = 1.1061 \ldots$ as given in Eq.~(\ref{c0}). For $\alpha = 1$, one recovers $\langle T_{\alpha=1}(t=1) \rangle = \sqrt{2/\pi}$ \cite{MC2005b} for the average area under a Brownian double meander (see Fig.~\ref{brown_meandre}). As a function of $\alpha$ it has an interesting non-monotonic behavior, diverging when $\alpha \to -2$ as $\langle T_\alpha(t=1) \rangle \sim 4/(2+\alpha)$ as well as when $\alpha \to \infty$ as $\langle T_\alpha(t=1) \rangle \sim 4\sqrt{2}\left(\alpha/e\right)^{\alpha/2}/\alpha$, exhibiting a minimum for $\alpha \approx 1.148$. Similarly, for the bridge one obtains
\begin{eqnarray}\label{intro_Talpha_BB_average}
\langle T_{\alpha,B}(t) \rangle  = \frac{t^{1+\alpha/2}}{2^{\alpha/2}}\Gamma\left(1+\frac{\alpha}{2}\right) \;.
\end{eqnarray}
In particular, for $\alpha = 1$ it yields back the first moment of the Airy-distribution, $\langle T_{\alpha=1,B}(t=1) \rangle = \sqrt{\pi/8}$ \cite{Takacs91,Takacs95} while for $\alpha = -1$ it gives the equivalent of $c_0$ in Eq. (\ref{def_I}) for the Bridge, $c_{0,B} = {T_{\alpha=-1,B}(t=1)}/{2} = \sqrt{\pi/2} = 1.25331 \ldots$. Interestingly, this means that, on average, the cost of Odlyzko's algorithm is higher for the BB than for the free BM. This can be roughly understood through the fact that the DOS close to the maximum is slightly higher for the BB, which is pinned to the origin on both sides of the time interval, than for the BM which is free on one side. As a function of $\alpha$, $\langle T_{\alpha,B}(t) \rangle$ is also non-monotonic diverging when $\alpha \to -2$ as $\langle T_{\alpha,B}(t=1) \rangle \sim 4/(2+\alpha)$ as well as when $\alpha \to \infty$ as $\langle  T_{\alpha,B}(t=1) \rangle \sim 2^{-\alpha} \left(\alpha/e\right)^{\alpha/2} \sqrt{\pi \alpha}$, with a minimum for $\alpha \approx 2.960$. 

\begin{figure}[ht]
\begin{center}
\resizebox{120mm}{!}{\includegraphics{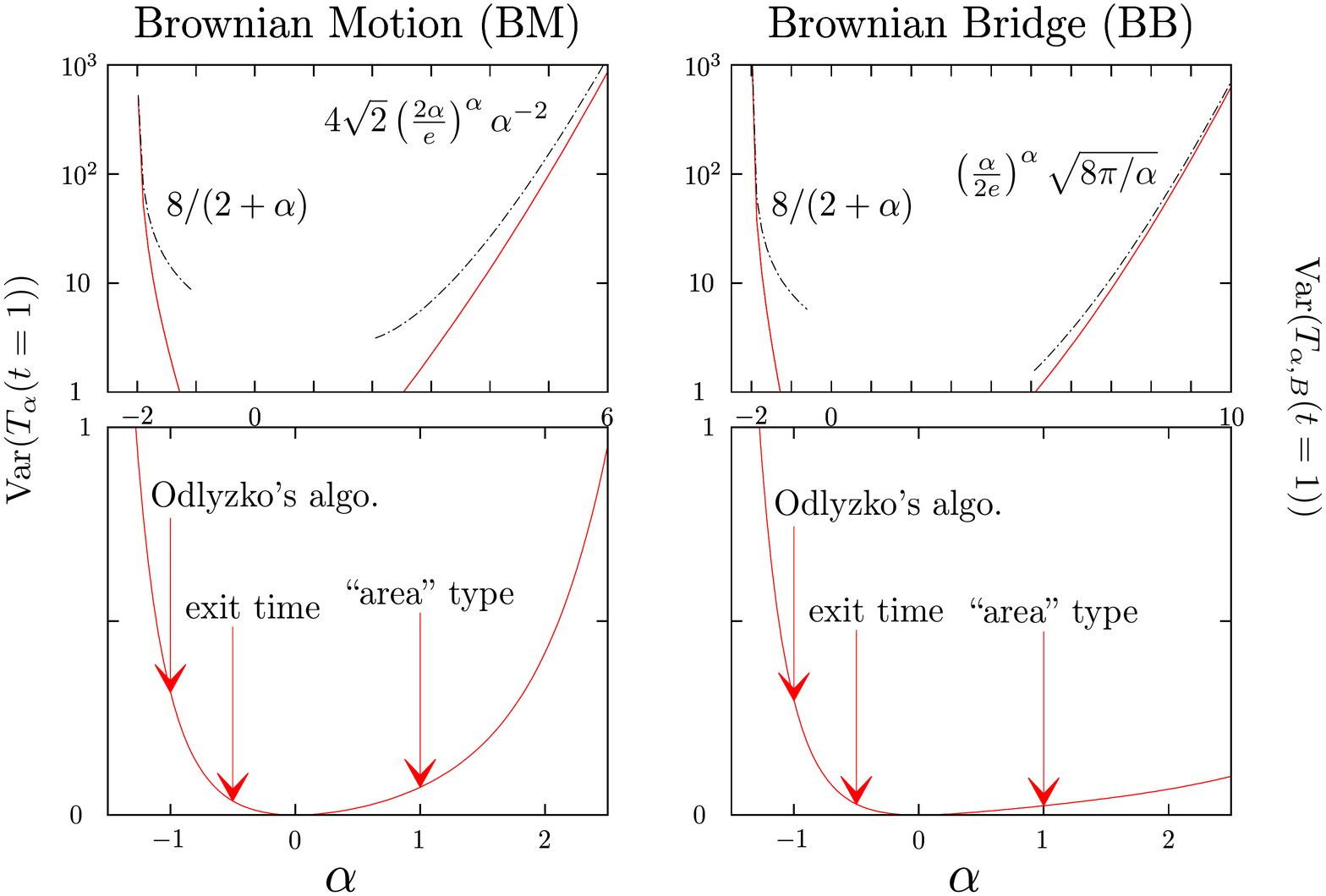}}
\caption{{\bf Left panel (bottom):} Plot of the variance ${\rm Var}(T_{\alpha}(t=1))=\langle T_{\alpha}^2(t=1) \rangle-\langle T_{\alpha}(t=1) \rangle^2$ as a function of $\alpha$ for the BM, as given in Eq. (\ref{second_momentBM}). {\bf Left panel (top):} Focus on the asymptotic behaviors of ${\rm Var}(T_{\alpha}(t=1))$ for $\alpha \to -2$ and $\alpha \to \infty$. The dashed lines indicate the asymptotic behaviors given in the text below Eq. (\ref{second_momentBM}). 
 {\bf Right panel (bottom):} same quantity for the Brownian bridge, ${\rm Var}(T_{\alpha,B}(t=1))$, as given in Eq. (\ref{second_momentBB}). {\bf Right panel (top):} Focus on the asymptotic behaviors of ${\rm Var}(T_{\alpha,B}(t=1))$ for $\alpha \to -2$ and $\alpha \to \infty$. The dashed lines indicate the asymptotic behaviors given in the text below Eq. (\ref{second_momentBB}). The functionals $T_{\alpha}(t)$ and $T_{\alpha,B}(t)$ interpolate between different interesting observables: for $\alpha = -1$ they describe the cost of Odlyzko's optimal algorithm to find the maximum of a discrete random walk, for $\alpha = -1/2$, they describe 
the largest exit time of a particle through a Brownian potential and for $\alpha = 1$, $T_{\alpha=1}(t)$ and $T_{\alpha=1,B}(t)$ are random variables of the  
``area'' type. For the bridge (right panel), this corresponds to the area under the BE (Airy random variable).}
\label{figure_DTalpha}
\end{center}
\end{figure}

If one is interested not only in the average of the functional $T_\alpha(t)$ but also in higher cumulants or even in its full distribution, the knowledge of the
average DOS is obviously not enough and these computations based on propagators become cumbersome. Instead, we present here a method,  based on path-integral which allows us to compute its Laplace transform, namely $\langle e^{- \lambda {\cal O}_{\max}(t) }\rangle$, from which the moments of arbitrary order can be obtained by differentiation with respect to (wrt) $\lambda$. In some cases, this can also allow us to compute the full distribution of the functional. We first illustrate this method on the DOS itself, for which $V(x) = \delta(x-r)$. The main results in this case were recently announced in Ref. \cite{Perret_PRL}, without any details, which we thus provide here. Then we use this general formalism to study the special functionals $T_\alpha(t)$ (\ref{T_alpha}) as well as $T_{\alpha,B}(t)$ in (\ref{T_alpha_B}) for the BB. In particular, we show how this method allows us to compute the second moment, and eventually the variance ${\rm Var}(T_\alpha(t)) = \langle T_\alpha^2(t) \rangle - \langle T_\alpha(t) \rangle^2$ for arbitrary $\alpha \in ]-2, +\infty[$ under the form:
\begin{eqnarray}\label{second_momentBM}
{\rm Var}(T_\alpha(t))=\frac{t^{2 + \alpha}}{(2^{3 \alpha} (\alpha+2)^2} &&\left(\frac{(2^\alpha-1)(2^{\alpha+1}-1)\Gamma(\alpha+3)}{(\alpha+1)^2}-\frac{2^{2\alpha+2}(2^{\alpha+1}-1)^2 \Gamma(\frac{\alpha+1}{2})^2}{\pi}  \right.\\
&&\hspace{0.cm}\left.
+\frac{(\alpha+2)\Gamma(2\alpha+2)(2^{2\alpha+2}-2^{\alpha+1}(\alpha+1)B_{1/2}(\alpha+2,-2(\alpha+1))-1)}{(\alpha+1)^2\Gamma(\alpha+1)}
\right) \,, \nonumber
\end{eqnarray}
where $B_z(a,b) = \int_0^z t^{a-1} (1-t)^{b-1} \,\rmd t$ is the incomplete beta function. Formula (\ref{second_momentBM}) yields in particular ${\rm Var}(T_{\alpha=-1}(t=1)) = \frac{\pi^2}{3}+ (4-32/\pi) \log(2)^2$, as obtained previously in Ref. \cite{chassaing_yor}, using probabilistic methods. For $\alpha =1$, Eq. (\ref{second_momentBM}) yields ${\rm Var}(T_{\alpha=1}(t=1)) = 17/24-2/\pi$ \cite{MC2005b}. Interestingly, as a function of $\alpha$, it has a non monotonic behavior. It is diverging when $\alpha \to -2$ as ${\rm Var}(T_\alpha(t=1))  \sim 8/(\alpha+2)$ as well as when $\alpha \to \infty$ as ${4\sqrt{2}}(2\alpha/e)^\alpha \alpha^{-2}$ exhibiting a single minimum for $\alpha = 0$. In Fig. \ref{figure_DTalpha} (left panel), we show a plot of ${\rm Var}(T_{\alpha}(t=1))$ as a function of $\alpha$.

Similarly, one can also compute the second moment in the case of the bridge, yielding the variance ${\rm Var}(T_{\alpha,B}(t)) = \langle T_{\alpha,B}^2(t)\rangle - \langle T_{\alpha,B}(t)\rangle^2$:
\begin{eqnarray}\label{second_momentBB}
{\rm Var}(T_{\alpha,B}(t)) = t^{\alpha+2}\left[\frac{\sqrt{\pi}(\Gamma(2\alpha+3)-\Gamma(\alpha+2)^2)}{(\alpha+1)^2\Gamma(\alpha+\frac32)2^{3\alpha+1}}-\frac{\Gamma(\frac{\alpha}{2}+1)^2}{2^\alpha}\right] \;.
%
%
\end{eqnarray}
In particular, we can check that ${\rm Var}(T_{\alpha=1,B}(t=1)) = {5}/{12} - \pi/8$ which is the variance of the Airy distribution \cite{MC2005,MC2005b,Takacs91,Takacs95}, while for $\alpha = -1$, one has ${\rm Var}(T_{\alpha=-1,B}(t=1)) = 2 \pi^2/3 - 2 \pi$. As a function of $\alpha$ it has also a non-monotonic behavior, diverging when $\alpha \to -2$  as ${\rm Var}(T_{\alpha,B}(t=1)) \sim 8/(\alpha+2)$ and when $\alpha \to \infty$ as $(\frac{\alpha}{2e})^\alpha \sqrt{8 \pi/\alpha}$, exhibiting a minimum for $\alpha = 0$. In Fig. \ref{figure_DTalpha} (right panel), we show a plot of ${\rm Var}(T_{\alpha,B}(t=1))$ as a function of $\alpha$.

Finally, in the special case $\alpha = -1$, which corresponds to the cost of the Odlyzko's algorithm, we are able to compute exactly the moments of arbitrary order, both for the free BM, $\langle T^k_{\alpha=-1}(t=1)\rangle$ and for the bridge $\langle T^k_{\alpha=-1,B}(t=1)\rangle$. One obtains indeed,
\begin{eqnarray}\label{moments_BM_arbitrary}
\langle T^k_{\alpha=-1}(t=1)\rangle = \Gamma\left(\frac{k+1}{2} \right) \frac{2^{\frac{k}{2} + 2}}{\sqrt{\pi}} \sum_{m=0}^k \tilde \zeta(m) \tilde \zeta(k-m) \;, \; \tilde \zeta(m) = (1 - 2^{1-m}) \zeta(m) = \sum_{n \geq 1} \frac{(-1)^{n+1}}{n^m} \;,\hspace{0.5cm}
\end{eqnarray}
recovering, using a completely different method, the result of Chassaing, Marckert and Yor \cite{chassaing_yor}. For $k=1$ and $k=2$, this formula (\ref{moments_BM_arbitrary}) yields back the results for the first moment (\ref{intro_Talpha_average}) and for the variance~(\ref{second_momentBM}).

For the bridge, we obtain the result, for any real $k$
 \begin{eqnarray}\label{moments_BB_arbitrary}
\langle T^k_{\alpha=-1,B}(t=1)\rangle = - 2^{1+\frac{k}{2}} \pi^{k-\frac{1}{2}} \, k \, \Gamma\left(\frac{3}{2} - \frac{k}{2} \right) \zeta(1-k) \;.
\end{eqnarray}
Of course for $k=1$ and $k=2$, this formula (\ref{moments_BB_arbitrary}) yields back the aforementioned results for the first moment (\ref{intro_Talpha_BB_average}) and the variance (\ref{second_momentBB}). In Ref. \cite{chassaing_yor}, the authors obtained the full probability distribution function (PDF) of $T_{\alpha=-1}(t=1)$ in terms of a convolution of two theta-functions. Here, we obtain the full PDF $p_B(s)$ of $T_{\alpha=-1,B}(t=1)$ in the case of the BB, as:
\begin{eqnarray}\label{pB_intro}
p_B(s) &=& \frac{d}{ds} \left(2\sum_{m=0}^\infty (1-m^2 s^2) e^{-\frac{m^2 s^2}{2}}\right) =  \frac{d}{ds} \left(\frac{8 \sqrt{2} \pi^{5/2}}{s^3} \sum_{m=1}^\infty m^2 e^{-\frac{2 m^2 \pi^2}{s^2}} \right) \;,
\end{eqnarray}  
where the two formulas are related to each other via the Poisson summation formula. Interestingly, in Eq. (\ref{pB_intro}), one actually recognizes the PDF of the maximum of a Brownian excursion on the unit time interval, $x_{\max,E} = \max_{0\leq \tau \leq 1} x_E(\tau)$. One has indeed
\begin{eqnarray}\label{relation_max_excursion}
p_B(s) = \frac{d}{ds} {\Pr}\left(x_{\max,E} \leq \frac{s}{2}\right) \;.
\end{eqnarray}
As explained below, one can show that $T_{\alpha=-1,B}(1) \overset{\rm law}{=} \int_0^1 {\rm d}\tau/x_E(\tau)$. Hence, Eqs. (\ref{pB_intro}) and (\ref{relation_max_excursion}) is a manifestation of a non-trivial identity in law for the Brownian excursion $x_E(\tau)$~\cite{BY87,CMY00}:
\begin{eqnarray}\label{identity_max}
\int_0^1 \frac{{\rm d} \tau}{x_E(\tau)} \overset{\rm law}{=} 2 \max_{0 \leq \tau \leq 1} x_E(\tau) \;.
\end{eqnarray}
Therefore, our result for $T_{\alpha=-1,B}(t)$ in Eq. (\ref{pB_intro}) provides a simple derivation of this non-trivial identity (\ref{identity_max}), which was proved in Refs. \cite{BY87,CMY00} using rather involved probabilistic tools. 

The paper is organized as follows. In section \ref{section_propagator}, we focus on a method based on "counting paths", using 
propagators of BM. We first illustrate this approach to compute the average DOS for BM $\langle \rho(r,t)\rangle$ in subsection \ref{DOS_BM_section} and then extend it to the Brownian bridge in section \ref{DOS_BB_section} as well as to other constrained BMs in 
sections \ref{DOS_BME_section} and \ref{DOS_oBM_section} and Appendix~\ref{DOS_BR_section}. In section \ref{DOS_oBM_section}, we also present a comparison between our exact results with numerical simulations, the details of which are given in Appendix \ref{simulation_BM_section}.
In section~\ref{pathintegral_section}, we develop a general path integral approach to compute functionals of the maximum of Brownian motion and Brownian bridge. Within this framework, which is presented in detail in section \ref{generalframework_section},  
we study more specifically the full statistics of the DOS in section \ref{DOSstat_section}, and the family of functional $T_\alpha(t)$ [see Eq.~(\ref{T_alpha})] in section \ref{Talpha_section}. We also analyze more precisely the case $\alpha=-1$ in section \ref{Tminus1_section} which is relevant to study Odlyzko's algorithm. Some technical details, including a description of the main ideas behind Odlyzko's algorithm, have been left in Appendices.



\section{The method of propagators}\label{section_propagator}

In this section, we present a rather simple method, based on "counting paths", to compute the average DOS $\langle \rho(r,t)\rangle$. In the cases of BB and BE (which turn out to be identical as we show below), this method allows us also to compute higher moments $\langle \rho^k_{B}(r,t)\rangle$ with $k$ an arbitrary integer. We illustrate the method in detail for the case of free BM and then apply it to various constrained BM: the Brownian bridge and Brownian excursion in section \ref{DOS_BB_section} and the Brownian meander in section \ref{DOS_BME_section}. In Appendix \ref{DOS_BR_section} we use the method, for completeness, for the reflected BM and for the reflected BB.

\subsection{The average DOS for free Brownian motion}\label{DOS_BM_section}

\begin{figure}[ht]
\begin{center}
\resizebox{100mm}{!}{\includegraphics{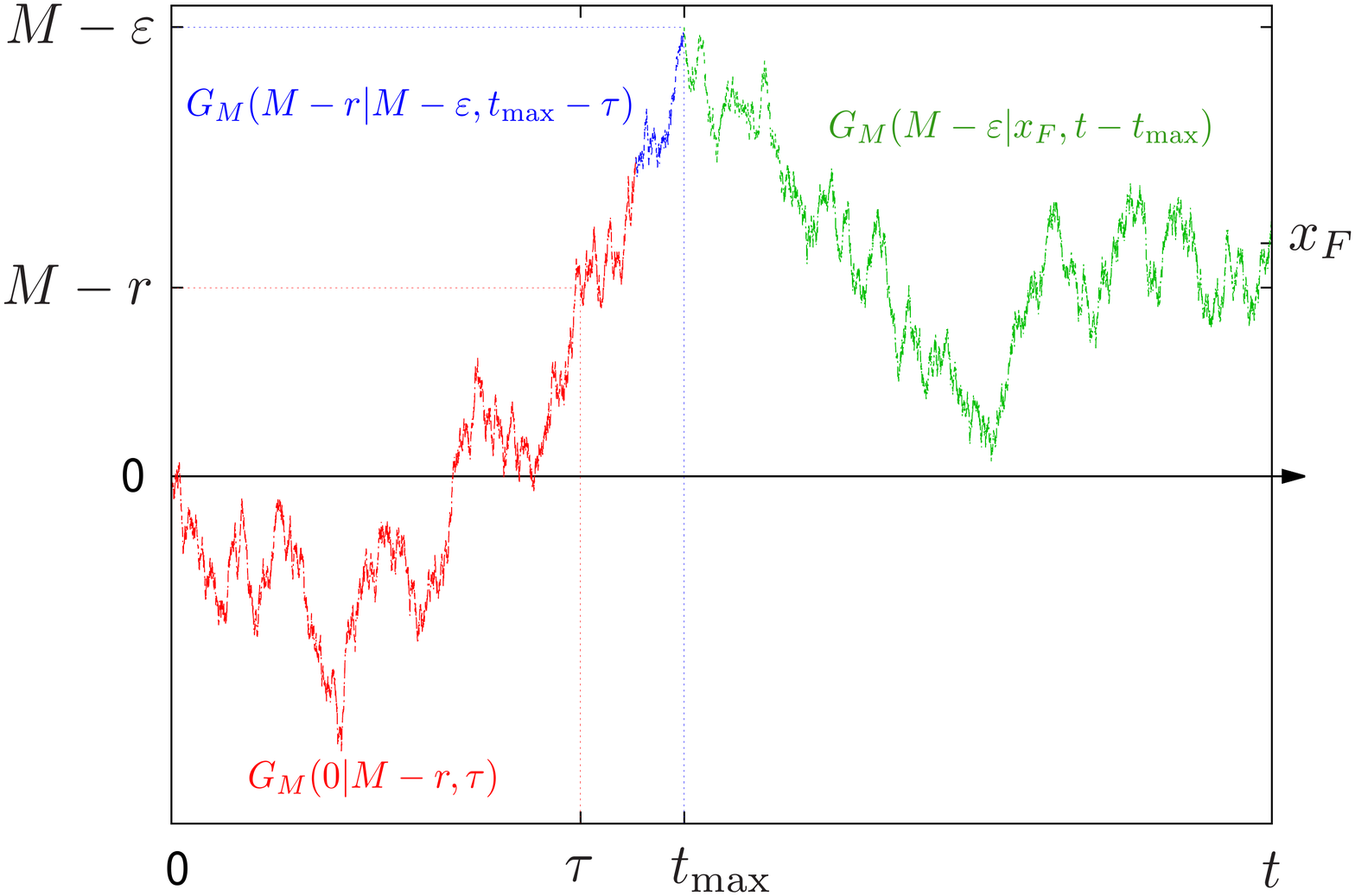}}
\caption{(Color online) Illustration of the method of propagators. The BM starts at 0 and visits the point $M-r$ at time $\tau$, passes through $M-\varepsilon$ at time $t_{\max}$ and ends at $x_F$ at time $t$.}
\label{figure_propagator}
\end{center}
\end{figure}

\begin{figure}[ht]
\begin{center}
\rotatebox{0}{\resizebox{100mm}{!}{\includegraphics{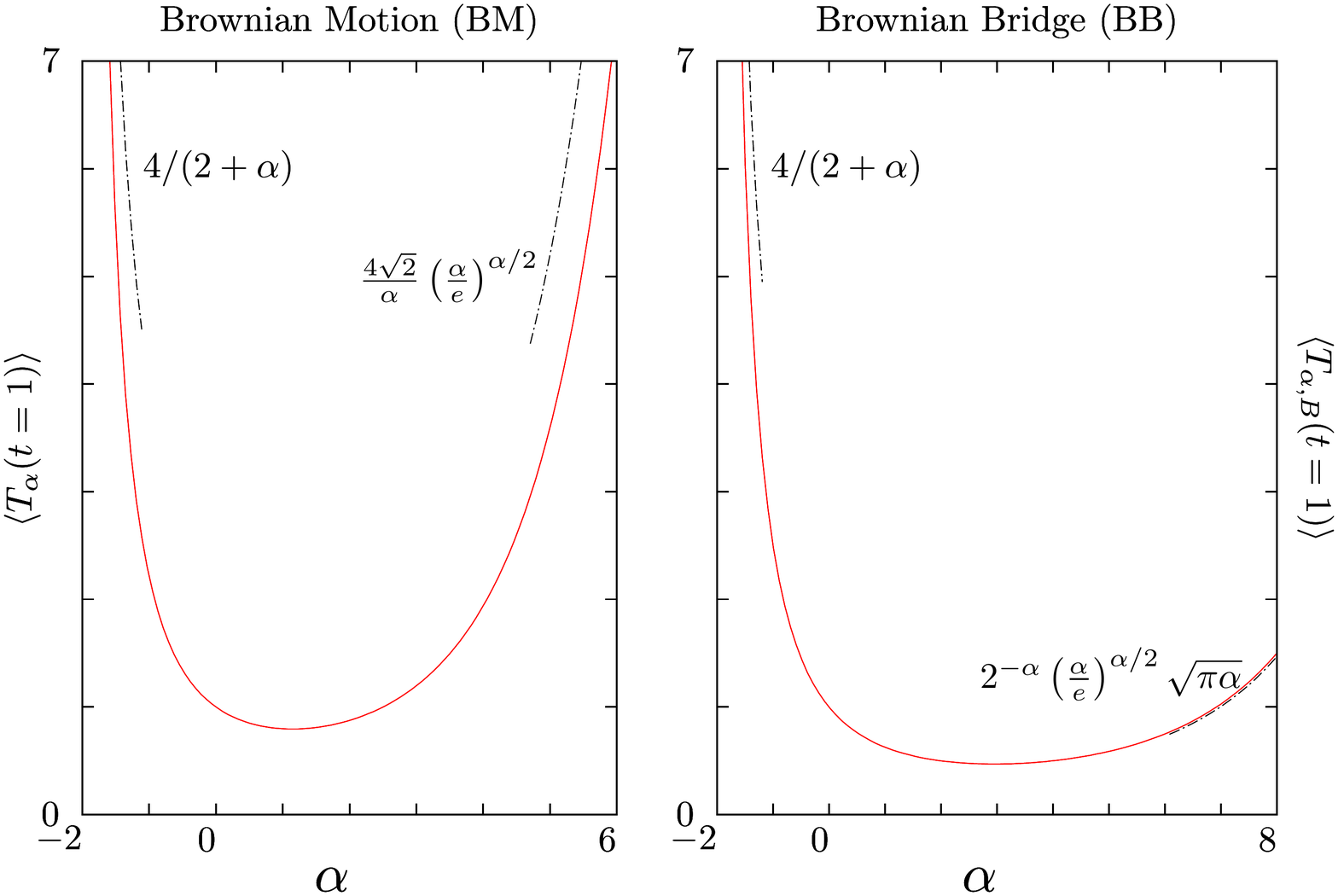}}}
\caption{{\bf Left panel:} Plot of $\langle T_{\alpha}(t=1) \rangle$, as a function of $\alpha$, for the BM, as given in Eq. (\ref{Talpha_average}). {\bf Right panel:} Plot of $\langle T_{\alpha,B}(t=1) \rangle$, as a function of $\alpha$, for the BB, as given in Eq. (\ref{first_moment_DOS}). In both panels, the dashed black lines indicate the asymptotic behaviors discussed in the introduction. In particular, for $\alpha = -1$ for the BM (corresponding to Odlyzko's algorithm), we recover $\langle T_{\alpha=-1}(t=1) \rangle = 4 \sqrt{2/\pi} \log 2$ \cite{Hwang1997,chassaing1999many}. For $\alpha = 1$ for the BB (corresponding to the Airy distribution), we recover $\langle T_{\alpha=1,B}(t=1) \rangle = \sqrt{\pi/8}$ \cite{MC2005,MC2005b}.}
\label{figure_Talpha}
\end{center}
\end{figure}

To compute $\langle \rho(r,t) \rangle$, we simply average Eq. (\ref{def_rho}) over the trajectories of the BM and write it as
\begin{eqnarray}\label{ave_rho_1}
\langle \rho(r,t) \rangle = \int_0^t  \langle \delta(x_{\max}-x(\tau)-r) \rangle \, {\rm d}\tau \;.
\end{eqnarray}
In Eq. (\ref{ave_rho_1}) we recognize that the integrand $\langle \delta(x_{\max}-x(\tau)-r) \rangle$ has a simple probabilistic interpretation. Indeed, one has 
\begin{eqnarray}\label{ave_rho_2}
\langle \delta(x_{\max}-x(\tau)-r) \rangle {\rm dr} = \Pr \left[ x(\tau) \in [x_{\max}-r-{\rm d}r, x_{\max}-r] \right] \;.
\end{eqnarray} 
The idea to compute the PDF $\langle \delta(x_{\max}-x(\tau)-r) \rangle$ in (\ref{ave_rho_2}) is to evaluate the "number" of paths that reach their maximum $M$ at time $t_{\max}$, pass through $M-r$ at time $\tau$, and end at $x_F \leq M$ at time $t$. The total number which we want to compute is then obtained by
integrating over $x_F, M$ and $t_{\max}$ (see Fig. \ref{figure_propagator}). In each of the three time intervals delimited by $\tau, t_{\max}$ and $t$, the BM is constrained to stay below $M$. This number of paths can thus be computed from the propagator $G_{M}(\alpha|\beta,t)$ of the BM, starting at time $0$ at $x= \alpha < M$ and arriving at time $t$ at $x = \beta<M$ and staying below $M$ during the whole time interval $[0,t]$. It can be easily computed using, for instance, the method of images:
\begin{eqnarray}\label{propag_free_below_M}
G_{M}(\alpha|\beta,t) = \frac{1}{\sqrt{2 \pi t}} \left( e^{-\frac{(\beta-\alpha)^2}{2t}}-e^{-\frac{(2M-\beta-\alpha)^2}{2t}} \right) \;.
\end{eqnarray}
For future purpose, it is useful to compute its Laplace transform (LT) wrt $t$, $\tilde G_{M}(\alpha|\beta,s)$. It is given by
\begin{eqnarray}
\tilde G_{M}(\alpha|\beta,s) = \int_0^\infty e^{-s t} G_{M}(\alpha|\beta,t) \, \rmd t  = \frac{1}{\sqrt{2s}} \left(e^{-\sqrt{2s} |\beta - \alpha|}-e^{-\sqrt{2s} (2M-\beta - \alpha)}\right) \;. \label{Propagator_Laplace}
\end{eqnarray}
When dividing the time interval $[0,t]$ into three parts as in Fig.~\ref{figure_propagator}, two cases may arise: $\tau < t_{\max}$ or $\tau > t_{\max}$. One can show that these two configurations give rise to the same contributions to $\langle \rho(r,t) \rangle$: this can be seen by making a global shift $x(\tau) \to x(\tau) - x_F/2$ (for each realization of BM) and using the time reversal symmetry. Note also that, when dealing with BM which is continuous both in space and time, one can not impose simultaneously $x(t_{\max}) = M$ and $x(t) < M$ right before and after $t_{\max}$. To circumvent this difficulty, one imposes instead $x(t_{\max}) =M-\varepsilon$, and take the limit $\varepsilon \to 0$ at the end of the calculation.
Finally, $\langle \rho(r,t) \rangle$ is given by the ratio of paths which have the analyzed properties to the normalization constant $Z(\varepsilon)$ which count the same paths without the condition to pass through $M-r$ at time $\tau$
\begin{eqnarray}
\langle \rho(r,t) \rangle \,  =\underset{\varepsilon \to 0}\lim \frac{2}{Z(\varepsilon)}\int_{0}^\infty \rmd M \int_0^t \rmd t_{\max} \int_{-\infty}^M\rmd x_F  \int_{0}^{t_{\max}} \rmd \tau && 
{\color{red}G_{M}(0 | M-r,\tau)}   {\color{blue}G_{M}(M-r | M -\varepsilon,t_{\max}-\tau )} \nn\\
&&\times {\color[rgb]{0.20,0.33,0.09}G_{M}(M-\varepsilon|x_F,t-t_{\max})},\,\, \label{eq:Tmax_via_Markov}
\end{eqnarray}
where we have used the Markov property of BM and where the factor of $2$ comes from the two aforementioned equivalent situations corresponding $\tau<t_{\max}$ and $\tau > t_{\max}$. In (\ref{eq:Tmax_via_Markov}) the normalization constant $Z(\varepsilon)$ is given by
\begin{eqnarray}
Z(\varepsilon)=\int_{0}^\infty \rmd M \int_0^t \rmd t_{\max} \int_{-\infty}^M\rmd x_F \,G_{M}(0 | M-\varepsilon, t_{\max}) \, G_{M}(M-\varepsilon | x_F,t-t_{\max}) \;.
\end{eqnarray} 
The normalization is easily computed as 
$Z(\varepsilon) \sim \, 2 \varepsilon^2$ when $\varepsilon \to 0$. In (\ref{eq:Tmax_via_Markov}), we recognize a convolution structure. Taking the LT wrt to $t$, we find
\begin{eqnarray}
&&\int_0^\infty {\rmd}t e^{-st}\langle \rho(r,t)\rangle \,  =\\
&&\lim_{\varepsilon \to 0}\frac{2}{{Z(\varepsilon)}} \int_0^{\infty} \rmd M \int_{-\infty}^{M} \rmd x_F \,
{\tilde G_{M}(0 | M-r,s)}   
 {\tilde G_{M}(M-r | M -\varepsilon,s)}  {\tilde G_{M}(M-\varepsilon|x_F,s)} \;.\hspace{0.0cm} \nonumber
\end{eqnarray}
Using (\ref{Propagator_Laplace}) and performing a small $\varepsilon$ expansion, one obtains straightforwardly  
\begin{eqnarray}
&&\int_0^\infty {\rmd}t e^{-st} \langle \rho(r,t)\rangle \,  = 8 
 \Big[ \int_0^{r} \rmd M   \frac{e^{-\sqrt{2s} r}}{\sqrt{2s}} \sinh{(\sqrt{2s}M)} e^{-\sqrt{2s}r} \int_{-\infty}^{M} \rmd x_F e^{-\sqrt{2s}(M-x_F)}  \nonumber \\
&&+ \int_r^{\infty} \rmd M  \frac{e^{-\sqrt{2s} M}}{\sqrt{2s}} \sinh{(\sqrt{2s}r)} e^{-\sqrt{2s}r} \int_{-\infty}^{M} \rmd x_F e^{-\sqrt{2s}(M-x_F)}\Big ]\,.
\end{eqnarray}
Performing the remaining integrals over $x_F$ and $M$ we obtain 
%
%
%
%
\begin{equation}\label{free_BM_LT}
\int_0^\infty {\rmd}t e^{-st} \langle \rho(r,t)\rangle \, = 8 \frac{e^{-\sqrt{2s} r}-e^{-2\sqrt{2s} r}}{(2s)^{3/2}} \,.
\end{equation} 
By inverting the above LT (\ref{free_BM_LT}) we finally obtain
\begin{eqnarray}\label{rho_BM_intro}
\langle \rho(r,t)\rangle = \sqrt{t} \overline{\rho} \left(\frac{r}{\sqrt{t}}\right) \,, \, \overline{\rho}(r) = 
8 \left(\Phi^{(2)}(r) - \Phi^{(2)}(2r) \right) \;, \, \Phi^{(2)}(r)=\frac{e^{-\frac{r^2}{2}}}{\sqrt{2\pi}} -\frac{r}{2} \text{erfc}\left(\frac{r}{\sqrt{2}}\right),\;\hspace{0.0cm}
\end{eqnarray}
where $\Phi^{(2)}$ belongs to a useful hierarchy of functions, $\Phi^{(j)}$, as explained in Appendix \ref{useful_functions}. From the average value of the DOS, one can compute the average value of any functional of the maximum, according to (\ref{property_average}). In particular, for the special case $V(x) = x^\alpha$, one obtains from (\ref{property_average}) and (\ref{rho_BM_intro}):
\begin{eqnarray}\label{Talpha_average}
\langle T_{\alpha}(t) \rangle &=& \langle \int_0^t [x_{\max} - x(\tau)]^{\alpha} {\rmd \tau} \rangle = \int_0^\infty \langle \rho(r,t)\rangle r^\alpha \,\rmd r  \nonumber \\
&=& 8t^{1+\alpha/2} \int_0^\infty r^{\alpha} \left(\Phi^{(2)}(r)-\Phi^{(2)}(2r)\right) \,{\rmd} r 
= \frac{(2t)^{1+\frac{\alpha}{2}}(2-2^{-\alpha})\Gamma(\frac{1+\alpha}{2})}{(2+\alpha)\sqrt{\pi}}\,,
\end{eqnarray}
as announced in the introduction in Eq. (\ref{intro_Talpha_average}). In Fig. \ref{figure_Talpha} (left panel), we show a plot of $\langle T_\alpha(t) \rangle$ as a function of $\alpha$. Specifying this formula (\ref{Talpha_average}) to the case $\alpha = - 1$, one obtains $c_0 = T_{\alpha=-1}(t=1)/2 = \sqrt{\frac{8}{\pi}} \log 2$ as announced in (\ref{c0}), recovering in a rather simple way the result of Refs.~\cite{Hwang1997,chassaing1999many}.
 
This method can in principle be adapted to compute higher moments of the DOS, $\langle \rho^k(r,t)\rangle$, but such computations involve a rather cumbersome combinatorial analysis, for $k > 1$. A more powerful approach amounts instead to compute the generating function of $\rho(r,t)$ \cite{Perret_PRL} -- as shown in section \ref{pathintegral_section}. As we show now, this method can also be easily extended to compute the average DOS for various constrained BM.


\subsection{The average DOS and its higher moments for Brownian bridges and Brownian excursions}\label{DOS_BB_section}

\begin{figure}[ht]
\begin{center}
\resizebox{100mm}{!}{\includegraphics{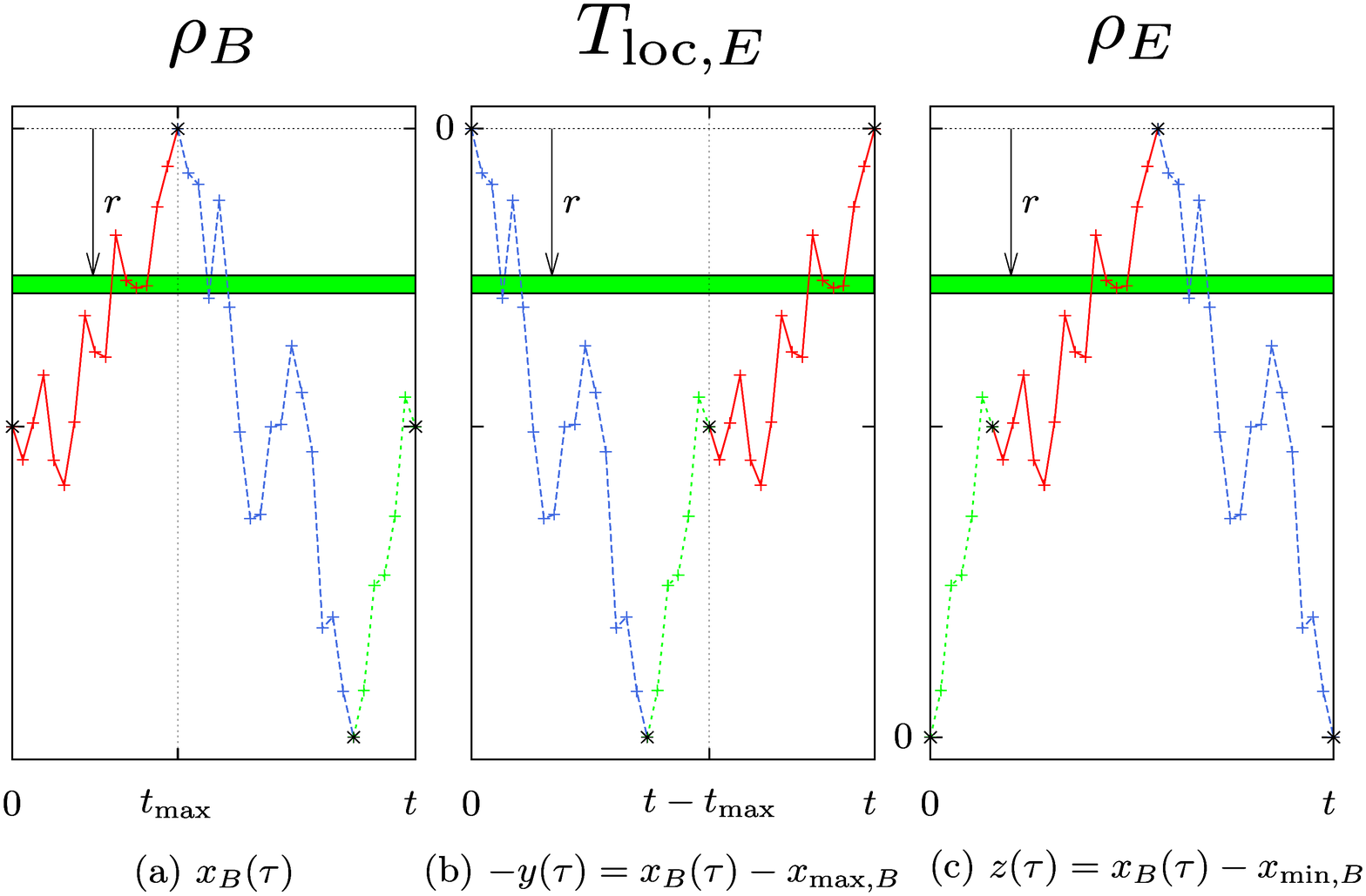}}
\caption{(Color online) Illustration of the identities in law given in Eq. (\ref{summary_vervaat}) in the text relating 
the DOS near the maximum for the BB to the local time, on the one hand, and to the DOS, on the other hand, for the BE. {\bf (a)} A typical trajectory of a BB, $x_B(\tau)$. The process spends a time $\rho_{B}(r,t)\rmd r$ in the green stripe $[x_{\max,B}-r-\rmd r,x_{\max,B}-r]$. {\bf (b)} The process $-y(\tau) = x_{\max,B} - x_B(\tau)$ after the Vervaat's transformation explained in the text is an excursion. The time spent in the green stripe is now given by $T_{{\rm loc},E}(r,t) \rmd r$ for this new process $y(\tau)$. {\bf (c)} The process $z(\tau) = x_B(\tau) - x_{\min,B}$ after the transformation explained in the text is also an excursion. The time spent in the green stripe is now given by $\rho_{E}(r,t) \rmd r$ for this new process.}
\label{Vervaat}
\end{center}
\end{figure}

For Brownian bridges, as well as Brownian excursions, the method based on propagators of constrained BM, allows us to compute arbitrary moments of the DOS, $\langle \rho_{B}^k(r,t) \rangle$. An expression for these moments can be written from the definition [see Eq. (\ref{def_rho}) where $x(\tau)$ is replaced by the BB $x_B(\tau)$] as
\begin{eqnarray}\label{DOS_starting_BB}
\langle \rho_{B}^k(r,t) \rangle=\int_0^t...\int_0^t  \langle \prod_{i=1}^{k} \rmd t_i\,  \delta(x_{\max,B}-x_B(t_i)-r) \rangle \;,
\end{eqnarray}
such that in (\ref{DOS_starting_BB}) the BB visits $k$ times the point $M-r$ at successive times $t_1, t_2 ,\cdots, t_k$ where $M$ is the value of the maximum on $[0,t]$.  
To compute this quantity (\ref{DOS_starting_BB}) for the BB, it is useful to invoke the Vervaat's construction to relate the DOS of the Brownian bridge to the {\it local time} of the Brownian excursion (see Fig. \ref{Vervaat}). 

The Vervaat's construction works as follows~\cite{Ver1979}: we denote by $t_{\max}$ the time at which the BB reaches its maximum on $[0,t]$ (see Fig.~\ref{Vervaat} (a)) and break the time interval $[0,t]$ into two parts, $[0,t_{\max}]$ and $[t_{\max},t]$ as illustrated in Fig.~\ref{Vervaat}~(a). Then we permute the two associated portions of the path, the continuity of the path being guaranteed by $x_B(t) = x_B(0) = 0$ for the BB. We can further transform the path by considering $y(\tau) = x_{\max,B} - x_B(\tau)$ and take finally the origin of times at $t_{\max}$: this yields the configuration shown in Fig.~\ref{Vervaat}~(b). Furthermore, if we initially break the time interval into $[0,t_{\min}]$ and $[t_{\min},t]$, where $t_{\min}$ denotes the time at which the BB reaches its minimum on $[0,t]$, we permute the two associated portions of the path and finally take the origin of times at $t_{\min}$, we obtain the configuration shown in Fig.~\ref{Vervaat} (c). 

These transformations show that $\rho_{E}(r,t)$ for an excursion is identical in law to the local time in $r$ for an excursion, $T_{{\rm loc},E}(r,t)$ and hence also identical in law to the DOS of the BB. Hence the Vervaat's construction allows to us write the following identities:
\begin{eqnarray}\label{summary_vervaat}
\rho_{B}(r,t)  \overset{\rm law}{=}  \rho_{E}(r,t) \overset{\rm law}{=} T_{{\rm loc},E}(r,t) = \int_0^t \delta(x_E(\tau) - r) \,\rmd\tau \;,
\end{eqnarray}  
where we emphasize that $x_E(\tau)$ is a Brownian excursion. Therefore, the moments of the DOS for the bridge or the excursion can be computed as
\begin{eqnarray}\label{moments_BB}
\langle \rho_{B}^k(r,t) \rangle= \langle \rho_{E}^k(r,t) \rangle= \langle \prod_{i=1}^{k} \int_0^t \rmd t_i\, \delta(x_E(t_i)-r)\rangle \;.
\end{eqnarray}
As done before for the free BM in Eq. (\ref{eq:Tmax_via_Markov}), the expression in (\ref{moments_BB}) can be computed using the propagator of the excursion, i.e. the propagator $G^+(\alpha|\beta,t)$ of a free BM starting at $\alpha > 0$ at time $t=0$ and reaching $\beta > 0$ at time $t$ and staying positive over the whole interval $[0,t]$. This propagator $G^+(\alpha|\beta,t)$ can also be computed simply by the method of images, yielding:
\begin{eqnarray}
G^{+}(\alpha|\beta,t)&=& \frac{1}{\sqrt{2\pi t}} \left(e^{-\frac{(\alpha-\beta)^2}{2 t}}-e^{-\frac{(\alpha+\beta)^2}{2 t}}\right) \;,
\end{eqnarray}
together with its LT transform wrt $t$, which will be useful in the following:
\begin{eqnarray}\label{LT_propag_BE}
\tilde G^+(\alpha|\beta,s) = \frac{1}{\sqrt{2s}} \left(e^{-\sqrt{2s} |\alpha-\beta| }-e^{-\sqrt{2s} (\beta+\alpha) }\right)\,.
\end{eqnarray}
Since we are studying here a Brownian excursion, we need, as before, to introduce a regulator such that $x_E(0) = x_E(t) = \varepsilon > 0$ and take the limit $\varepsilon \to 0$ of a suitably defined quantity, as done in Eq. (\ref{eq:Tmax_via_Markov}) -- since we can not impose simultaneously $x_E(0) = 0$
 and $x_E(0^+) > 0$. One has here:
\begin{equation}\label{moments_BB_2}
 \langle \rho_{B}^k(r,t) \rangle= \langle \rho_{E}^k(r,t) \rangle = \lim_{\varepsilon \to 0} \frac{k!}{Z_B(\varepsilon)} \int_{0<t_1<t_2<\cdots<t_k<t} \hspace*{-2cm}\rmd t_1 \cdots \rmd t_k \; \; G^+(\varepsilon|r,t_1) G^+(r|r,t_2-t_1) \cdots G^+(r|\varepsilon,t-t_k) \;,
\end{equation} 
where the combinatorial factor $k!$ comes from the different permutations of the intermediate times $t_1, \cdots, t_k$ where the Brownian excursion $x_E(\tau)$ reaches the value $r$. The denominator $Z_B(\varepsilon)$ is given by
\begin{eqnarray}\label{ZBB}
Z_B(\varepsilon) = G^+(\varepsilon|\varepsilon,t) \sim \varepsilon^2 \sqrt{\frac{2}{\pi}}t^{-3/2} \;, \; \varepsilon \to 0 \;.
\end{eqnarray}
To compute the multiple integral over the times $t_1, \cdots, t_k$ in the right hand side of Eq. (\ref{moments_BB_2}), we recognize, as before for the free BM, a convolution structure. Hence its LT wrt $t$ is given by:
\begin{eqnarray}
&&\int_0^\infty \rmd t \, e^{-st} \int_{0<t_1<t_2<\cdots<t_k<t} \hspace*{-2cm}\rmd t_1 \cdots \rmd t_k \; \; G^+(\varepsilon|r,t_1) G^+(r|r,t_2-t_1) \cdots G^+(r|\varepsilon,t-t_k) = \tilde G^{+}(\varepsilon|r,s)^2 \tilde G^{+}(r|r,s)^{k-1} \nn\\
&&\hspace{2cm} \sim 4 \varepsilon^2  e^{-2 \sqrt{2s}r}\frac{(1-e^{-2 \sqrt{2s}r})^{k-1}}{(2s)^{(k-1)/2}} \;, \; \varepsilon \to 0 \;,\label{expr_LT_BB}
\end{eqnarray}
where we have used the explicit expression of the LT of the propagator in (\ref{LT_propag_BE}). Using the above expression (\ref{expr_LT_BB}) together with (\ref{ZBB}), one obtains finally after Laplace inversion:
\begin{equation}
\langle \rho_{B}^k(r,t=1) \rangle= \langle \rho_{E}^k(r,t=1) \rangle =\langle T_{{\rm loc}, E}^k(r,t=1) \rangle=2 \sqrt{2 \pi} k! \sum_{l=1}^{k} (-1)^{l+1} \tbinom{k-1}{l-1}\Phi^{(k-2)}(2r l), \label{mu_B}
\end{equation}
with the convention $\Phi^{(-1)} = - d\Phi^{(0)}/dr$ and where the $\Phi^{(j)}$'s are defined in Appendix \ref{useful_functions}, see Eq. (\ref{phi_Lap}). Thus we recover in (\ref{mu_B}) the result obtained by Takacs \cite{takacs1995} (see his Eq. (38)) by a probabilistic method (note the correspondence between the functions $J_k$'s in \cite{takacs1995} and the functions $\Phi^{(k)}$'s: $J_k(r)=r^{1+k} \int_1^{\infty}\rmd x\, e^{-(x r)^2/2}(x-1)^k=\sqrt{2\pi}k!\,\Phi^{(k+1)}(r)$.

{For $k=1$, one finds the mean DOS for the BB on the unit time interval, $\langle \rho_{B}(r,t=1) \rangle= 4 r e^{-2r^2}$, as found in Ref. \cite{takacs1995}. Note that   it coincides in this case with the PDF of the maximum of a BB, which is a generic property for periodic signals such that $x(t) = x(0)$~\cite{burkhardt2007}. One has indeed}
\begin{eqnarray} \label{rho_pont}
\langle \rho_{B}(r,t=1) \rangle= \langle \rho_{E}(r,t=1) \rangle\, &= & \int_{0}^{1} \rmd \tau \langle \delta( x_{{\rm max},B} - x_B(\tau) -r ] \rangle \nonumber\\
&= & \langle \delta[ x_{{\max},B} -r ]\rangle\nonumber\\
&= &4 r  e^{-2 r^2} \, ,
\end{eqnarray}
where we have used the periodicity of the bridge and the possibility of adding to it an arbitrary constant. From the average DOS, we can compute, using (\ref{property_average}), the average value of any functional of the maximum of a BB. In particular, for the interesting family of functionals $V(x) = x^\alpha$, one obtains 
\begin{eqnarray}\label{first_moment_DOS}
\langle T_{\alpha,B}(t)\rangle = \langle \int_0^t (x_{\max,B} - x_B(\tau))^\alpha \,\rmd \tau \rangle &=&  \int_0^\infty \langle \rho_{B}(r,t) \rangle r^\alpha \,\rmd r  \\
&=& t^{1+\frac{\alpha}{2}} \frac{\Gamma(1+\frac{\alpha}{2})}{2^{\alpha/2}} \;,
\end{eqnarray}
as announced in the introduction in Eq. (\ref{intro_Talpha_BB_average}). In Fig. \ref{figure_Talpha} (right panel), we show a plot of $\langle T_{\alpha,B}(t) \rangle$ as a function of $\alpha$.




\subsection{The average DOS for the Brownian meander}\label{DOS_BME_section}

Using the same method based on propagators (\ref{eq:Tmax_via_Markov}), we can also compute the average DOS for the Brownian meander (see Fig. \ref{fig_Brownian} (d)). To this purpose, we need to know the propagator of a Brownian particle confined in a given interval $[0,M]$ with absorbing boundary conditions both in $x=0$ and $x=M$. Denoting by $G_M^+(\alpha|\beta,t)$ the propagator of such a particle starting at $\alpha$ and ending, at time $t$, at $\beta$, one has
\begin{eqnarray}
G_M^+(\alpha|\beta,t) = \sum_{n=1}^{\infty} \frac{2}{M} \sin{\left(\frac{\pi n}{M} \alpha\right)} \sin{\left(\frac{\pi n}{M} \beta\right)} e^{-\frac{\pi^2}{2 M^2} n^2 t } \;.
\end{eqnarray} 
Its LT wrt $t$, $\tilde G_M^+(\alpha|\beta,s)$ reads
\begin{eqnarray}\label{propag_me_laplace}
\tilde G_M^+(\alpha|\beta,s) = \frac{2 \sinh{\left[ \sqrt{2s}(M-\max(\alpha,\beta))   \right]} \sinh{\left[ \sqrt{2s}\min(\alpha,\beta) \right]}  }{\sqrt{2s} \sinh{\left( \sqrt{2s}M  \right)} } \;,
\end{eqnarray}
where we have used the identity
\begin{eqnarray}
\sum_{k=1}^{\infty} \frac{ \cos{\left( k x\right)}}{a^2 +k^2} =\frac{1}{2a} \left(\frac{\pi \cosh{\left((\pi-x)a \right)}}{\sinh{\left(\pi a \right)}} -\frac{1}{a}\right) \;.
\end{eqnarray}

As done before in (\ref{eq:Tmax_via_Markov}), we introduce the two times $t_{\max}$ and $\tau$ such that $x(t_{\max}) = M-\epsilon$ and $x(\tau) = M-r$. These two times break the interval into three sub-intervals (see Fig. \ref{figure_propagator}). As shown in Fig. \ref{figure_propagator}, two cases may arise: 0<$\tau < t_{\max}$ or $t>\tau > t_{\max}$. In this case, for the Brownian meander (BMe), these two configurations do not give the same contributions to the average DOS $\langle \rho_{Me}(r,t) \rangle$. Using the same type of regularization procedure as used before (see Eq. (\ref{eq:Tmax_via_Markov})), one has
\begin{eqnarray}\label{rho_Me}
\langle \rho_{Me}(r,t) \rangle   &=&\underset{\varepsilon \to 0}\lim \frac{1}{Z_{Me}(\varepsilon)} \int_r^\infty \rmd M \int_0^t \rmd t_{\max} \int_0^{M-\varepsilon}\rmd x_F\\  
&&\Big(\int_{0}^{t_{\max}}\rmd \tau \, G^{+}_{M}(\varepsilon | M-r,\tau)   G^{+}_{M}(M-r|M-\varepsilon, t_{\max}-\tau)G^{+}_M{}(M-\varepsilon|x_F,t-t_{\max})\nonumber\\
&&+ \int_{t_{\max}}^{t}\rmd \tau \, G^{+}_M(\varepsilon | M-\varepsilon,t_{\max})   G^{+}_M(M-\varepsilon | M-r,\tau-t_{\max})G^{+}_M(M-r|x_F,t-\tau)\Big)\;,\nn
\end{eqnarray}
where we have used the Markov property of BM and where the normalization constant $Z_{Me}(\varepsilon)$ is given by
\begin{eqnarray}
Z_{Me}(\varepsilon)&=&\int_0^\infty \rmd M \int_0^t \rmd t_{\max} \int_0^{M-\varepsilon} \rmd x_F \,G^{+}_M{}(0 | M-\varepsilon, t_{\max}) \, G^{+}_M{}(M-\varepsilon | x_F, t-t_{\max}) \nn \\
&\sim& 2 \varepsilon^3 \sqrt{\frac{2}{\pi t} } \;, \; \varepsilon \to 0 \;.\hspace{0.cm}
\end{eqnarray} 
To compute the numerator in (\ref{rho_Me}), we take advantage of its convolution structure and perform its LT wrt $t$, using the expression of the LT of the propagator (\ref{propag_me_laplace}). After some manipulations, we obtain finally
\begin{equation}\label{Form_rho_ME}
\langle \rho_{Me}(r,t=1) \rangle \,  =   2\sqrt{2 \pi} \left(  \sum_{n=1}^{\infty} \frac{4n (-1)^n}{2n^2+3(-1)^n-5}  \Phi^{(1)}\left(n r\right)  -\Phi^{(1)}\left(2 r\right) \right) \;,
\end{equation}
where $\Phi^{(1)}(x) = {\rm erfc}(x)/2$ belongs to the family of functions studied in Appendix \ref{useful_functions}.

\subsection{Average DOS of constrained Brownian motions: summary and comparison}\label{DOS_oBM_section}

\begin{table}[tt]
\centering
\begin{tabular}{|c|c|c|}
\hline 
  &  $r \to 0$ &  $r \to \infty$ \\
\hline 
Brownian (BM) & $4r$ & $4 \sqrt{\frac{2}{\pi}} \frac{1}{r^2} e^{-\frac{r^2}{2}}$ \\ 
\hline 
Bridge/Excursion (BB/BE) & $4r$ & $4 r e^{-2 r^2}$  \\ 
\hline 
Meander (BMe) & $4r$ & $\frac{4}{3} \frac{1}{r} e^{-\frac{r^2}{2}}$  \\ 
\hline 
Reflected Brownian & $4r$ &  $\frac83 \sqrt{\frac{2}{\pi}}\frac{1}{r^2} e^{-\frac{r^2}{2}}$  \\ 
\hline 
Reflected Bridge & $4r$ &  $3 \frac{1}{r} e^{-2 r^2}$  \\ 
\hline 
\end{tabular} 
\caption{Asymptotic behaviors of average DOS for various constrained Brownian motions, both for small and large argument.}\label{summary_rho_asympt}
\end{table}

\begin{figure}[ht]
\begin{center}
\resizebox{120mm}{!}{\includegraphics{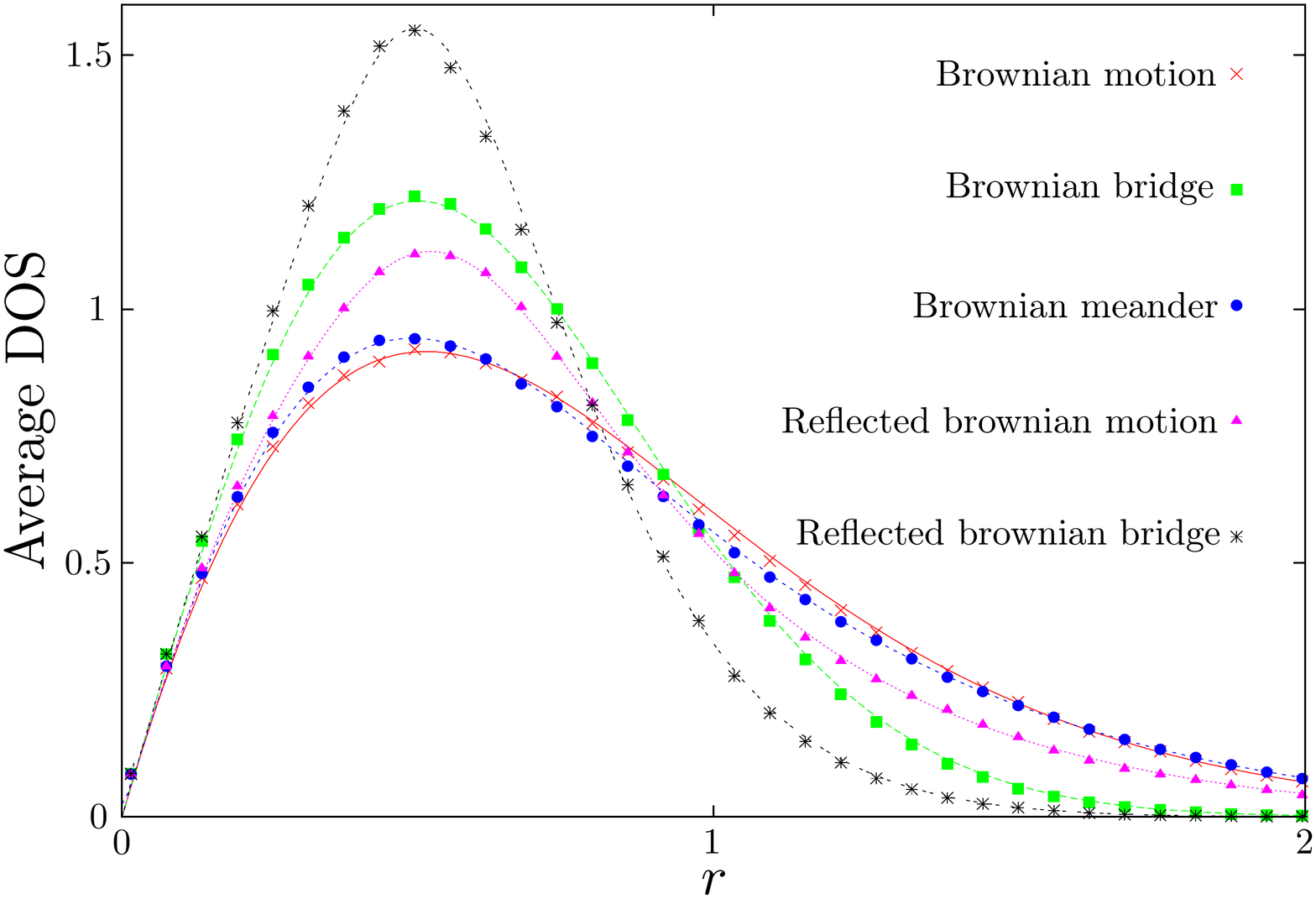}}
\caption{(Color online) Plot of the average DOS for Brownian motion and its variants on the unit time interval. The symbols indicate the results of our numerical simulations (obtained by averaging over $10^4$ realizations of independent RWs of $10^4$ steps). The lines correspond to our exact analytical results given in Eqs.~ (\ref{rho_BM_intro}, \ref{mu_B}, \ref{Form_rho_ME}, \ref{Form_rho_BMR}, \ref{Form_rho_BBR}). We refer the reader to Appendix \ref{simulation_BM_section} for a description of the algorithms which were used here.}\label{tous_les_Browniens}
\end{center}
\end{figure}

It is useful to summarize and compare the results for the DOS for the BM and its variants which we have studied here, using a method based on propagators. In Fig. \ref{tous_les_Browniens}, we have plotted (in lines) the results for the DOS for the Brownian motion (\ref{rho_BM_intro}), Brownian bridge and Brownian excursion (\ref{mu_B}), the Brownian meander (\ref{Form_rho_ME}) as well as for the reflected Brownian motion (\ref{Form_rho_BMR}) and reflected Brownian bridge (\ref{Form_rho_BBR}), the study of which has been left in Appendix \ref{DOS_BR_section}. We have computed numerically the DOS for these different constrained BMs -- which have been simulated using the constructions described in detail in Appendix \ref{simulation_BM_section}. It is useful to remind that $\langle \rho(r,t)\rangle/t = t^{-1/2}\langle \rho(r t^{-1/2},t=1)\rangle$ has the interpretation of a PDF, as ${\rmd}r\langle \rho(r,t)\rangle/t$ is the probability that the walker lies in the interval $[x_{\max}-r-{\rmd}r,x_{\max}-r]$.

\begin{table}[bb]
\centering
\begin{tabular}{|c|c|c|c|c|}
\cline{4-5} \multicolumn{3}{c|}{} & \multicolumn{2}{c|}{numerics} \\
\hline 
  & $\langle r \rangle$ & $r_{\rm typ}$ & $\langle r \rangle$ & $r_{\rm typ}$\\
\hline 
Brownian (BM) & $\sqrt{\frac{2}{\pi}}= 0.7979...$ & 0.5145...& $0.79 \pm 0.01$ & $0.52 \pm 0.02$ \\ 
\hline 
Bridge/Excursion (BB/BE) & $\frac12 \sqrt{\frac{\pi}{2}}=0.6267...$ & $0.5$ & $0.62 \pm 0.01$ & $0.48 \pm 0.02$ \\ 
\hline 
Meander (BMe)  & $ \frac14 \sqrt{\frac{\pi}{2}} (8 \log{2} - 3)=0.7975...$ & 0.4907... & $0.79 \pm 0.01$ & $0.48 \pm 0.02$ \\ 
\hline 
Reflected Brownian & $\frac{3 \pi-4}{3\sqrt{2\pi}}=0.7214...$& 0.5212...& $0.72 \pm 0.01$ & $0.52 \pm 0.02$ \\ 
\hline 
Reflected Bridge  &$ \frac14 \sqrt{\frac{\pi}{2}} (4 \log{2} - 1)=0.5554...$ & 0.4907... & $0.55 \pm 0.01$ & $0.48 \pm 0.02$ \\ 
\hline 
\end{tabular} 
\caption{Main characteristics of the average DOS for various constrained Brownian motions: the average value $\langle r \rangle$ [see Eq. (\ref{average_r})] and the typical value $r_{\rm typ}$ which is the location of the peak of $\langle \rho(r,t=1)\rangle$. The numerical result are obtained by averaging over $4.10^4$ realizations of independent RWs of $10^4$ steps. We refer the reader to Appendix \ref{simulation_BM_section} for a description of the algorithms which were used here.}\label{summary_rho}
\end{table}

The main characteristics of these DOS are summarized in Tables \ref{summary_rho_asympt} and \ref{summary_rho}. A first interesting feature is that the small $r$ behavior of the DOS is, at leading order, the same for all these constrained BMs, $\rho(r,t=1) \sim 4r$. This indicates that the local vicinity of the maximum of the BM is insensitive to the boundary conditions in space (the presence of a wall at $x=0$ either absorbing or reflecting does not affect it) as well as in time (the free BM and the BB, corresponding to periodic boundary conditions in the time direction, share the same local properties). On the other hand, for large argument, the DOS exhibits, in all the cases, a leading order Gaussian decay (see Table \ref{summary_rho_asympt}) but with different rates. In particular, for bridges it decays much faster, $\propto e^{-2 r^2}$, than for free processes, for which it decays as $e^{-r^2/2}$.

This also implies that the mean value 
\begin{eqnarray}\label{average_r}
\langle r \rangle = \int_{0}^{\infty} \rmd r\, r\,  \langle \rho(r,t=1) \rangle \, 
\end{eqnarray}
is larger for the bridge process than for the one which is free at the extremity of the time interval. Interestingly enough, although the mean value might differ notably from one process to another (see Table \ref{summary_rho}), the typical value $r_{\rm typ}$, which is the location of the peak of $\langle\rho(r,t=1)\rangle$, does not vary too much, with a typical value $r_{\rm typ} \approx 1/2$ for all the processes (see Table \ref{summary_rho} and Fig. \ref{tous_les_Browniens}).

\section{Path integral approach}\label{pathintegral_section}

As mentioned in the introduction, the average DOS is useful to compute the average of any functional of the maximum ${\cal O}_{\max}(t) = \int_0^t V(x_{\max} - x(\tau)) \rmd\tau$, see Eq. (\ref{property_average}). However, if one is interested in higher moments of ${\cal O}_{\max}(t)$, the knowledge of the average DOS is not enough. A useful approach to study the statistics of ${\cal O}_{\max}(t)$, beyond the first moment, is to compute the Laplace transform of the full PDF of ${\cal O}_{\max}(t)$, namely $\langle e^{-\lambda \int_0^t V(x_{\max} - x(\tau)) \rmd\tau}\rangle$, from which the moments of ${\cal O}_{\max}(t)$ can be obtained by successive derivation wrt $\lambda$. In this section, we present a general approach, based on path integral techniques to compute this Laplace transform, for any function $V(x)$. We develop this general framework both for the free BM and then for the Brownian bridge -- which turns out to be easier to study, thanks to the Vervaat's construction (see Fig. \ref{Vervaat}). We then apply these general methods to study the full statistics of $\rho_{\rm DOS}(r,t)$ and then to the study of the functionals $T_\alpha(t)$, corresponding to the case $V(x) = x^\alpha$, for the free BM and for the BB. Then, we focus on the special case $\alpha = -1$, corresponding to Odlyzko's algorithm, which, as we show below, leads to a quantum mechanical problem which is exactly solvable.

\subsection{General framework}\label{generalframework_section}

\subsubsection{Free Brownian motion}\label{generalframeworkBM_section}

\begin{figure}[h]
\begin{center}
\resizebox{100mm}{!}{\includegraphics{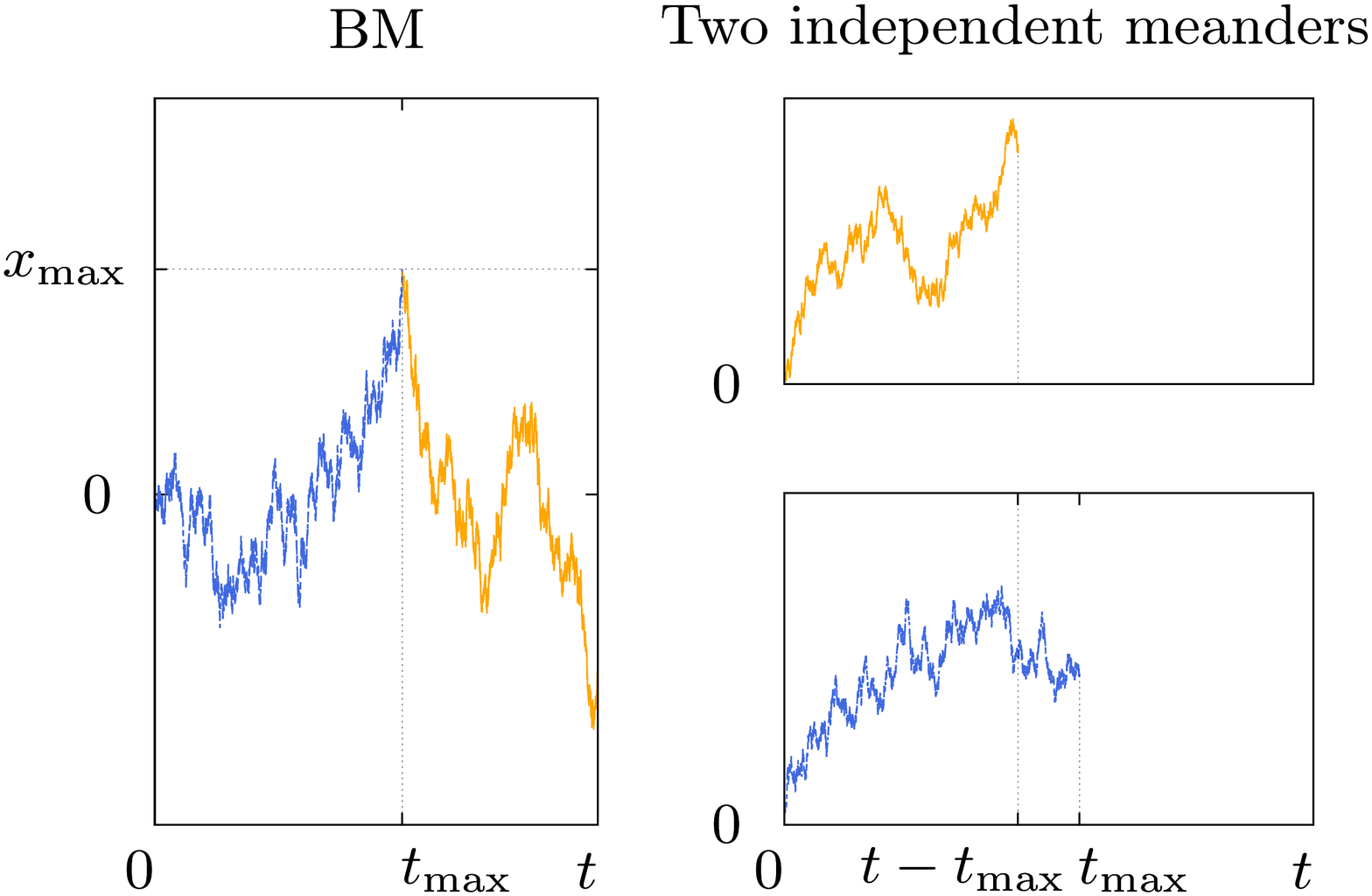}}
\caption{(Color online) Illustration of the path decomposition used in the path integral method (\ref{eq:convolution}): the BM is built from two independent realizations of a Brownian meander, one of duration $t_{\max}$ and the other one of duration $t-t_{\max}$.}
\label{brown_meandre}
\end{center}
\end{figure}

To study analytically $\langle \exp[- \lambda \int_0^t \rmd\tau V(x_{\max} - x(\tau)) ] \rangle$, with $\lambda >0$, for an arbitrary function $V(x)$, we first decompose the time interval $[0,t]$ into two subintervals $[0, t_{\max}]$ and $[t_{\max}, t]$ where 
$t_{\max}$ is the time at which the maximum is reached. These two intervals $[0, t_{\max}]$ and $[t_{\max}, t]$ are statistically independent (as BM is Markovian), and the PDF of $t_{\max}$ is given by the arcsine law~\cite{Lev40}, 
\begin{eqnarray}\label{arcsine}
P(t_{\max}) = \frac{1}{\pi \sqrt{t_{\max}(t-t_{\max})}} \;.
\end{eqnarray}
The process $y(\tau) = x_{\max} - x(\tau)$ is obviously a BM which stays positive on $[0, t]$. By reversing the time arrow in the interval $[0,t_{\max}]$ and taking $t_{\max}$ as the new origin of time, we see that $y(\tau)$ is built from two independent Brownian meanders (BMe): one of duration $t_{\max}$ and the other (independent) one of duration $t-t_{\max}$  (see Fig. \ref{brown_meandre}). Therefore one has
\begin{eqnarray}
\hspace*{-0.25cm}\langle e^{- \lambda \int_0^t \rmd\tau V(x_{\max} - x(\tau))}\rangle &=& \int_0^t \rmd t_{\max} \varphi(t_{\max}) \varphi(t-t_{\max}) \label{eq:convolution} \\
\varphi(\tau)&=&\frac{1}{\sqrt{\pi \tau}} \langle e^{- \lambda \int_0^\tau \rmd u \, V(y(u))}\rangle_+ \;, \label{eq:def_phi}
\end{eqnarray}
where $\langle \cdots \rangle_+$ denotes an average over the trajectories of a BMe $y(\tau)$. In (\ref{eq:def_phi}) the prefactor ${1}/{\sqrt{\pi \tau}}$ comes from the PDF of $t_{\max}$ (\ref{arcsine}). 

On the other hand the convolution structure in (\ref{eq:convolution}) suggests to compute its LT wrt $t$: 
\begin{eqnarray}
&&\int_0^\infty e^{-s\,t} \langle e^{- \lambda \int_0^t  V(x_{\max} - x(\tau)) \rmd\tau}\rangle {\rmd}t = [\tilde \varphi(s)]^2 \;, \label{product_laplace} \\
&& \tilde \varphi(s) = \int_0^\infty e^{-st} \varphi(t) \, \rmd t \;. \label{eq:def_phi_laplace}
\end{eqnarray}
The next step is to compute $\tilde \varphi(s)$ in (\ref{eq:def_phi_laplace}) using a path-integral method. For a Brownian meander, which is
continuous both in space and time, this path integral method must be handled with care: as noticed repeatedly in the previous section, one can not impose simultaneously $y(0) = 0$ and $y(0^+) >0$. This can be circumvented \cite{MC2005} by introducing a cut-off $\varepsilon > 0$ such that $y(0) =\varepsilon$ so that the Feynman-Kac formula reads: 

\begin{eqnarray}
\langle e^{- \lambda \int_0^\tau V(y(u))  \rmd u}\rangle_+ = \underset{\varepsilon \to 0} {\lim} \frac{ \int_0^{\infty} \langle y_F | e^{-H_\lambda \tau} | \varepsilon \rangle {\rmd}y_F }{\int_0^{\infty} \langle y_F | e^{-H_0 \tau} | \varepsilon \rangle {\rmd}y_F} \;,
\end{eqnarray}
with
\begin{eqnarray}\label{schrodinger_operator} 
H_\lambda=-\frac{1}{2} \frac{\rmd^2}{\rmd y^2} + \lambda V(y) + V_{\rm wall}(y) \;,
\end{eqnarray}
where $V_{\rm wall}(y)$ is a hard-wall potential, $V_{\rm wall}(y) = 0$ for $y\geq 0$ and $V_{\rm wall}(y) = +\infty$ for $y<0$, which guarantees that the walker stays positive, as it should for a meander. We can easily compute the eigenfunctions of $H_0 = -\frac{1}{2} \frac{\rmd^2}{\rmd y^2} + V_{\rm wall}(y)$: 
\begin{eqnarray} 
\psi_{k}(y)=\sqrt{\frac{2}{\pi}} \sin(k y),
\end{eqnarray}
and 
\begin{eqnarray}\label{denominator}
\langle y_F | e^{-H_0 \tau} | \varepsilon \rangle = \frac{2}{\pi} \int_0^{\infty} \rmd k\, \sin(k y_F) \sin (k \varepsilon) e^{-\tau \frac{k^2}{2}} \;.
\end{eqnarray}
A straightforward computation yields
\begin{eqnarray}
\int_0^{\infty} {\rmd}y_F \,\langle y_F | e^{-H_0 \tau} | \varepsilon \rangle \underset{\varepsilon \to 0}{\sim} \frac{2 \varepsilon}{\sqrt{2 \pi \tau}}.
\end{eqnarray}
We denote by $\Psi_n$ the wave functions of $H_\lambda$ associated to the energy $E_n$ (we will assume that the spectrum of $H_\lambda$ is discrete but the computation can easily be extended to a continuum spectrum): 

\begin{eqnarray}
\langle e^{- \lambda \int_0^\tau \rmd u \, V(y(u))}\rangle_+
&=& \underset{\varepsilon \to 0} {\lim} \frac{\sqrt{2 \pi \tau}}{2 \varepsilon} \int_0^{\infty} {\rmd}y_F\, \sum_{n=0}^{\infty} \Psi^{*}_n(y_F) \Psi_n(\varepsilon) e^{-E_n \tau} \nonumber \\
&=& \underset{\varepsilon \to 0} {\lim} \sqrt{\frac{\pi \tau}{2}} \int_0^{\infty} {\rmd}y_F\, \sum_{n=0}^{\infty} \Psi^{*}_n(y_F) \frac{\Psi_n(\varepsilon)}{\varepsilon} e^{-E_n \tau} \nonumber \\
&=& \sqrt{\frac{\pi \tau}{2}} \int_0^{\infty} {\rmd}y_F\, \sum_{n=0}^{\infty} \Psi^{*}_n(y_F) \Psi'_n(0)e^{-E_n \tau} \nonumber \\
&=& \sqrt{\frac{\pi \tau}{2}} \int_0^{\infty} {\rmd}y_F\, \partial_x {\rm G}_\tau(0,y_F) \;, \label{the_formula}
\end{eqnarray}
where we have used $\Psi_n(0)=0$, because of the absorbing wall in $0$, and where we denote 
\begin{eqnarray}\label{Green_general}
{\rm G}_\tau(x,y)=\sum_{n=0}^{\infty} \Psi^{*}_n(y) \Psi_n(x)e^{-E_n \tau} \;.
\end{eqnarray}
We can use these formulae (\ref{the_formula}, \ref{Green_general}) to calculate the Laplace transform $\tilde \varphi(s)$ in (\ref{eq:def_phi_laplace})
\begin{eqnarray}
\tilde \varphi(s) = \int_0^\infty e^{-s\tau} \varphi(\tau) \, {\rmd}t &=& \frac{1}{\sqrt{2}} \int_0^\infty {\rmd}\tau \,e^{-s\tau} \int_0^{\infty} {\rmd}y_F\, \partial_x {\rm G}_\tau(0,y_F) = \frac{1}{\sqrt{2}} \int_0^{\infty} \rmd y_F\, \sum_{n=0}^{\infty} \frac{\Psi^{*}_n(y_F) \Psi'_n(0)}{s+E_n} \nonumber \\
&=&\frac{1}{\sqrt{2}} \int_0^{\infty} {\rmd}y_F\, \partial_x  \tilde{{\rm G}}_s(0,y_F) \;, \label{the_formula_Laplace}
\end{eqnarray}
where $\tilde {\rm G}_s(x,y)=\sum_{n=0}^{\infty} \frac{\Psi^{*}_n(y) \Psi_n(x)}{s+E_n}$ is the Laplace transform of ${\rm G}_{\tau}(x,y)$ wrt $\tau$. One thus recognizes that $\tilde {\rm G}_s(x,y)$ is the Green's function satisfying
\begin{eqnarray}\label{Green_equation}
\left[H_{\lambda} + s \right] \tilde {\rm G}_s(x,y) = \delta(x-y) \;,
\end{eqnarray}
such as $\tilde {\rm G}_s(x,0) = \tilde {\rm G}_s(0,y) = 0$. 
To compute $\tilde{\rm G}_s(x,y)$, we look for two complementary functions $u_s(y)$ and $v_s(y)$ solution of the homogeneous equation $\left[H_\lambda + s \right] \psi (y) = 0$ with $u_s(0)=0$ and $v_s(y \to \infty) = 0$. From $u_s(y)$ and $v_s(y)$, we can compute $\tilde{{\rm G}}_s(x,y)$ as \cite{Kri03}
\begin{eqnarray}\label{green_wronskian}
\tilde {\rm G}_s(x,y) = 
\begin{cases}
&\dfrac{2}{W} \, u_s(x) \,v_s(y) \mbox{ if } x \le y \;,\\
&\\
&\dfrac{2}{W} \, u_s(y) \,v_s(x) \mbox{ if } x \ge y \;, 
\end{cases}
\end{eqnarray}
where 
\begin{eqnarray}
W=u'_s(x) v_s(x) - u_s(x) v'_s(x)
\end{eqnarray} 
is the Wronskian associated to $u_s(x)$ and $v_s(x)$, which is here independent of $x$. Finally we obtain from (\ref{the_formula_Laplace}) and (\ref{green_wronskian})
\begin{eqnarray}\label{last_gen_wronskian}
\tilde{\varphi}(s)=\int_0^\infty e^{-s\tau} \varphi(\tau) \, {\rmd}\tau
&=&\frac{\sqrt{2}}{W} \int_0^{\infty} {\rmd}y_F\, u'_s(0) v_s(y_F) \;. \label{green_wronskian2}
\end{eqnarray}
This formula (\ref{last_gen_wronskian}) together with (\ref{product_laplace}) and (\ref{eq:def_phi_laplace}) allows one to compute $\langle e^{-\lambda\int_0^t V(x_{\max} - x(\tau))\rmd \tau}\rangle$ for any function $V(x)$. To obtain explicit results from these general formulas, we need to analyze in more detail the Schr\"odinger operator in Eq. (\ref{schrodinger_operator}). This will be done, for some special cases, in section \ref{application_section}.

\subsubsection{Brownian bridge}\label{generalframeworkBB_section}

In the case of a BB, the method presented above can be straightforwardly adapted to compute the Laplace transform of the PDF of functional of the maximum of the Brownian bridge, ${\cal O}_{\max, B}(t)$, namely $\langle e^{- \lambda \int_0^t V(x_{\max,B}-x_B(\tau)) \rmd\tau}\rangle$. In principle, one could use the same reasoning as before, i.e. break the time interval $[0,t]$ into $[0, t_{\max}]$ and $[t_{\max},t]$ where $t_{\max}$ is the time at which the maximum is reached (see Fig. \ref{brown_meandre}). The main difference is that, for the BB, the PDF of $t_{\max}$ is uniform $P(t_{\max}) = 1/t$, and not given by the arcsine law (\ref{arcsine}) -- this is a consequence of periodic boundary conditions in the time direction. There is however a simpler way to proceed, which makes use of the Vervaat's construction (see Fig. \ref{Vervaat}), which allows us to map any functional of the maximum of a BB onto a (standard) functional of the Brownian excursion $x_E(\tau)$, from which path-integral techniques have already been developed \cite{Maj05a,MC2005,MC2005b}. Hence, generalizing the relation in (\ref{summary_vervaat}) to more general functionals, we have
\begin{eqnarray}\label{general_vervaat}
\int_0^t V(x_{\max,B} - x_B(\tau)) \rmd\tau \overset{\rm law}{=} \int_0^t V(x_E(\tau)) \rmd\tau \;.
\end{eqnarray}  
Hence, using that identity in law (\ref{general_vervaat}), the LT of the PDF of any functional of the maximum of the BB, can be written as a path-integral~\cite{Maj05a,MC2005,MC2005b}. Because we are dealing with a Brownian excursion, which prevents us to impose simultaneously 
$x_E(0) = 0$ and $x_E(0^+) > 0$, this path integral method needs to be suitably regularized. This can be done, as explained before [see Eq. (\ref{moments_BB_2})], by using a cutoff $\varepsilon$ such that $x_E(0) = x_E(t) = \varepsilon$ and computing the statistics of these observables (\ref{general_vervaat}) by using a limiting procedure:
\begin{eqnarray}\label{gen_bridge}
\langle e^{- \lambda \int_0^t V(x_{\max,B}-x_B(\tau)) \rmd\tau}\rangle =  \underset{\varepsilon \to 0}\lim \frac{\langle \varepsilon| e^{-H_\lambda t} | \varepsilon \rangle}{\langle \varepsilon| e^{-H_0 t} | \varepsilon \rangle} \;,
\end{eqnarray}
where $H_\lambda$ is the Schr\"odinger operator defined above (\ref{schrodinger_operator}). Note that the denominator in (\ref{gen_bridge}) does not depend on the functional at hand, i.e. it is independent on $V(x)$, and it is readily computed from (\ref{denominator}) to be
\begin{eqnarray}\label{denominator_bridge}
\langle \varepsilon| e^{-H_0 t} | \varepsilon \rangle \sim \varepsilon^2 \sqrt{\frac{2}{\pi t^3}} \;, \; \varepsilon \to 0 \;.
\end{eqnarray}
This formula (\ref{gen_bridge}) is quite general and it allows us to compute the statistics of a wide class of functionals of the maximum of the Brownian bridge, as we illustrate it below.

 \subsection{Applications to some specific functionals of the maximum}\label{application_section}
 
 In this section, we present some concrete applications of the above path integral formalism, both for the free BM as well as for the BB. We first illustrate the method on the computation of the full statistics of the DOS, providing a detailed derivation of the results recently announced in Ref. \cite{Perret_PRL}. Then, we study the special family of functionals corresponding to $V(x) = x^{\alpha}$, whose applications where discussed in the introduction. Finally, we provide a full detailed study of the special case $\alpha = -1$, i.e. $V(x) = 1/x$, which corresponds to the analysis of Odlyzko's algorithm.

 \subsubsection{Full statistics of the DOS}\label{DOSstat_section}
 
 The main body of results for the statistics of the DOS [see e.g., Eqs. (\ref{mu_n_BM}) and (\ref{Lap_Tmax_BM}) below] were recently announced by us in a short Letter~\cite{Perret_PRL}. In this section we provide a detailed derivation of these results using the path-integral framework presented above. 
 
 {\bf The case of free BM}. To compute the statistics of $\rho(r,t) = \int_{0}^t \delta(x_{\max}-x(\tau) - r)\rmd \tau$, we apply the above formalism (\ref{product_laplace}, \ref{eq:def_phi_laplace}, \ref{last_gen_wronskian}) to the special case $V(x) = \delta(x-r)$. In this case, the two independent solutions are $u_s(x), v_s(x)$ of $[H_\lambda + s] \psi(x) = 0$, with $H_\lambda = -(1/2)\rmd^2/\rmd x^2 + \lambda \delta(x-r) + V_{\rm wall}(x)$ which reads simply here, for $x \neq r$:
\begin{eqnarray}
 -\frac{1}{2} \psi''(x) + s \, \psi(x) = 0 \;, {\rm for \;} x \in [0,r[ \cup ]r,+\infty[
 \end{eqnarray}
with the following boundary conditions: 
 \begin{eqnarray}
 &\left\{
     \begin{array}{lr}
			u_s(0)=0 \\
			u_s(r^+)=u_s(r^-)\\
			\frac{1}{2} (u'_s(r^+)-u'_s(r^-))=\lambda u_s(r)
     \end{array}
     \right. \;, \;
 &\left\{
     \begin{array}{lr}
			\underset{y \to \infty}{\lim}v_s(y) = 0\\
			v_s(r^+)=v_s(r^-)\\
			\frac{1}{2} (v'_s(r^+)-v'_s(r^-))=\lambda v_s(r) \;,
     \end{array}
     \right.
\end{eqnarray}
where the boundary conditions in $r$ result from the presence of the delta peak in $H_\lambda = -\frac{1}{2}{\rmd}^2/{\rmd} x^2 + \lambda \delta(x-r)$. Hence, $u_s(x)$ and $v_s(x)$ are given by
 \begin{eqnarray}
u_s(x) &=  &\left\{
     \begin{array}{lr}
			A \sinh(\sqrt{2s} x) &  x \le r\\
			 A\left( \sinh(\sqrt{2s} x) +\frac{2 \lambda \sinh(\sqrt{2s}r)}{\sqrt{2s}} \sinh(\sqrt{2s}( x-r))\right)&  x>r \label{explicit_u} \:,
     \end{array}
     \right.\\
v_s(x) &=  &\left\{
     \begin{array}{lr}
			B \left(  e^{-\sqrt{2s} (x-r)} + \frac{2 \lambda}{ \sqrt{2s}} \sinh{\left(\sqrt{2s} (r-x)\right)}\right) &  x \le r\\
			 B e^{-\sqrt{2s} (x-r)}&  x>r, \label{explicit_v}
     \end{array}
     \right.
\end{eqnarray}
where $A$ and $B$ are normalization constants which are irrelevant here. The Wronskian is thus given by
\begin{eqnarray}
W=u'_s(x) v_s(x) - u_s(x) v'_s(x) = A B e^{\sqrt{2s} r }\left[\left(1-e^{-2\sqrt{2s} r}\right) \lambda+ \sqrt{2s}\right] \;. \label{wronskian}
\end{eqnarray}
With these formulas (\ref{the_formula_Laplace}, \ref{green_wronskian2}, \ref{explicit_u}, \ref{explicit_v}, \ref{wronskian}), we finally find
\begin{eqnarray} \label{varphi_s}
\tilde{\varphi}(s)
&=&\dfrac{1}{\sqrt{s}}\frac{\sqrt{2s}+\lambda\left(1-e^{-\sqrt{2s} r}\right)^2  }{\sqrt{2s}+\lambda\left(1-e^{-2 \sqrt{2s} r }\right)} \;,
\end{eqnarray} 
which, combined with Eq. (\ref{product_laplace}), yields finally the formula :
\begin{equation} \label{Wronskien_final}
\int_{0}^{\infty} \rmd t \, e^{-st} \langle e^{-\lambda \rho(r,t)} \rangle =\frac{1}{s}\left(
\frac{\sqrt{2s}+{\lambda} (1-e^{-\sqrt{2s}r})^2}
{\sqrt{2s}+{\lambda} (1-e^{-2\sqrt{2s}r})}\right)^2.
\end{equation} 

From this expression (\ref{Wronskien_final}), one can obtain the moments of arbitrary order $\langle \rho^k(r,t)\rangle$, for $k \in \mathbb{N}$ by expanding it in powers of $\lambda$. It yields:
\begin{eqnarray}\label{expansion_lambda}
\int_{0}^{\infty} \rmd t \, e^{-st} \langle e^{-\lambda \rho(r,t)} \rangle =
1+\sum_{k=1}^{\infty} (-\lambda)^k \frac{8\,e^{-\sqrt{2s}r}}{({2s})^{\frac{k}{2}+1}} (1-e^{-\sqrt{2s}r})^2 (1+k e^{-\sqrt{2s}r}) (1-e^{-2\sqrt{2s}r})^{k-2} \hspace*{0.cm} \;.\hspace{0.4cm}
\end{eqnarray}
This expansion (\ref{expansion_lambda}) in powers of $\lambda$ yields the LT of the moments $\langle \rho^k(r,t)\rangle$ wrt $t$ as:
\begin{eqnarray}
\int_0^\infty \langle \rho^k(r,t)\rangle e^{-st} {\rmd}t= \frac{8 k!}{({2s})^{\frac{k}{2}+1}} \sum_{l=0}^{k-1} (-1)^l \tbinom{k-1}{l}
\left(  (2l-1) e^{-(2l+1)\sqrt{2s}r} + (k-2(l+1))e^{-2(l+1)\sqrt{2s}r}\right) \hspace*{-0.17cm}\,.\hspace{0.3cm}
\end{eqnarray}
It is then possible to invert this LT using the functions $\Phi^{(j)}$'s presented in Appendix \ref{useful_functions} to obtain
\begin{eqnarray} \label{mu_n_BM}
\langle \rho^k(r,t=1) \rangle=8 k! && \sum_{l=0}^{k-1} (-1)^l \tbinom{k-1}{l} [ (2l+1)\Phi^{(k+1)}((2l+1)r) + (k-2(l+1))\Phi^{(k+1)}(2(l+1)r)] \,.\hspace{0.2cm}
\end{eqnarray}
For $k=1$, this yields back the result obtained in~(\ref{rho_BM_intro}). 

By inverting the LT wrt $\lambda$ in (\ref{Wronskien_final}), we obtain the full PDF $P_t(\rho,r)$ of the DOS (\ref{def_rho}), as a function of $\rho$, for different values of the parameter $r$.
\begin{eqnarray}\label{Lap_Tmax_BM}
\int_0^\infty  e^{-st} P_t(\rho,r) {\rmd}t =
\delta(\rho) \dfrac{(e^{-\sqrt{2s}r}-1)^2}{s(1+e^{-\sqrt{2s}r})^2} \label{delta} +\dfrac{e^{-\frac{ \rho \sqrt{2s}e^{\sqrt{2s} r }}{2\sinh\left(\sqrt{2s} r \right)}}}{\cosh^3\left(\frac{r \sqrt{2s}}{2}\right)}
\left(\dfrac{ e^{\frac{r \sqrt{2s}}{2}}}{\sqrt{2s}}+\dfrac{\rho e^{\sqrt{2s} r}}{4 \sinh\left({r \sqrt{2s}}\right) \sinh\left(\frac{r \sqrt{2s}}{2}\right) }\right) \hspace*{-0.1cm} \;.\hspace{0.35cm}
\end{eqnarray}
It has an unusual form with a peak $\propto \delta(\rho)$
at $\rho=0$, in addition to a non trivial continuous background density $p_t(\rho,r)$ for $\rho>0$. Hence one has
\begin{eqnarray} \label{loi_Tmax}
P_t(\rho,r)= F_W(r,t) \delta(\rho) + p_t(\rho,r) \, , 
\end{eqnarray}
where $F_W(r,t) =  {\rm Prob.} [W(t) \leq r]$, given below in (\ref{Pwidth_int}), is the probability that the 
width $W(t) = \max_{\tau \in [0,t]} x(\tau) - \min_{\tau \in [0,t]} x(\tau)$ is smaller than $r$. This can be understood because if $W(t)$ is smaller than $r$, the amount of time spent by the process at a distance within $[r, r + {\rmd}r]$ from the maximum is $0$ (see Fig. \ref{fig_intro}), yielding the delta peak at $\rho = 0$.
Indeed we can check that the coefficient of the term $\propto \delta(\rho)$ in (\ref{delta}) is the LT wrt $t$ of 
\begin{eqnarray} \label{Pwidth_int}
F_W(r,t) = 1+\sum_{l=1}^\infty 4 l (-1)^l {\rm erfc}({lr}/{\sqrt{2t}}) \,,
\end{eqnarray}
which corresponds precisely to the distribution of the width (or the span) of BM~\cite{Fel1951,KMS13}. 

\begin{figure}[ht]
\begin{center}
\resizebox{90mm}{!}{\includegraphics{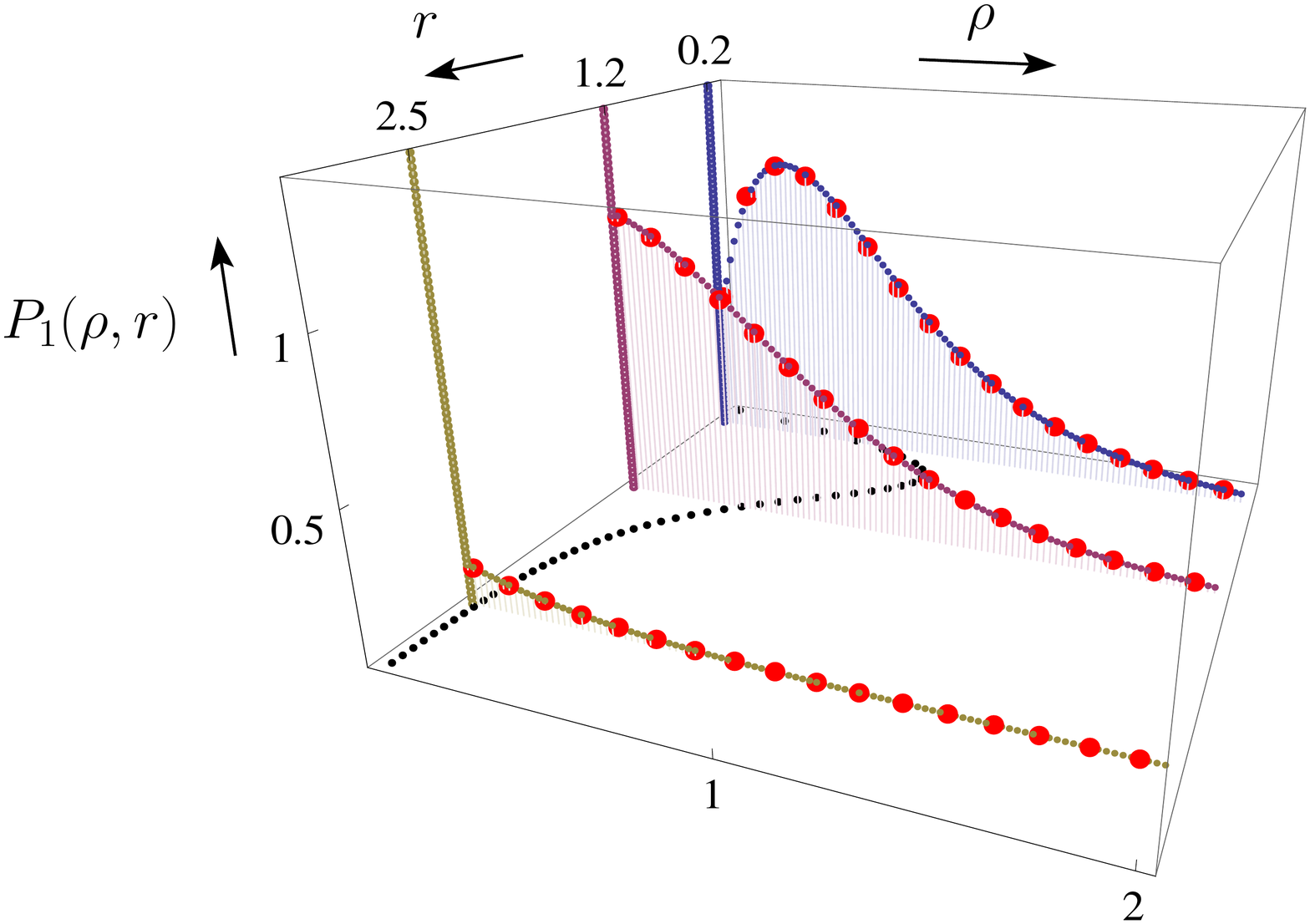}}
\caption{(Color online) Plot of $P_1(\rho,r)$ as a function of $\rho$ for different values of $r$. The solid lines for $\rho=0$ represent the $\delta(\rho)$ peak, $\propto \delta(\rho)$ in Eq. (\ref{Lap_Tmax_BM}). The dotted lines correspond to our exact analytical results for the background density $p_1(\rho,r)$ in (\ref{Lap_Tmax_BM}) -- where the inverse LT wrt $s$ has been performed numerically (in green for $r = 2.5$, purple for $r=1.2$ and blue for $r=0.2$) -- while the red dots indicate the results of simulations. On the $z=0$ plane, we have plotted the exact mean DOS in Eq. (\ref{rho_BM_intro}).}\label{3DPlot}
\end{center}
\end{figure}

 On the other hand, in (\ref{loi_Tmax}), $p_t(\rho,r) =p_1(\rho/\sqrt{t},r/\sqrt{t})/\sqrt{t}$ is a regular function of $\rho$, for $r>0$ (see Fig. \ref{3DPlot}), and has a more complicated structure. We obtain an explicit expression of its LT wrt $t$ given by the second term of Eq.~(\ref{Lap_Tmax_BM}). 
In Fig. \ref{3DPlot} we show the results of $p_1(\rho,r)$ obtained from numerical simulations (averages are performed over $10^7$ samples) for three different values of $r$. We see that they are in perfect agreement with our exact formula (\ref{Lap_Tmax_BM}). The analysis of the distribution $p_1(\rho,r)$, including its asymptotic behaviors when $\rho \to 0$ and $\rho \to \infty$ was carried out in Ref. \cite{Perret_PRL} and we refer the interested reader to the supplementary material of Ref. \cite{Perret_PRL} for more details.

{\bf The case of a Brownian bridge.} In this case, we remind that, as it can be seen using the Vervaat's construction, the DOS for the BB coincides with the DOS for the Brownian excursion, which in turn, coincides also with the local time of the excursion [see the identity in law in Eq. (\ref{summary_vervaat})]. The full PDF of the $\rho_{B}(r,t)$ can be obtained from the general formula (\ref{gen_bridge}), applied to the special case $V(x) = \delta(x-r)$: 
\begin{eqnarray}\label{LT_DOS_BB}
\langle e^{- \lambda \rho_{B}(r,t)}\rangle =  \underset{\varepsilon \to 0}\lim \frac{\langle \varepsilon| e^{-H_\lambda t} | \varepsilon \rangle}{\langle \varepsilon| e^{-H_0 t} | \varepsilon \rangle} \; , \;  H_\lambda=-\frac{1}{2} \frac{\rmd^2}{\rmd x^2} + \lambda \delta(x-r) + V_{\rm wall}(x) \;,
\end{eqnarray}
where we recall that $V_{\rm wall}(x)$ is a hard-wall potential, $V_{\rm wall}(x) = 0$ for $x\geq 0$ and $V_{\rm wall}(x) = +\infty$ for $x<0$ -- hence $H_{\lambda}$ is defined on $[0,+\infty[$. The numerator in (\ref{LT_DOS_BB}) can be computed by expanding the matrix element on the eigenfunctions $\psi_{k}(x)$ of $H_\lambda$ which are given by
\begin{eqnarray}
H_\lambda \psi_k(x) = \frac{k^2}{2} \psi_k(x) \;,\;
\psi_k(x)= \left\{
    \begin{array}{cc}
        0 \:\:\: &x \le 0 \\
        a(k,\lambda) \sin(kx) \:\:\:  &0 \le x \le r \\
        a(k,\lambda) \left(\sin(kx)+ \frac{2\lambda}{k}\sin(kr)\sin(k(x-r))\right)\:\: & r \le x,
    \end{array}
\right.
\end{eqnarray}
where the amplitude $a(k,\lambda)$ ensures the normalization condition:
\begin{eqnarray}
|a(k,\lambda)|^{-2}= \frac{\pi}{2} \left[1+\frac{4\lambda}{k} \sin(kr) \left(\cos(kr)+\frac{\lambda}{k}\sin(kr)\right)\right] \;. 
\end{eqnarray}
We finally obtain 
\begin{equation}\label{LT_Bridge}
\langle e^{-\lambda \rho_{B}(r,t)} \rangle=\sqrt{\frac{2 t^3}{\pi}} \int_0^{\infty}  
\dfrac{ \,  k^2 e^{-\frac{k^2 t}{2}}}
{1+\frac{4\lambda}{k}\sin{(kr)}\left(\frac{\lambda}{k}\sin{(kr)} + \cos{(kr)}\right)} {\rmd}k \;.
\end{equation} 
Note that from the identity in law in (\ref{summary_vervaat}) this expression (\ref{LT_Bridge}) yields also an interesting
relation for the local time of the Brownian excursion, which we have not seen in the literature. Furthermore, by expanding
this formula (\ref{LT_Bridge}) in powers of $\lambda$, it is possible to obtain, using yet another method, the moments $\langle [\rho_{B}(r,t)]^k\rangle$ of arbitrary order $k \in \mathbb{N}$ and recover the expression given above (\ref{mu_B}).

On the other hand, by studying the large $\lambda$ behavior of (\ref{LT_Bridge}), which is of order ${\cal O}(\lambda^{0})$, 
we can show that the PDF of $\rho_{B}(r,t)$ has an expression similar to, albeit different from, the one for BM in (\ref{loi_Tmax}):
\begin{eqnarray}
P_{t,B}(\rho,r)= F_{W,B}(r,t) \delta(\rho) + p_{t,B}(\rho,r) \;, 
\end{eqnarray}
where $F_{W,B}(r,t)$ is the distribution function of the width of the BB \cite{Chu1976}: 
\begin{eqnarray}
F_{W,B}(r,t) =  1+ 2 \sum_{l=1}^\infty e^{-\frac{2 l^2 r^2}{t}} \left(1-\frac{4 l^2 r^2}{t}\right) \;,
\end{eqnarray}
while $p_{t,B}(\rho,r)$ is now a different distribution. In particular, one can show that~(\ref{mu_B}) yields back the
complicated though explicit formula for $p_{t,B}(\rho,r)$ found in Ref.~\cite{takacs1995,takacs1992} using a completely different method:
\begin{eqnarray}
p_{t,B}(\rho,r) = 1-2 \sum_{j=1}^\infty \sum_{k=0}^{j-1} \binom{j-1}{k} e^{-(\rho+2r j)^2/2}(-\rho)^k H_{k+2}(\rho+2r j)/k!\;,
\end{eqnarray}
where the $H_n$'s are Hermite polynomials defined by
\begin{eqnarray}
H_{n}(x)=n! \sum_{i=0}^{\lfloor n/2 \rfloor }\frac{(-1)^i x^{n-2i}}{2^i\, i! (n-2i)!}\;,
\end{eqnarray}
where $\lfloor x \rfloor$ denotes the largest integer not larger than $x$. We refer the interested reader to Ref. \cite{Perret_PRL} for more details on the distribution $p_{t,B}(\rho,r)$, including its asymptotic behaviors.

\subsubsection{The case $V(x) = x^\alpha$: first and second moments of the functionals}\label{Talpha_section}

We now apply this path integral formalism to the functionals $T_\alpha(t)$, for free BM (\ref{T_alpha}), and $T_{\alpha, B}(t)$ the associated observable for the BB (\ref{T_alpha_B}).  
We showed previously that the first moment can be obtained directly from the corresponding average DOS. This simple method can not be easily adapted to compute higher moments of $T_{\alpha}(t)$ or $T_{\alpha, B}(t)$. We show here how to compute these moments using the path integral formalism developed above. We will treat separately the case of free BM and BB.   

{\bf The case of free BM.} It is convenient to start from the following formula, obtained from the combination of Eqs. (\ref{product_laplace}, \ref{eq:def_phi_laplace}, \ref{the_formula_Laplace}): 
\begin{eqnarray}\label{BM_Gs}
\int_0^{\infty}\rmd t e^{-st} \langle e^{-\lambda \int_0^t \rmd \tau V\left(x_{\max}-x(\tau)\right)}\rangle  = \frac12 \left(\int_0^\infty \rmd y \partial_x \tilde{\rm G}_s(0,y) \right)^2  \,,
\end{eqnarray}
where we recall that $\tilde{\rm G}_s(x,y)$ is the Green's function of the following Schr\"odinger equation
\begin{eqnarray}\label{Green}
\left[-\frac{\rmd^2}{{\rmd} x^2}+\lambda V(x) +V_{\rm wall}(x)+s \right] \tilde{\rm G}_s(x,y) =\delta(x-y)\,.
\end{eqnarray}
Of course, there exist very few instances of potential $V(x)$ for which this Eq. (\ref{Green}) can be solved exactly (see below). However, the first moments of 
$T_{\alpha}(t)$ can be extracted, in principle, for a generic $V(x)$. Indeed, from (\ref{BM_Gs}), we see that the moments of $T_{\alpha}(t)$ are obtained by successive derivations of the right hand side of (\ref{BM_Gs}) wrt $\lambda$, evaluated in $\lambda = 0$. These successive derivatives can in turn be expressed as combinations of the successive derivatives of $\tilde{\rm G}_s(x,y)$ wrt $\lambda$. Indeed, if we write the following expansion:
\begin{eqnarray}\label{def_fn}
\tilde {\rm G}_{s}(x,y) = \sum_{n=0}^\infty \frac{\lambda^n}{n!} f_n(x,y) \;,
\end{eqnarray}
one has from (\ref{BM_Gs}):
\begin{eqnarray}\label{gen_formula_moment}
\int_0^\infty \langle T_{\alpha}^k(t) \rangle e^{-s t} \rmd t = (-1)^k \frac{k!}{2} \sum_{n=0}^k b_n b_{k-n} \;, \; b_n = \frac{1}{n!}\int_0^\infty \partial_x f_n(x,y) \big|_{x=0} \, \rmd y \;.
\end{eqnarray}
Hence, we need to solve perturbatively the equation for the Green's function (\ref{Green}) in powers of $\lambda$ to compute the functions $f_n(x,y)$ in (\ref{def_fn}). First it is easy to obtain $f_0(x,y)$ as
\begin{eqnarray} \label{f_0}
f_0(x,y) = \frac{1}{\sqrt{2s}}(e^{-\sqrt{2s}|x-y|}-e^{-\sqrt{2s}(x+y)}) \,.
\end{eqnarray}
Furthermore, one can show that the functions $f_n(x,y)$ satisfy the following recursion relation:
\begin{eqnarray}\label{recursion}
f_n(x,y)= - n\int_0^\infty {\rmd} z f_0(x,z) V(z) f_{n-1}(z,y) \;, \; n \geq 1 \,.
\end{eqnarray}
The recursion relation (\ref{recursion}) can be solved formally in the closed form
\begin{eqnarray} \label{f_n}
f_n(x,y)= (-1)^n n!   \prod_{i=1}^{n}\int_0^\infty{\rmd} z_i \, f_0(x,z_1) \prod_{i=1}^{n-1} V(z_i) f_0(z_i,z_{i+1})  \,  V(z_n) f_0(z_n,y)\,.
\end{eqnarray}
Hence, from Eq. (\ref{gen_formula_moment}) together with (\ref{f_n}) one can compute the moments of arbitrary order of any functional of the maximum of the free BM. Note that this technique is a generalization of the method based on propagators which we used before for the computation of the average DOS, corresponding to a particular potential $V(x) = \delta(x-r)$. Here we apply this formalism to the case $V(x) = x^\alpha$. 

We first compute the first moment, using Eq. (\ref{gen_formula_moment}) for $k=1$, together with (\ref{f_n}). We obtain:
\begin{eqnarray}\label{1ermoment_BM}
\int_0^{\infty}\rmd t e^{-st} \langle T_\alpha(t)\rangle  = - \left(\int_0^\infty \rmd y \partial_x f_1(x,y)\big|_{x=0} \right)\left( \int_0^\infty \rmd y \partial_x f_0(x,y)\big|_{x=0}  \right)\,.
\end{eqnarray}
Using the expressions of $f_0(x,y)$ in (\ref{f_0}) and of $f_1(x,y)$, obtained from Eq. (\ref{f_n}) together with the fact that $\partial_x f_0(0,y) = 2 e^{-\sqrt{2s} y}$, we obtain (performing the change of variable $z\to \sqrt{2s}z$): 
\begin{eqnarray}
\int_{0}^\infty {\rmd}y \partial_x f_1(0,y) &=& - \int_0^\infty {\rmd}y \int_0^\infty{\rmd}z \, \partial_x f_0(0,z) z^\alpha f_0(z,y) \nn \\
&=& -\frac{2}{\sqrt{2s}^{\alpha+3}}\int_0^\infty {\rmd}y \int_0^\infty{\rmd}z \, e^{-z} z^\alpha (e^{-|z-y|}-e^{-(z+y)}) \nn \\
&=&-\frac{2}{{(2s)}^{\frac{\alpha+3}{2}}} \Gamma(\alpha+1)(2-2^{-\alpha})\,. \label{intf_1}
\end{eqnarray}
Hence, combining (\ref{1ermoment_BM}, \ref{intf_1}) and using that $\int_0^\infty \rmd y \partial_x f_0(x,y)\big|_{x=0} = \sqrt{2/s}$ we arrive at
\begin{eqnarray}
\int_0^{\infty}\rmd t e^{-st} \langle T_\alpha(t)\rangle  = \frac{4}{({2s})^{\frac{\alpha+4}{2}}} \Gamma(\alpha+1)(2-2^{-\alpha}) \,,
\end{eqnarray}
which after Laplace inversion, yields
\begin{eqnarray} 
\langle T_{\alpha}(t)\rangle  = \frac{(2t)^{1+\frac{\alpha}{2}}(2-2^{-\alpha})\Gamma\left(\frac{1+\alpha}{2}\right)}{(2+\alpha)\sqrt{\pi}} \;.
\end{eqnarray}
This formula coincides, as it should, with the formula obtained directly from the average DOS in (\ref{Talpha_average}). 

We now compute the second moment, using Eq. (\ref{gen_formula_moment}) for $k=2$:
\begin{eqnarray}\label{2ndmoment_BM}
\int_0^{\infty}\rmd t e^{-st} \langle T^2_\alpha(t)\rangle  =\left(\int_0^\infty \rmd y \partial_x f_1(0,y) \right)^2+ \left(\int_0^\infty \rmd y \partial_x f_0(0,y)  \right)\left(\int_0^\infty \rmd y \partial_x f_2(0,y)  \right)\,.
\end{eqnarray}
We then have, using (\ref{f_0}, \ref{f_n}), and performing the change of variables with the substitution $u\to \sqrt{2s}\,z_1$, $v\to \sqrt{2s}\,z_2$:
\begin{eqnarray}
&&\int_{0}^\infty {\rmd}y \partial_x f_2(0,y) = 2\int_0^\infty {\rmd}y \int_0^\infty {\rmd}z_1\int_0^\infty {\rmd}z_2 \, \partial_x f_0(0,z_1) z_1^\alpha f_0(z_1,z_2)  z_2^\alpha f_0(z_2,y) \nn \\
&=& \frac{4}{\sqrt{2s}^{5+2\alpha}} \int_0^\infty {\rmd}y \int_0^\infty {\rmd}u \int_0^\infty {\rmd}v \,  e^{-u} u^\alpha (e^{-|u-v|}-e^{-(u+v)})  v^\alpha (e^{-|v-y|}-e^{-(v+y)})\nn \\
&=& \frac{8}{\sqrt{2s}^{5+2\alpha}} C_\alpha \;, \; \; {\rm with} \;\; C_\alpha =  \int_0^\infty {\rmd}u \int_0^\infty {\rmd}v \,  e^{-u} u^\alpha (e^{-|u-v|}-e^{-(u+v)})  v^\alpha (1-e^{-v}) \label{f2_int}
\end{eqnarray}
where $C_\alpha$ can be explicitly computed as
\begin{eqnarray}
C_\alpha &=&\sum_{n=1}^{\infty}  \frac{\Gamma(2+2\alpha+n)}{n! 2^{2+2\alpha+n}(1+\alpha+n)}+\frac{(4^{1 + \alpha}-1) \Gamma(3 + 2 \alpha)}{2^{3+2 \alpha}(1 + \alpha)^2}- \frac{2^{1+\alpha}-1}{2^{2+2 \alpha}} \Gamma(1 + \alpha)^2 \;.
\end{eqnarray}
Finally, using (\ref{f_0}, \ref{intf_1}, \ref{2ndmoment_BM}, \ref{f2_int}), we obtain
\begin{eqnarray}
\int_0^{\infty}\rmd t e^{-st} \langle T^2_\alpha(t)\rangle  =\frac{4}{(2s)^{3+\alpha}} \left(\Gamma(\alpha+1)^2 (2-2^{-\alpha})^2 +4C_\alpha
\right) \,,
\end{eqnarray}
which after Laplace inversion yields:
\begin{eqnarray}\label{T2_text}
\langle T^2_\alpha(t)\rangle = \frac{t^{2 + \alpha}}{2^{3\alpha}\Gamma(3+\alpha)} &&\left(\Gamma(\alpha+1)^2 (2^\alpha-1)(2^{\alpha+1}-1) +\frac{\Gamma(3+2\alpha)(4^{\alpha+1}-1)}{4(1+\alpha)^2} \right. \nn \\ 
&&\left. +\sum_{n=1}^{\infty}  \frac{\Gamma(2+2\alpha+n)}{n! 2^{1+n}(1+\alpha+n)}\right) \;.\hspace{1cm}
\end{eqnarray}
The last sum over $n$ can be finally expressed in terms of incomplete beta function, which yields the formula (\ref{second_momentBM}). In particular, by taking (carefully) the limit $\alpha \to -1$ in the above expression (\ref{T2_text}), we recover the result of \cite{chassaing_yor}
\begin{eqnarray}
\lim_{\alpha \to -1} \langle T^2_\alpha(t=1)\rangle  = \frac{\pi^2}{3}+ 4 \log(2)^2 \;,
\end{eqnarray}
which was obtained by the authors of \cite{chassaing_yor} using a completely different method.
\begin{figure}[ht]
\begin{center}
\rotatebox{-90}{\resizebox{90mm}{!}{\includegraphics{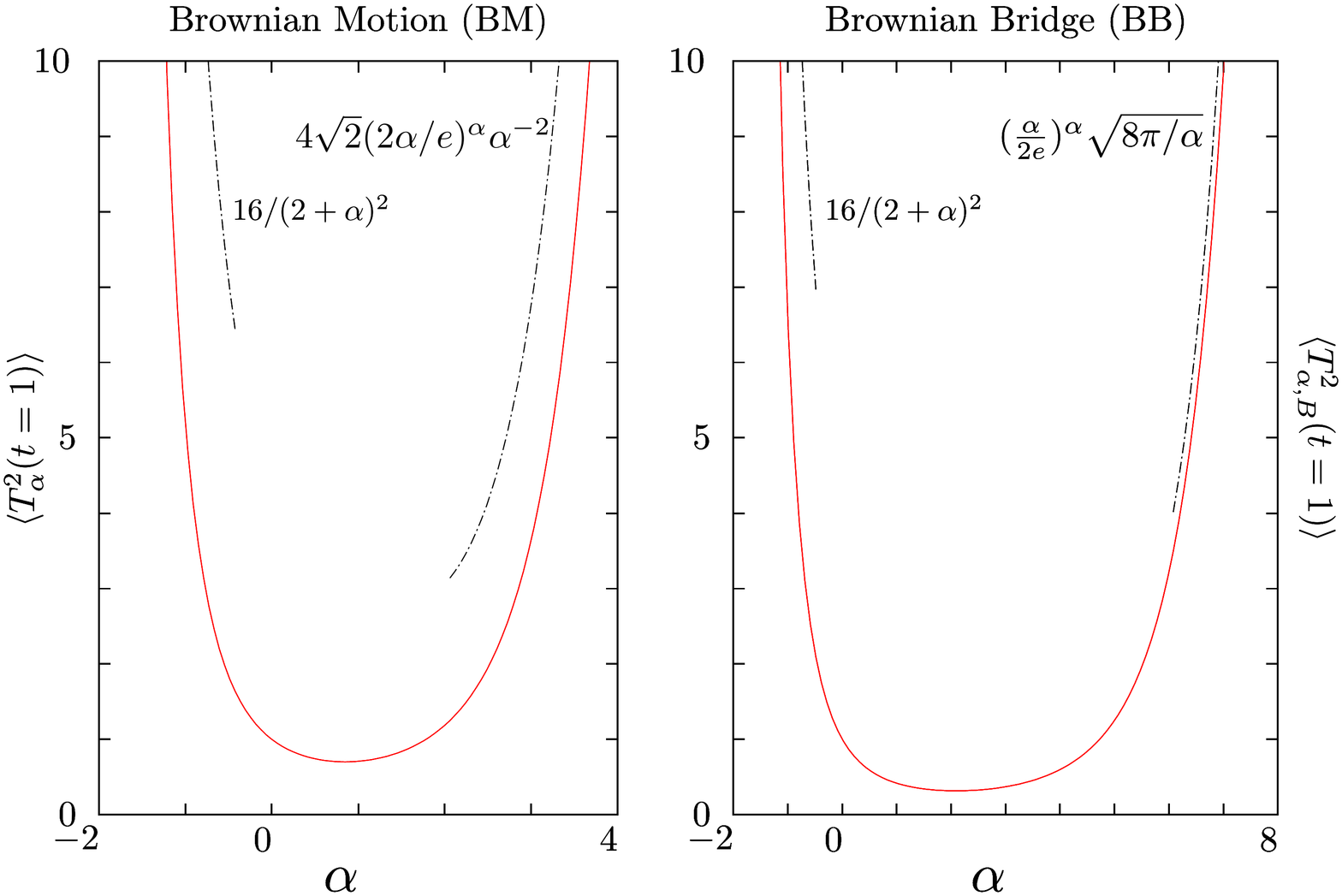}}}
\caption{{\bf Left panel:} Plot of $\langle T_{\alpha}^2(t=1) \rangle$, as a function of $\alpha$, for the BM, as given in Eq. (\ref{T2_text}). {\bf Right panel:} Plot of $\langle T_{\alpha,B}^2(t=1) \rangle$, as a function of $\alpha$, for the BB, as given in Eq. (\ref{Talpha2_BB}). In both panels, the dashed black lines indicate the asymptotic behaviors discussed in the introduction. In particular, for $\alpha = -1$ for the BM (corresponding to Odlyzko's algorithm), we recover $\langle T_{\alpha=-1}^2(t=1) \rangle = \frac{\pi^2}{3}+4 \log(2)^2$ \cite{chassaing_yor}. For $\alpha = 1$ for the BB (corresponding to the Airy distribution), we recover $\langle T^2_{\alpha=1,B}(t=1) \rangle = \frac{5}{12}$ \cite{MC2005,MC2005b}.}
\label{figure_Talpha2}
\end{center}
\end{figure}

{\bf The case of BB.} In this case, the equivalent of Eq. (\ref{BM_Gs}) is the formula derived in Eqs. (\ref{gen_bridge}) and (\ref{denominator_bridge}) which reads
\begin{eqnarray}
\langle e^{-\lambda \int_0^t \rmd \tau V\left(x_{\max}-x(\tau)\right)}\rangle  = \sqrt{\frac{\pi}{2}} t^{3/2} \partial_x \partial_y {\rm G}_t(0,0)  \,.
\end{eqnarray}
Taking the LT of the above relation wrt $t$, one obtains:
\begin{eqnarray} \label{BB_Gs}
\int_0^{\infty}\rmd t e^{-st} t^{-3/2} \langle e^{-\lambda \int_0^t \rmd \tau V\left(x_{\max}-x(\tau)\right)}\rangle  = \sqrt{\frac{\pi}{2}} \partial_x \partial_y \tilde{\rm G}_s(x,y) \Big |_{x=y=0} \,,
\end{eqnarray} 
where $\tilde {\rm G}_s(x,y)$ is the Green's function of the operator in (\ref{Green}). From (\ref{BB_Gs}), one deduces the following relation 
\begin{eqnarray}\label{gen_BB}
\int_0^\infty e^{-st} t^{-3/2} \langle \left[\int_0^t V(x_{\max,B} - x_B(\tau))  \rmd \tau\right]^k  \rangle \rmd t = (-1)^k \sqrt{\frac{\pi}{2}} \frac{\partial^2}{\partial x \partial y} f_k(x,y) \Big|_{x=y=0} \;,
\end{eqnarray}
where $f_k(x,y)$ is given in Eq. (\ref{f_n}). From (\ref{gen_BB}), specified to the case $V(x) = x^\alpha$ and $k=1$, we obtain the first moment $\langle T_{\alpha, B}(t)\rangle$ as
\begin{eqnarray} 
\int_0^{\infty}\rmd t e^{-st} t^{-3/2} \langle T_{\alpha,B}(t)\rangle  = -\sqrt{\frac{\pi}{2}} \frac{\partial^2}{\partial x \partial y} f_1(x,y) \Big |_{x=y=0}\,.
\end{eqnarray}
Hence, using (\ref{f_0}, \ref{f_n}) together with the change of variable $z\to \sqrt{2s}z$, we obtain
\begin{eqnarray} 
\int_0^{\infty}\rmd t e^{-st} t^{-3/2} \langle T_{\alpha,B}(t)\rangle  
&=&\sqrt{\frac{\pi}{2}} \int_0^{\infty} {\rmd}z \, \partial_x f_0(0,z) z^\alpha \partial_y f_0(z,0)\\
&=& \sqrt{\frac{\pi}{2}} \frac{4}{\sqrt{2s}^{\alpha+1}}\int_0^{\infty} {\rmd}z \, e^{-2z} z^\alpha = \sqrt{\frac{\pi}{2}} \frac{2^{1-\alpha}}{\sqrt{2s}^{\alpha+1}} \Gamma(1+\alpha) \,,
\end{eqnarray}
which, after Laplace inversion, yields immediately 
\begin{eqnarray} 
\langle T_{\alpha,B}(t)\rangle  = \frac{t^{1+\alpha/2}}{2^{\alpha/2}}\Gamma\left(1+\frac{\alpha}{2}\right) \;,
\end{eqnarray}
which coincides with the formula obtained above from the average DOS (\ref{first_moment_DOS}).

We can also compute the second moment $\langle T^2_{\alpha,B}(t)\rangle$ by 
using Eq. (\ref{BB_Gs}) for the case $k=2$, yielding
\begin{eqnarray} 
\int_0^{\infty}\rmd t e^{-st} t^{-3/2} \langle T^2_{\alpha,B}(t)\rangle  &=& \sqrt{\frac{\pi}{2}} \partial_x \partial_y f_2(0,0)\,,
\end{eqnarray}
where $f_2(x,y)$ can be obtained from (\ref{f_n}) with $V(x)=x^\alpha$. Using  (\ref{f_0}, \ref{f_n}), we obtain, performing the changes of variables 
$u\to \sqrt{2s}z_1$, $v\to \sqrt{2s}z_2$, and after some manipulations 
\begin{eqnarray} 
\int_0^{\infty}\rmd t e^{-st} t^{-3/2} \langle T^2_{\alpha,B}(t)\rangle 
&=&\sqrt{2\pi}\int_0^{\infty} {\rmd}z_1 \int_0^{\infty} {\rmd}z_2 \, \partial_x f_0(0,z_1) z_1^\alpha f_0(z_1,z_2) z_2^{\alpha} \partial_y f_0(z_2,0)\\
&=& \frac{\sqrt{2\pi}}{2^{2\alpha}\sqrt{2s}^{2\alpha+3}}\left(\frac{2\Gamma(2+2\alpha)}{1+\alpha} - \Gamma(1+\alpha)^2\right)\,,
\end{eqnarray}
which, after inverse Laplace transform, yields
\begin{eqnarray}\label{Talpha2_BB}
\langle T^2_{\alpha,B}(t)\rangle  = t^{\alpha+2} \frac{\sqrt{\pi}}{2^{3\alpha+1}\Gamma(\alpha+\frac32)}\left(\frac{2\Gamma(2+2\alpha)}{1+\alpha} - \Gamma(1+\alpha)^2\right)\,.
\end{eqnarray}
In particular, we can check that $\langle T^2_{\alpha=1,B}(t=1)\rangle = \frac{5}{12}$ coincide with the known result for the area under a Brownian excursion, i.e.the second moment of the Airy distribution~\cite{MC2005,MC2005b}.

\subsubsection{The exactly solvable case $V(x) \propto 1/x$ and application to the Odlyzko's algorithm}\label{Tminus1_section}

In this section, we apply our general formalism to the special case $V(x) = 1/x$, which is relevant to analyze the distribution of the cost of the optimal algorithm to find the maximum of a random walk (see Eq. (\ref{def_I})), as shown in Refs. \cite{odlyzko,chassaing_yor}. We apply this formalism separately both to the free Brownian motion and then to the Brownian bridge. 

{\bf The case of free BM.} In this case, we want to construct the Green's function $\tilde{\rm G}_s(x,y)$ in (\ref{green_wronskian}) which can be built from two independent solutions $u_s(x)$ and $v_s(x)$ of the following Schr\"odinger equation
\begin{eqnarray}\label{schrod_odlyzko}
\left[-\frac{1}{2} \frac{{\rmd}^2}{{\rmd} x^2} + \frac{\lambda}{x} + s \right] \psi(x) = 0 \;, \; s > 0
\end{eqnarray}
with the asymptotic behaviors
\begin{eqnarray}\label{bc}
u_s(0) = 0 \;\; \& \;\; v_s(y \to \infty) = 0 \;. 
\end{eqnarray}
It turns out that the above Schr\"odinger equation (\ref{schrod_odlyzko}) can be solved in terms of hypergeometric functions. The solutions $u_s(x)$ and $v_s(x)$ satisfying (\ref{bc}) read
\begin{eqnarray}
&&u_s(x) = {\cal A} \; e^{-\sqrt{2s} x} x \; {}_1F_1\left(1+ \frac{\lambda}{\sqrt{2 s}}, 2 , 2 \sqrt{2s} x \right) \;,  \label{us_odlyzko} \\
&&v_s(x) = {\cal B} \; e^{-\sqrt{2s} x} x \;  U\left(1+ \frac{\lambda}{\sqrt{2 s}}, 2 , 2 \sqrt{2s} x \right)\;, \label{vs_odlyzko}
\end{eqnarray}
where ${}_1F_1(a,b,x)$ and $U(a,b,x)$ are confluent hypergeometric functions (respectively Kummer's and Tricomi's hypergeometric function) and where 
${\cal A}$ and ${\cal B}$ are two irrelevant numerical constants (note that $u_s(x)$ and $v_s(x)$ are also known as Whittaker functions). The Wronskian $W$ reads
\begin{eqnarray}
W = u'_s(x) v_s(x) - u_s(x) v_s'(x) = \frac{{\cal A\, B}}{2 \lambda \Gamma(\lambda/\sqrt{2s})} \;.
\end{eqnarray}
Note that one can check from (\ref{us_odlyzko}) that $u_s(0) = 0$, as it should [see Eq.~(\ref{bc})] and that $u_s'(0) = {\cal A}$. The function $\tilde \varphi(s)$ in Eq. (\ref{last_gen_wronskian}) reads in this special case (\ref{us_odlyzko}, \ref{vs_odlyzko}):
\begin{eqnarray}
\tilde \varphi(s) = 2^{3/2} \lambda \Gamma\left(\frac{\lambda}{\sqrt{2s}} \right) \int_0^\infty e^{-\sqrt{2s} y} \, y \, U\left(1+ \frac{\lambda}{\sqrt{2 s}}, 2 , 2 \sqrt{2s} y \right)  \, \rmd y \;. \label{integral_phi_odlyzko_free}
\end{eqnarray}
After some manipulations, the integral over $y$ in Eq. (\ref{integral_phi_odlyzko_free}) can be evaluated as (see Appendix \ref{hypergeometric_section_first})
\begin{eqnarray}\label{explicit_tildephi}
\tilde \varphi(s) &=& \frac{1}{\sqrt{s}} G\left(\frac{\lambda}{\sqrt{2s}} \right) \;,\nn \\
G(x) &=& 2 \sum_{k=0}^\infty (-1)^{k+1} \tilde \zeta(k) x^k \;, \; {\rm with} \; \tilde \zeta(k) = (1-2^{-1-k})\zeta(k) = \sum_{n=1}^\infty \frac{(-1)^{n+1}}{n^k} \;.\hspace{1cm}
 \end{eqnarray}
Hence, from the general formulae (\ref{product_laplace}, \ref{eq:def_phi_laplace}), together with (\ref{explicit_tildephi}), one obtains the explicit formula for the case $V(x)~=~1/x$:
\begin{eqnarray}\label{explicit_LT}
\int_0^\infty e^{-s\,t} \langle e^{- \lambda \int_0^t  \frac{\rmd\tau}{x_{\max} - x(\tau)}}\rangle {\rmd}t = \frac{4}{s} \sum_{n=0}^\infty (-1)^n \frac{\lambda^n}{(\sqrt{2s})^n} \sum_{k=0}^n \tilde \zeta(k) \tilde \zeta(n-k) \;.
\end{eqnarray}
Expanding the left hand side of Eq. (\ref{explicit_LT}) in powers of $\lambda$ we obtain the moments of $T_{\alpha=-1}(t=1)$ of arbitrary order, as announced in the introduction (\ref{moments_BM_arbitrary}),  
\begin{eqnarray}
\langle T^k_{\alpha=-1}(t=1)\rangle = \Gamma\left(\frac{k+1}{2} \right) \frac{2^{\frac{k}{2} + 2}}{\sqrt{\pi}} \sum_{m=0}^k \tilde \zeta(m) \tilde \zeta(k-m) \;.
\end{eqnarray} 
This allows us to recover in a completely different manner the results obtained in \cite{chassaing_yor} by probabilistic tools. For completeness, we mention that the authors of \cite{chassaing_yor} obtained an explicit expression of the PDF $p(s)$ of $T_{\alpha=-1}(t=1)$ as (see Theorem 4.2 of that paper)
\begin{eqnarray}
p(s) = \frac{8}{s^3} \Theta\left(\frac{4}{s^2}\right)  \;, \; \Theta(x) = \int_0^x \theta_1(y) \theta_2(x-y) {\rmd}y \;,  
\end{eqnarray}
where the functions $\theta_1(x)$ and $\theta_2(x)$ are given by
\begin{eqnarray}
&&\theta_1(x) = \sum_{n=1}^\infty \int_x^\infty \frac{\rmd u}{u} \exp{\left(-\tilde n^2 \frac{u}{2} \right)} \;, \; \tilde n = \pi \left(n-\frac{1}{2}\right) \;, \; \\
&&\theta_2(x) = \frac{\partial}{\partial x} \sum_{n=-\infty}^{+\infty} (1-n^2 \pi^2 x) e^{-n^2 \pi^2 x/2} \;.
\end{eqnarray}

{\bf The case of the BB.} In this case, the starting point of our analysis is the general formula given in (\ref{gen_bridge}):
\begin{eqnarray}\label{gen_bridge_odlyzko}
\left\langle \exp{\left[- \lambda \int_0^t \frac{\rmd\tau}{(x_{\max,B}-x_B(\tau))}\right]} \right \rangle =  \underset{\varepsilon \to 0}\lim \frac{\langle \varepsilon| e^{-H_\lambda t} | \varepsilon \rangle}{\langle \varepsilon| e^{-H_0 t} | \varepsilon \rangle} \;,
\end{eqnarray}
where $H_\lambda$ is the Schr\"odinger operator defined in (\ref{schrodinger_operator}), which reads here
\begin{eqnarray}\label{Schrod_odlyzko}
H_\lambda = -\frac12\frac{\rmd^2}{\rmd x^2} + \frac{\lambda}{x} + V_{\rm wall}(x) \;, \; \lambda \geq 0 \;,
\end{eqnarray}
while the small $\varepsilon$ behavior of the denominator of (\ref{gen_bridge_odlyzko}) is given in (\ref{denominator_bridge}). To compute the numerator in Eq. (\ref{gen_bridge_odlyzko}), we expand the matrix element on the eigenfunctions $|\phi_E \rangle$ of $H_\lambda$, which satisfy
\begin{eqnarray}\label{Eq_Schrod_odlyzko}
H_\lambda |\phi_E \rangle = E  |\phi_E \rangle \;,
\end{eqnarray}
where the eigenvalues $E > 0$ form a continuous spectrum, as there are no bound states here [we recall that $\lambda \geq 0$ in (\ref{Schrod_odlyzko})] and where the eigenvectors satisfy the boundary condition
\begin{eqnarray}\label{bc_BB}
\lim_{x\to 0^+} \phi_E(x) = \lim_{x\to 0^+} \langle x | \phi_E\rangle  = 0\;.
\end{eqnarray}
The general solution of the Schr\"odinger equation (\ref{Schrod_odlyzko}, \ref{Eq_Schrod_odlyzko}) reads
\begin{eqnarray}\label{linear_hypergeo}
\phi_E(x) = c_E x \, e^{-i \sqrt{2 E} x} \;_1 F_1\left(1 - i \frac{s}{\sqrt{2E}}, 2, 2 i x \sqrt{2E} \right) + d_E x \, e^{-i \sqrt{2 E} x} U \left(1 - i \frac{s}{\sqrt{2E}}, 2, 2 i x\sqrt{2E} \right) \;,\hspace{0.5cm}
\end{eqnarray}
where we recall that ${}_1F_1(a,b,x)$ and $U(a,b,x)$ are confluent hypergeometric functions. The boundary condition at $x=0$ (\ref{bc_BB}) imposes that $d_E=0$. The remaining task is to compute the normalization constant $c_E$ such that
\begin{eqnarray}\label{ortho_condition}
\int_0^\infty \phi^*_{E'}(x) \phi_E(x) \rmd x= \delta(E-E') \;. 
\end{eqnarray} 
Using a formula given in Landau-Lifshitz (see Appendix \ref{LL-appendix}), one can show that 
\begin{eqnarray}\label{c_E}
|c_E|^2  = \frac{4 \lambda}{\exp{\left(\dfrac{2 \pi \lambda}{\sqrt{2E}}\right)} - 1} \;.
\end{eqnarray}
Therefore, Eq. (\ref{gen_bridge_odlyzko}), together with Eq. (\ref{c_E}) yields
\begin{eqnarray}\label{final_LT}
\left\langle \exp{\left[- \lambda \int_0^t \frac{\rmd\tau}{(x_{\max,B}-x_B(\tau))}\right]} \right \rangle = 4 \sqrt{\frac{\pi}{2}} t^{3/2}\lambda \int_0^\infty \frac{e^{-Et}}{\left(\exp{\left[{2 \pi \lambda}/{(\sqrt{2E})}\right]} - 1\right)} \rmd E \;.
\end{eqnarray}
It is then possible, using residue theorem, to invert the LT wrt $\lambda$ in (\ref{final_LT}) and then perform the remaining integral over $E$. This finally yields the PDF of $T_{\alpha=-1,B}(t=1)$ given in Eq. (\ref{pB_intro}). The expression for the moments given in Eq. (\ref{moments_BB_arbitrary}) then follows straightforwardly.

\section{Conclusion}\label{Conclusion_section}

In this paper, we have presented several analytical tools to study functionals of the Brownian motion and its variants, in particular Brownian bridge and Brownian excursion. These tools include (i) a ``paths counting'' method, relying on propagators of BM with appropriate boundary conditions and (ii) a suitably adapted path-integral method, which allows us to recast the study of functionals of the maximum of BM into the study of quantum mechanical problems. The first method (i) is conceptually quite simple and allows us to obtain in a rather simple manner the mean value of any functional of the BM, while the second method (ii) is better adapted to compute the full PDF of such functionals. We have used these methods to calculate the statistics of the density of near-extremes, or density of states (DOS), for Brownian motion $\rho(r,t)$ and its variants. In particular, from the mean DOS $\langle \rho(r,t)\rangle$, one can compute the average value of any functional of the maximum of BM. Then, we provided a thorough study of functionals of the form $T_\alpha(t) = \int_0^t (x_{\max}-x(\tau))^\alpha {\rmd}\tau$, with $\alpha \in ]-2, +\infty[$. As $\alpha$ is varied, $T_{\alpha}(t)$ interpolates between various physical observables, as discussed above. We have obtained an exact expression for the two first moments $\langle T_\alpha(t) \rangle$ and $\langle T_{\alpha}^2(t) \rangle$ both of which exhibit a non-trivial, non-monotonic behavior as a function of $\alpha$. Thanks to the path-integral method, when the associated quantum problem can be solved exactly, it is possible to obtain an explicit expression of the Laplace transform for the full PDF of $T_\alpha(t)$ or $T_{\alpha,B}(t)$, from which moments of arbitrary order and in some cases the full PDF can be computed. We have worked out in detail the case $\alpha = -1$, corresponding to $V(x) \propto 1/x$, which corresponds to the cost of the optimal algorithm (due to Odlyzko's) to find the maximum of a discrete RW. In this case, we 
provided an explicit expression for the moments of arbitrary order $\langle T^k_{\alpha = - 1}(t)\rangle$, recovering by physical methods the results obtained in Ref. \cite{chassaing_yor} by completely different probabilistic approaches. Furthermore, we have generalized these results to functionals of the Brownian bridge, $T_{\alpha,B}(t)$. In particular, we argued that, for $\alpha = -1$, the random variable $T_{\alpha=-1,B}(t)$ describes the cost of the optimal algorithm (i.e. Odlyzko's algorithm) for the search of the maximum of a RW in a bridge configuration and computed explicitly its PDF as well as its moments of arbitrary order. 

Several interesting questions are left open. For instance, here we have studied the case of a single Brownian motion and
it would be interesting to extend this study to the case of multi-particle systems, where there are $N > 1$ walkers, which could be independent or instead interacting, as in the case of non-intersecting (vicious) walkers, whose extreme value statistics have recently attracted some attention \cite{SMCR08}. 
Finally, we have treated the case of Brownian motion, which is continuous both in space and time, and it would be interesting to extend these results to random walks, which are discrete in time. This would, in particular, allow us to study the DOS of L\'evy flights, whose behavior is expected to be qualitatively different from Brownian motion.

\acknowledgement
We acknowledge support by the Indo-French 
Centre for the Promotion of Advanced Research under Project~$4604-3$. We acknowledge a useful correspondence with Philippe Chassaing. 

\newpage

\begin{appendix}

\section{Some useful functions}\label{useful_functions}

We introduce the family of functions
$\Phi^{(j)}$, $j \in {\mathbb N}$, which satisfy
\begin{eqnarray} \label{phi_Lap}
\frac{e^{-\sqrt{2s}u}}{(\sqrt{2s})^{j+1}}=
\int_0^{\infty} {{t}}^{\frac{j-1}{2}} \Phi^{(j)}\left(\frac{u}{\sqrt{t}}\right)  e^{-s t} {\rmd}t \,.
\end{eqnarray}
These functions can be obtained explicitly by induction, using~\cite{chassaing2002}
\begin{eqnarray}
\Phi^{(0)}(x)=\frac{1}{\sqrt{2\pi}}e^{-\frac{x^2}{2}}, \Phi^{(j+1)}(x)=\int_{x}^{\infty} \Phi^{(j)}(u) {\rmd}u \,.
\end{eqnarray}
The first functions can easily be computed as
\begin{eqnarray}
\Phi^{(0)}(x) &=&\frac{e^{-\frac{x^2}{2}}}{\sqrt{2 \pi }} \label{Def_Phi0} \;,\\
\Phi^{(1)}(x) &=& \frac{1}{2} \, \text{erfc}\left(\frac{x}{\sqrt{2}}\right)  \label{Def_Phi1} \;,\\
\Phi^{(2)}(x) &=& \frac{e^{-\frac{x^2}{2}}}{\sqrt{2 \pi }}-\frac{1}{2} x \,  \text{erfc}\left(\frac{x}{\sqrt{2}}\right)  \label{Def_Phi2} \;.
\end{eqnarray}
More generally, one can show \cite{chassaing2002} that they can be written in the form
\begin{eqnarray}
\Phi^{(j)}(x)=p_j(x) \frac{1}{\sqrt{2\pi}}e^{-\frac{x^2}{2}}+q_j(x) \mathrm{erfc}(\frac{x}{\sqrt{2}}) \;,
\end{eqnarray}
where $p_j(x)$ and $q_j(x)$ are rational polynomials of degree $j-2$ and $j-1$, respectively, for $j\geq 2$ \cite{chassaing2002}. We refer the interested reader to Ref. \cite{chassaing2002} for efficient algorithms, which can be implemented numerically, to compute these polynomials in a systematic way.

\section{Average DOS for reflected Brownian motion}\label{DOS_BR_section}

Using the method based on propagators, explained in section \ref{DOS_BM_section}, see Eq. (\ref{eq:Tmax_via_Markov}), we can also compute the average DOS for the reflected Brownian motion $x_{R}(\tau)$ which is the absolute value of the Brownian motion, $x_R(\tau) = |x(\tau)|$. The expression in (\ref{eq:Tmax_via_Markov}), see also Fig. \ref{figure_propagator}, indicates that we need to compute 
the propagator of the reflected Brownian motion such that $x_R(\tau) \leq M$ or equivalently $-M \leq x(\tau) \leq M$. Therefore, we compute the propagator of a Brownian particle confined in a given interval $[-M,M]$ with absorbing boundary conditions both in $x=-M$ and $x=M$. Denoting by $G_M^R(\alpha|\beta,t)$ the propagator of such a particle starting at $\alpha$ and ending, at time $t$, at $\beta$, its LT wrt $t$ is given by
\begin{eqnarray}\label{propag_RBB_laplace}
\tilde{G}^{R}_M(\alpha|\beta,s) = \frac{2 \sinh{\left( \sqrt{2s}(M-\max(\alpha,\beta))   \right)} \sinh{\left( \sqrt{2s}(M+\min(\alpha,\beta)) \right)}  }{\sqrt{2s} \sinh{\left( \sqrt{2s}2M  \right)} } \;.
\end{eqnarray}
In order to compute the average DOS $\langle \rho_R(r,t) \rangle$ for the reflected BM, we evaluate  the ``number'' of Brownian
trajectories satisfying the following constraints: the process reaches its maximum $M$ or its minimum $-M$ at time $t_{\rmext}$, passes through $M-r$ or $-M+r$ at time $\tau$ and end in $x_F \in [-M,M]$  at time $t$. The total number of such trajectories is then obtained by integrating over $x_F, M$ and $t_{\rmext}$. When dividing the time interval $[0,t]$ into three parts delimited by $\tau$ and $t_{\rmext}$, 8 different cases may arise: $\tau < t_{\rmext}$ or $\tau > t_{\rmext}$, $x( t_{\rmext})=\pm M$ and $x(\tau)=\pm(M-r)$. Using the invariance of the process under the reflection symmetry 
$x \to - x$ we have to consider only four different cases (each one with a multiplicity of 2): 
\begin{eqnarray}
\langle \rho_R(r,t) \rangle \,  &=&\underset{\varepsilon \to 0}\lim \frac{2}{Z_R(\epsilon)}\int_{r}^\infty \rmd M \int_0^t \rmd t_{\rmext}\int_{-M}^{M} \rmd x_F \nonumber\\
&\Big[& \int_{0}^{t_{\rmext}} \rmd \tau 
G^{R}_M(0 | M-r,\tau)  G^{R}_M(M-r|M-\varepsilon,t_{\rmext}-\tau)G^{R}_M(M-\varepsilon|x_F,t-t_{\rmext})\nonumber\\
&+&\int_{t_{\rmext}}^{t} \rmd \tau 
G^{R}_M(0 | M-\varepsilon,t_{\rmext})   G^{R}_M(M-\varepsilon | M-r,\tau-t_{\rmext})G^{R}_M(M-r|x_F,t-\tau) \nonumber\\
&+& \int_{0}^{t_{\rmext}} \rmd \tau 
G^{R}_M(0 | r-M,\tau)   G^{R}_M(r-M|M-\varepsilon,t_{\rmext}-\tau)G^{R}_M(M-\varepsilon|x_F,t-t_{\rmext})\nonumber\\
&+&\int_{t_{\rmext}}^{t} \rmd \tau 
G^{R}_M(0 | M-\varepsilon,t_{\rmext})   G^{R}_M(M-\varepsilon | r-M,\tau-t_{\rmext})G^{R}_M(r-M|x_F,t-\tau) \Big] \;, \label{dos_reflected_bm}
\end{eqnarray}
where we have used the Markov property of BM and where $Z_R(\epsilon)$ is the normalization constant (such that $\int_0^\infty \rmd r \, \langle \rho_R(r,t)\rangle = t$)
\begin{eqnarray}
Z_R(\varepsilon)=2 \int_{0}^\infty \rmd M \int_0^t \rmd t_{\rmext}\int_{-M}^{M} \rmd x_F 
G^{R}_M(0 |M-\varepsilon,t_{\rmext})G^{R}_M(M-\varepsilon|x_F,t-t_{\rmext}) \;.
\end{eqnarray} 
The normalization is easily computed as $Z_R(\epsilon) \sim 2 \varepsilon^2$, when $\varepsilon \to 0$. Using the same kind of calculations as in section \ref{DOS_BM_section} -- exploiting the convolution structure of the integrals in Eq.~(\ref{dos_reflected_bm}) -- we find, after some manipulations
\begin{eqnarray}\label{Form_rho_BMR}
\langle \rho_{R}(r,t=1) \rangle \,  =8 \sum_{n=0}^{\infty} 
\frac{(-1)^{n+1} }{-3-2 n+12 n^2+8 n^3}
 \left(\sum_{k=0}^1(3+(-1)^k (2n+k)^{k+1})(2n+k)^2 \Phi^{(2)}((2n+k)r)\right) \hspace*{-0.1cm}\;,\hspace*{0.1cm}
\end{eqnarray}
where $\Phi^{(2)}(x)$ is given in Eq. (\ref{Def_Phi2}).

Similarly, we can study the DOS of the reflected Brownian bridge $x_{RB}(\tau)$, which is the absolute value of a Brownian bridge $x_{RB}(\tau) = |x_{BB}(\tau)|$. The calculation of the DOS in this case is very similar to the case of the free reflected BM in (\ref{dos_reflected_bm})  
without the integral over $x_F$ which is set to $x_F = 0$. Using time reversal symmetry, we can show that the average DOS $\langle \rho_{RB}(r,t) \rangle$ is given by
\begin{eqnarray}
\langle \rho_{RB}(r,t) \rangle \,  &=&
\underset{\varepsilon \to 0}\lim \frac{4}{Z}\int_{r}^\infty \rmd M \int_0^t \rmd t_{\rmext} \int_{t_{\rmext}}^{t} \rmd \tau \nonumber\\
&\Big[&
 G_M^R(0 | M-\varepsilon, t_\rmext)   G_M^R(M-\varepsilon | r-M,\tau-t_{\rmext})G_M^R(r-M|0,t-\tau)\nonumber\\
&+& 
G_M^R(0 | M-\varepsilon,t_{\rmext})   G_M^R(M-\varepsilon | M-r,\tau-t_{\rmext})G_M^R(M-r|0,t-\tau)\Big] \;,
\end{eqnarray}
where $Z_{RB}(\epsilon)$ is the normalization constant, given by
\begin{eqnarray}
Z_{RB}(\varepsilon)=2 \int_{0}^\infty \rmd M \int_0^t \rmd t_{\rmext}\,
G_M^R(0 | M-\varepsilon,t_{\rmext})   G_M^R(M-\varepsilon | 0,t-t_{\rmext}) \;.
\end{eqnarray} 
The normalization is easily computed as $Z_{RB}(\varepsilon) \sim {2 \varepsilon^2}/{(\sqrt{2 \pi t})}$, as $\varepsilon \to 0$ and eventually the average DOS $\langle \rho_{RB}(r,t) \rangle $ is obtained as:
 \begin{equation}\label{Form_rho_BBR}
\langle \rho_{RB}(r,t=1) \rangle \,  =  2\sqrt{2 \pi} \left(  4\sum_{n=0}^{\infty} n (-1)^{n+1} \Phi^{(1)}(2nr)  - \Phi^{(1)}(2r)\right) 
\end{equation}
where $\Phi^{(1)}(x)$ is given in Eq. (\ref{Def_Phi1}).

\section{Odlyzko's algorithm}\label{Odlyzko_section}

\subsection{Main ideas behind Odlyzko's algorithm}

\begin{figure}[t]
\begin{center}
\resizebox{100mm}{!}{\includegraphics{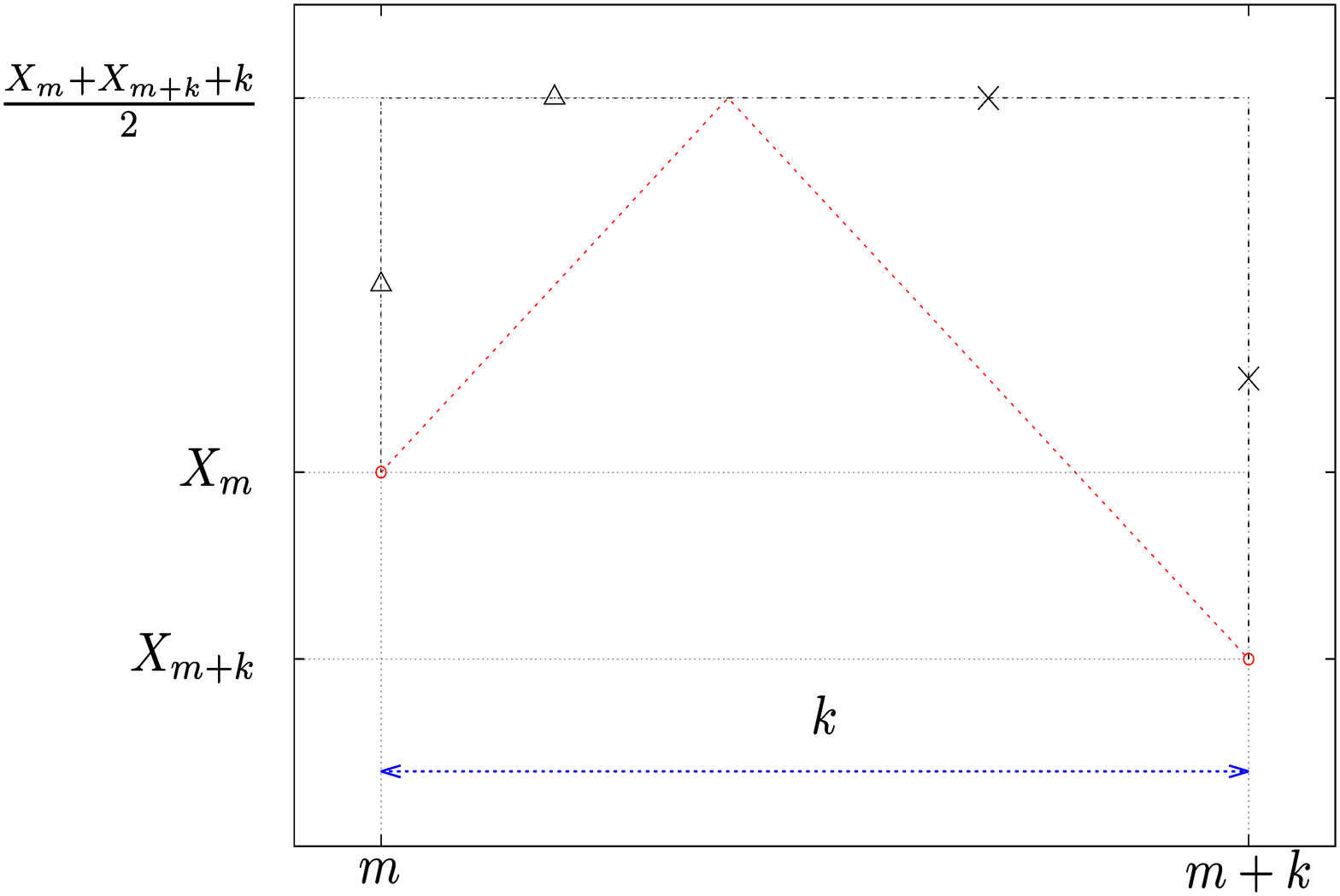}}
\caption{Illustration of the main idea of Odlyzko's optimal algorithm. The RW can not exceed $(X_m+X_{m+k}+k)/2$ between $m$ and $m+k$. If this quantity is smaller than $M^\#$, a new probe between $m$ and $m+k$ is useless.}
\label{passage_region_useless}
\end{center}
\end{figure}

\begin{figure}[hh]
\begin{center}
\resizebox{100mm}{!}{\includegraphics{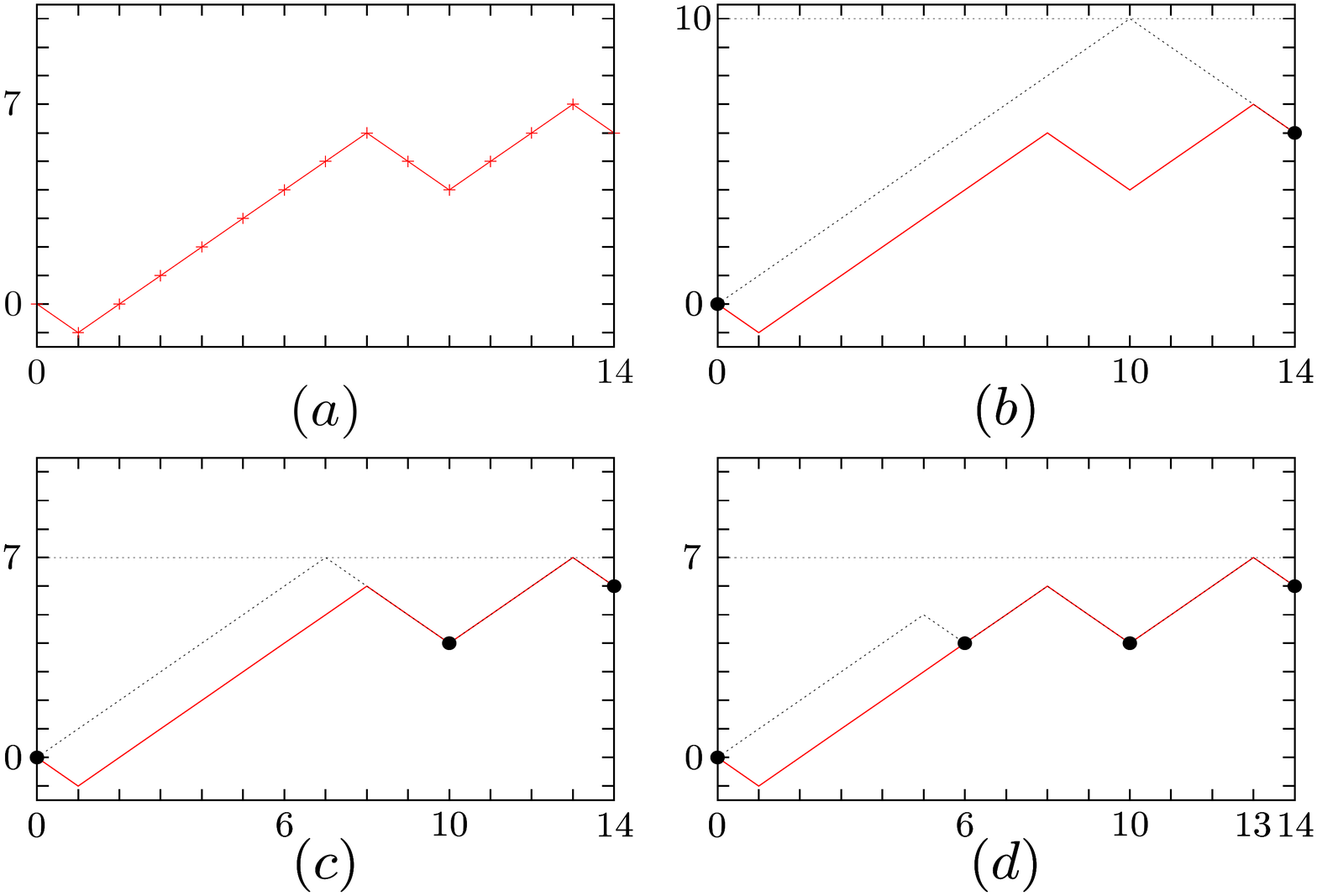}}
\caption{An example of the algorithm $u$ for finding $M_{14}=7$ for a RW in $4$ probes. \textbf{(a)} Typical realization of a 14 steps RW, for which we want to find the maximum. We know without any probe that $0\le M_{14}\le 14$, and if $M_{14}=14$, the maximum would be at position $14$ (RW with only $+1$ jumps) so we probe the position $14$.  \textbf{(b)} The first probe shows $X_{14}=6$ and we know now (see Fig.~\ref{passage_region_useless}) that $6\le M_{14}\le 10$, and if $M_{14}=10$, the maximum would be at position $10$ (dashed line) so we probe the position $10$. \textbf{(c)} The second probe shows $X_{10}=4$ and we know now that $6\le M_{14}\le 7$,  and if $M_{14}=7$, the maximum would be at position $6$ or $13$ (dashed line) so we probe the position $6$. \textbf{(d)} The third probe shows $X_6=4$ and we know now that $6\le M_{14}\le 7$, and if $M_{14}=7$, the maximum would be at position $13$ (dashed line) so we probe position $13$ and find the maximum $M_{14}=X_{13}=7$ in $4$ probes.}
\label{exemple_algo}
\end{center}
\end{figure}

To get familiar with this algorithm, it is useful to consider a simpler search algorithm, denoted by $u$, belonging to $A_n$ (that denotes the ensemble of the algorithms that find the maximum $M_n$ of a random walk of $n$ steps), which proceeds as follows: $u$ probes always the random walk at the step where the upper envelope of the (still) possible trajectories reaches its maximum. This algorithm $u$ is based on the idea that, as illustrated in Fig. \ref{passage_region_useless}, if $X_m$ and $X_{m+k}$ have been probed, then the searcher knows for sure that, between step $m$ and step $m+k$, the position of the random walker can not exceed $(X_{m}+X_{m+k}+k)/2$. This can be shown as follows. Let us denote by $n_+$ the number of up-steps ($+1$) and $n_-$ the number of down-steps ($-1$) between step $m$ and step $m+k$. Then $n_+$ and $n_-$ satisfy the equations
\begin{eqnarray}
&&n_+ + n _- = k \\
&&n_+-n_- = X_{m+k} - X_{m} \;.
\end{eqnarray}
Hence one has
\begin{eqnarray}
&&n_+ = \frac{X_{m+k} - X_{m} + k}{2} \\
&&n_- = \frac{X_{m} - X_{m+k} + k}{2} \;.
\end{eqnarray}
Therefore the position of the random walker can not exceed $X_{m} + n_+ = (X_{m}+X_{m+k}+k)/2$, as shown in Fig.~\ref{passage_region_useless}. 
%
This simple algorithm is illustrated in Fig. \ref{exemple_algo} on a realization of the RW for $n=14$ steps.  
This basic idea is at the heart of the algorithm proposed by Odlyzko.

Here we also want to explain briefly the occurrence of this particular functional of the maximum ${I}$ in (\ref{def_I}), in the analysis of this optimal algorithm, following the line of reasoning of \cite{chassaing1999many,chassaing_yor}. To understand this, let us consider a traveler, moving on a line, its position being denoted by $y$. Suppose that its velocity $v(y)$ at position $y$ is bounded by some function $z(y)$, such that $0<v(y) \leq z(y)$. Then the time $t$ to reach the point $x$ starting from the origin satisfies the bound
\begin{eqnarray}\label{velocity_bound}
t  = \int_0^x \frac{{\rmd}y}{v(y)} \geq \int_0^x \frac{{\rmd}y}{z(y)} \;.
\end{eqnarray}   
Now let us consider an algorithm $a$, its cost being $C(a)$ and denote by ${m_1, \ldots, m_{C(a)}}$ the steps at which the RW has been probed
by the searcher -- which has eventually found the maximum $M_n$ after $C(a)$ probes. To be sure that the maximum is not in the interval $[m_i, m_{i+1}]$, the potential maximum of the RW between these two steps, which is $(X_{m_i} + X_{m_{i+1}} + m_{i+1} - m_{i})/2$ (see Fig. \ref{passage_region_useless}), must be smaller than $M_n$ (by definition of the maximum). Hence this yields the following inequality
\begin{eqnarray}\label{discrete_velocity}
m_{i+1} - m_i \leq 2 M_n - X_{m_i} - X_{m_{i+1}} \;.
\end{eqnarray} 
Notice that $m_{i+1} - m_i$ can be seen as the velocity $v(m_i)$ of the algorithm at point $m_i$. One can further argue \cite{odlyzko}, using the fact
most of the RWs are ``slowly varying'' [see Eq. (\ref{slow_variation}) below], that $2 M_n - X_{m_i} - X_{m_{i+1}} \sim 2(M_n - X_{m_i})$ when $n$ is large. Hence 
\begin{eqnarray}
Z_k = 2(M_n-X_k) 
\end{eqnarray}
can be viewed as the speed limit at step $k$ of the random walk. Finally, by analogy with (\ref{velocity_bound}), $C(a)$ satisfies
\begin{eqnarray}\label{Ca_app}
C(a) \geq \sum_{k=1}^n \frac{1}{Z_k} = \frac{1}{2} \sum_{k=1}^n \frac{1}{M_n-X_k} \;,
\end{eqnarray}
which in the continuum limit yields the functional of the maximum $I$ in Eq. (\ref{def_I}). It is rather clear that these heuristic arguments leading to Eq. (\ref{Ca_app}) can be straightforwardly extended to the case of the Random Walk bridge, $X_{i,B}$, which is a RW conditioned to start and end at the origin $X_{0,B} = X_{n,B} = 0$. Of course in this case the maximum $M_n$ in (\ref{Ca_app}) is then replaced by the maximum of the Brownian bridge $M_{n,B} = \max_{1\leq i \leq N} X_{i,B}$.

\subsection{Description of the Odlyzko's algorithm}\label{section:description}

Here we describe in more detail Odlyzko's algorithm which finds the maximum of a random walk $X_{i+1} = X_i \pm 1$ with equal probability $1/2$ (starting from $X_0=0$). Let $c$ be a positive real number, which is sufficiently large. The algorithm is essentially based on the fact that most of the RWs has ``slow variations'' (SV), i.e., check the identity \cite{odlyzko}:
\begin{eqnarray}
|X_{i+k}-X_i| \le c \sqrt{k \log{n}}\; , \; \forall i, k \; {\rm with} \; i + k \leq n \;. 
\label{slow_variation}
\end{eqnarray}
Indeed if $c$ is large enough, the probability that a realization of the  RW does not satisfy the SV property (\ref{slow_variation}) decays as $n^{-1}$. This statement can be easily shown, as in \cite{odlyzko,chassaing_yor}, by using that for fixed $j$, ${\rm Pr} (|X_j| > x) \leq 2 \exp(-x^2/(2j))$ [the so called Chernoff's bound, see \cite{bollobas} p. 12]. Although the realizations of the RW that do not satisfy (\ref{slow_variation}) necessitates a large number of probes $\sim n$, their contribution to the average cost of the algorithm turns out to be negligible as they occur with a very small probability $\propto 1/n$. On the other hand, as we shall see below, it is relatively easy to find the maximum of a RW which satisfies the ``SV'' property. 

The algorithm proposed by Odlyzko consists in two steps:

\vspace*{0.5cm}

$\bullet$ In a first stage, one searches a good estimate $M^*$of $M_n$. This is done by probing $X_N$, $X_{2N}$, $X_{3N}$,... where $N=\lfloor \sqrt{n} \log n \rfloor$, where $\lfloor x \rfloor$ denotes the largest integer not larger than $x$. If the algorithm  finds, here or later, a violation of the SV inequality (\ref{slow_variation}), one has to probe all the positions of the RW (but this happens very rarely). We denote by $M'=\max \{X_0,X_N,X_{2N},X_{3N},...\}\le M_n$. If the RW satisfies SV (\ref{slow_variation}), then 
\begin{eqnarray}\label{ineq_1}
M_n-M' \le c \sqrt{N \log n} = c n^{1/4} \log n \;.
\end{eqnarray}
Indeed, if we denote by $k_{\max}$ such that $X_{k_{\max} N}\leq M_n \leq X_{(k_{\max}+1) N}$ then $M_n - \max(X_{k_{\max} N},X_{(k_{\max}+1) N}) \leq c \sqrt{N \log n}$, which follows from (\ref{slow_variation}), and which implies (\ref{ineq_1}) as $M' \geq \max(X_{k_{\max} N},X_{(k_{\max}+1) N})$. As we discuss it below, it turns out that this estimate $M'$ of $M_n$ (\ref{ineq_1}) is however not precise enough for the forthcoming steps of the algorithm. It is indeed necessary to scan the 
neighborhood of the large $X_{rN}$'s on a finer window. If for some integer $r$ one finds 
\begin{eqnarray}
X_{rN} \ge M' - c n^{1/4} \log n\,,
\label{scan_neighbor}
\end{eqnarray}
we probe $X_{rN \pm j K}$, $j=1,2,...\lfloor N/K \rfloor$, $K=\lfloor n^{1/4}\rfloor$. If the RW has SV, then any $k$ with $X_k=M_n$ must be as close as of a $rN$ for some $r$ for which (\ref{scan_neighbor}) is true.
We now denote $M^*$ the maximum of all probes found until now. Because we scan with intervals $\le n^{1/4} \log n$ around the maximum, the SV inequality (\ref{slow_variation}) give
\begin{eqnarray}\label{bound_sixth}
0 \le M_n-M^* \le c \sqrt{n^{1/4} \log^2 n} \le n^{1/6}\,.
\end{eqnarray}
One can prove \cite{odlyzko} that the average cost of this first phase of the algorithm is of order ${\cal O}(\sqrt{n}/\log n)$ negligible compared to the cost of the second phase, that we now describe, and which is of order ${\cal O}(\sqrt{n})$.

\vspace*{0.5cm}

$\bullet$ With this estimate $M^*$ of the actual maximum $M_n$, the second phase will eventually find $M_n$ in a number of probes that is of order ${\cal O}(\sqrt{n})$, which is the leading contribution to the cost of this algorithm.
To do this, we will scan the sample path from left to right as follows. We introduce $m$ the index of the RW position $X_m$ which is currently probed by the algorithm. We start with $m=0$ and we denote by $M^\#$ the greatest position probed so far by the algorithm including $M^*$. At each step of this phase, two cases may occur: 
\begin{itemize}
\item[(i)] If $M^{\#}-X_m \le n^{1/6}$, the algorithm will probe the right neighbor of $X_m$ and $m$ is incremented by $1$, $m \to m+1$.
\item[(ii)] If $M^{\#}-X_m > n^{1/6}$, this means that the algorithm is still far from the maximum, because we know that $M_n-M^{\#}\leq n^{1/6}$. In this case, the immediate vicinity of $X_m$ 
does not need to be explored and the strategy is to jump from $X_{m}$ to $X_{m+k}$, where $k$ is still to be determined. In order to be sure that the RW does not exceed $M^\#$ between $m$ and $m+k$, we must have in mind the upper envelope of the RW on that interval $[m, m+k]$ (see Fig. \ref{passage_region_useless}). Hence we impose the following bound

\begin{eqnarray}\label{sv_app}
k \le 2 (M^\#-X_m) +(X_{m}-X_{m+k}) \;. \\
\nonumber
\end{eqnarray}

The first term in the right hand side of this inequality (\ref{sv_app}), $2 (M^\#-X_m)$, is larger than $2 n^{1/6}$, while the second term, $(X_{m}-X_{m+k})$ is bounded by $c \sqrt{k \log n}$, thanks to SV (\ref{slow_variation}) -- as stated above, if $X_{m+k}-X_m$ does not satisfy the SV inequality (\ref{slow_variation}), we abort this approach and probe every position. Hence we can choose $k$ slightly smaller than $2(M^\#-X_m)$. 
If $m+k >n$, we probe $X_n$ and stop. When the full path has been scanned, the maximum $M_n$ of the RW has been found by the algorithm. 
\end{itemize}

For a RW which satisfies SV (\ref{slow_variation}), one can show \cite{odlyzko} that the major contribution to the cost of the algorithm is when $M^{\#}-X_m > n^{1/6}$. Indeed, one can show that the contributions of the probes of the type (i) to the cost of the algorithm is of the order ${\cal O}(n^{1/3})$. In fact, one can show that if the estimate $M^*$ of $M_n$ is such that $M_n - M^* < n^{\alpha}$ then the cost of these contributions is of order ${\cal O}(n^{2\alpha})$. If we want that the cost of this part of the algorithm to be smaller than the cost of the last one, which is of order ${\cal O}(\sqrt{n})$, then this requires $2\alpha < 1/2$, for instance $2\alpha = 1/3$, hence the choice $\alpha = 1/6$ made by Odlyzko \cite{odlyzko} [see Eq. (\ref{bound_sixth})].

The step size $k$ is slightly smaller than $2(M_n-X_m)$ and we need only one probe to control the $k$ positions between $m$ and $m+k$. Since $k$ can be interpreted as the velocity of the algorithm [see Eq. (\ref{velocity_bound})], the average cost of the algorithm is, at leading order when $n$ goes to infinity, $\langle C({\rm Od})\rangle$ given by
\begin{eqnarray}
\langle C({\rm Od})\rangle=\frac12 \big\langle \sum_{i=0}^{n} \frac{1}{M_n-X_i+1}\big\rangle\,,
\label{def_vn}
\end{eqnarray}
where we recall that $\langle...\rangle$ denotes an average over the different realizations of the RW $X_i$'s. When $n$ goes to infinity, the RW becomes a BM and
\begin{eqnarray}
\frac{C({\rm Od})}{\sqrt{n}} \underset{n \to \infty}{\to} I = \frac{1}{2} \int_0^1 \frac{\rmd \tau}{ x_{\max}-x(\tau)}\,,
\end{eqnarray}
as described in the text in (\ref{def_I}).

\subsection{Odlyzko's algorithm for the Bridge}\label{odl_bridge}

It is easy to check that the arguments presented above can be easily transposed to the case of a random walk bridge. In particular, given that the bridge is pinned at both extremities $X_{0,B} = X_{n,B}=0$, its variations are typically smaller than the one of the free walk and hence the property of ``slow variations'' (\ref{slow_variation}), which plays a crucial role in this algorithm, would follow naturally. 
Therefore we conjecture that Odlyzko's algorithm would be the optimal one to find the maximum $M_{n,B}$ and its cost would be given by $(1/2) T_{\alpha=-1}^B(t)$ given in Eq. (\ref{T_alpha_B}).

\section{Some useful integrals involving confluent hypergeometric functions relevant for the case $V(x) = 1/x$}\label{hypergeometric_section}

\subsection{An integral involving a single confluent hypergeometric function}\label{hypergeometric_section_first}

For the analysis of the functional $T_{\alpha = -1}(t)$ [see Eq. (\ref{integral_phi_odlyzko_free})], a useful integral involving the confluent hypergeometric function $U(a,2,z)$ is the following (see \cite{grad} as well as Mathematica):

\begin{eqnarray}\label{int_1}
\tilde \varphi(s) &=& 2^{3/2} \lambda \Gamma(\lambda/\sqrt{2 s}) \int_0^\infty \; e^{-\sqrt{2s} y} \, y \,U\left(1+\frac{\lambda}{\sqrt{2s}},2,2 \sqrt{2s} y\right)  {\rmd}y \\
&=& \frac{1}{\sqrt{2}s} \left(\sqrt{2s} - 2 \pi\,\lambda\, {\rm csc}\left(\frac{\pi \lambda}{\sqrt{2s}} \right) - \lambda \, H\left(-\frac{1}{2} - \frac{\lambda}{2 \sqrt{2s}} \right) + \lambda \, H\left(- \frac{\lambda}{2\sqrt{2s}}  \right) \right) \;,
\end{eqnarray}
where ${\rm csc}(x) = 1/\sin{x}$ and $H(x)$ are harmonic numbers, $H(x) = \psi(x) + \gamma_E$ where $\psi(x) = \Gamma'(x)/\Gamma(x)$ is the di-gamma function and $\gamma_E$ the Euler constant. The function $H(x)$ admits the following series expansion
\begin{eqnarray}\label{def_harmonic}
H(x) = \sum_{j=0}^\infty (-1)^j \zeta(j+2) \, x^{j+2} \;,
\end{eqnarray}
where $\zeta(x)$ is the Riemann zeta function. By combining (\ref{int_1}), together with (\ref{def_harmonic}) one arrives straightforwardly at the formula given in Eq. (\ref{explicit_tildephi}) in the text.

\subsection{An integral involving the product of two confluent hypergeometric functions}\label{LL-appendix}

To compute the amplitudes $c_E$ such that the functions $\phi_E(x)$ in (\ref{linear_hypergeo}) with $d_E =0$ satisfy the orthogonality condition in Eq. (\ref{ortho_condition}) we used the following relation, derived by Landau and Lifshitz \cite{LL81} (see formula (f.9) in Appendix f):
\begin{eqnarray}\label{formula_Landau}
J &=& \int_0^\infty e^{-\lambda z} z^{\gamma-1} \,_1 F_1(\alpha,\gamma,k z) _1 F_1(\alpha',\gamma',k' z) {\rmd} z \nonumber \\
&=& \Gamma(\gamma) \lambda^{\alpha+\alpha'-\gamma} (\lambda-k)^{-\alpha}(\lambda-k')^{-\alpha'} \,_2F_1\left(\alpha,\alpha',\gamma,\frac{k k'}{(\lambda-k)(\lambda-k')}\right) \;,
\end{eqnarray}
where $\,_2F_1(\alpha,\alpha',\gamma,z)$ is a generalized hypergeometric series. Such integrals (\ref{formula_Landau}) arise naturally in the study of certain matrix elements of quantum Hamiltonian involving Coulomb interactions. In our case (\ref{linear_hypergeo}), one has $\alpha = 1- i s/\sqrt{E}$, $\alpha' = 1- i s/\sqrt{E'}$, $\gamma = \gamma' = 2$, $k = 2i\sqrt{2E}$, $k' = 2i\sqrt{2E'}$ and $\lambda = 2\sqrt{2 s}$. Hence the desired formula in our case (\ref{ortho_condition}) can be obtained by differentiating (\ref{formula_Landau}) once wrt $\lambda$ and analyzing in detail the limit $k \to k'$ of the resulting formula (\ref{formula_Landau}). These somewhat cumbersome manipulations yield the expression for $c_E$ given in (\ref{c_E}).

\section{Numerical simulations of constrained Brownian motion}\label{simulation_BM_section}

In this appendix, we describe the algorithms that we have used here to simulate various constrained Brownian motions. We refer the interested reader to \cite{Devroye10} for an extended discussion of these algorithms.

\begin{figure}[ht]\label{fig_Brownian_app}
\begin{center}
\resizebox{110mm}{!}{\includegraphics{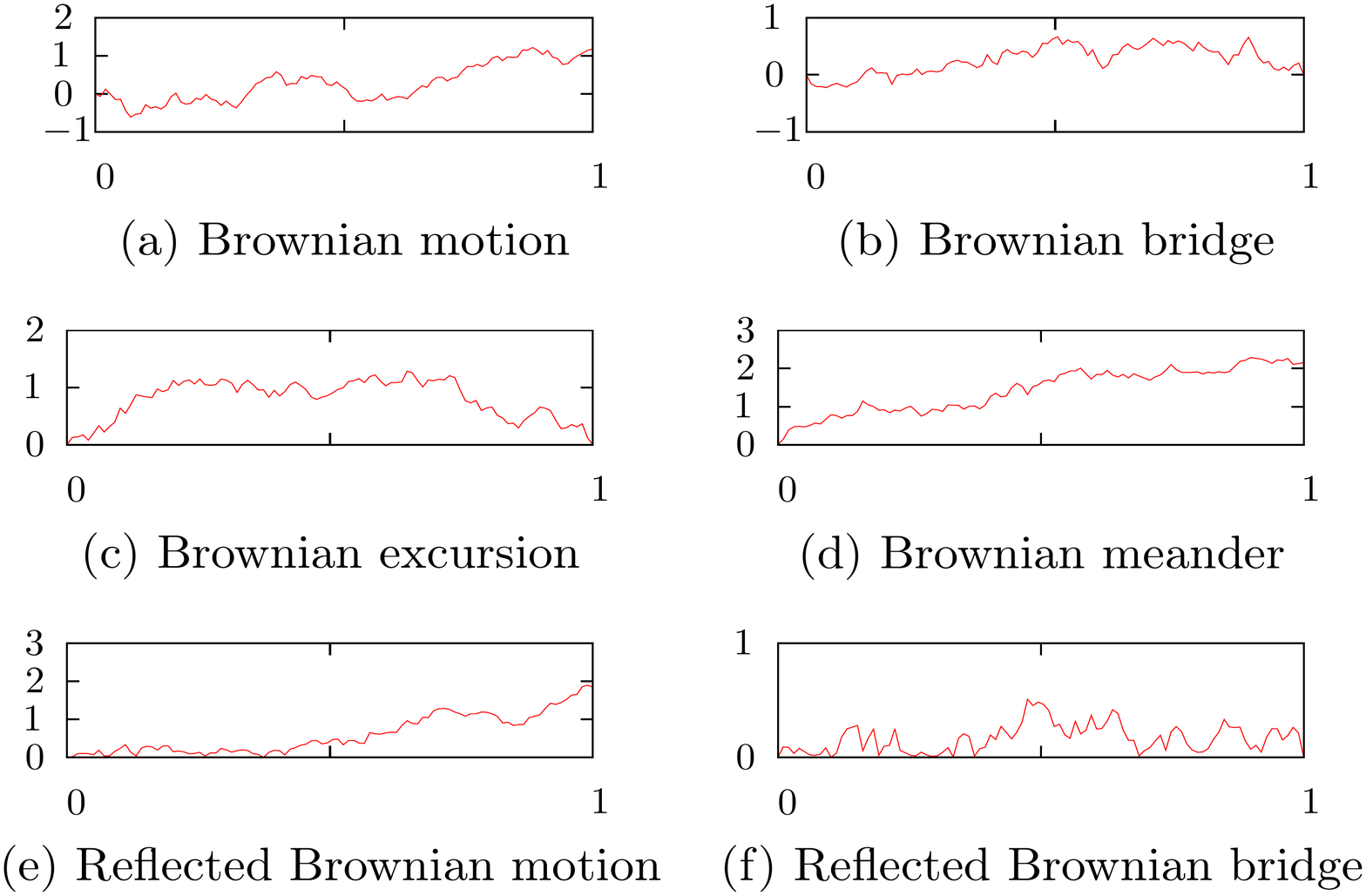}}
\caption{Example of different constrained Brownian motions studied in the present paper.}
\end{center}
\end{figure}

\subsection{Brownian motion}

In order to simulate a Brownian motion $x(\tau)$, we consider the discrete random walk
\begin{eqnarray}\label{BM}
\left\{
     \begin{array}{ll}
			X_0 = & 0\\
			X_i = & X_{i-1} + \frac{\eta_i}{\sqrt{N}}  \, , \, i\in [1,N]
     \end{array}
     \right.
\end{eqnarray}
where $\eta_i$'s are identical and independent Gaussian standard variables of variance unity. When $N$ goes to infinity, $X_{[\tau N]} \to x(\tau)$ where $x(\tau)$ is a Brownian motion, with $\tau \in [0,1]$ : $\dot x(\tau)= \zeta(\tau)$, where $\zeta(\tau)$ is a Gaussian white noise $\langle \zeta(\tau) \zeta(\tau') \rangle = \delta(\tau-\tau')$. This is the building block (\ref{BM}), to simulate different constrained Brownian motions.

\begin{verbatim}
//generation of Brownian Motion
void BM ( int N, double *X, gsl_rng * r)
{
//r is the 'seeds' of the random number generator.
  int i;
    X[0]=0;
    for(i=1; i<N; i++)
    {
        X[i]=X[i-1]+gsl_ran_gaussian (r, 1)/sqrt((double)N);
    }
}
\end{verbatim}

\subsection{Brownian bridge}

For a Brownian bridge $x_B(\tau)$, which is a Brownian motion starting and ending at the origin $x_B(0)=x_B(1)=0$, we use the identity $x(\tau)-\tau x(1)=x_B(\tau)$
\begin{eqnarray}\label{BB}
Y_i = X_i-\frac{i}{N}X_N , \, i\in [0,N],
\end{eqnarray}
where $X_i$'s are generated by (\ref{BM}). One can show that $Y_{[\tau N]}$ converges to a Brownian bridge $x_B(\tau)$.

\begin{verbatim}
//generation of Brownian Bridge
void BB ( int N, double *X, gsl_rng * r)
{
  BM(N,X,r);
    int i;
    for(i=1; i<N; i++)
    {
        X[i]=X[i]-(double)i/(N-1)*X[N-1];
    }
}
\end{verbatim}

\subsection{Brownian excursion}

For a Brownian excursion $x_E(\tau)$, which is a Brownian motion that starts and ends at the origin $x_E(0)=x_E(1)=0$ and staying positive in the interval $[0,1]$, we use the identity $\sqrt{[x_{B,1}(\tau)]^2 + [x_{B,2}(\tau)]^2+[x_{B,3}(\tau)]^2}=x_E(\tau)$ where $x_{B,1}, x_{B,2}$ and $x_{B,3}$ are three independent Brownian bridges~\cite{Wil70,Imh84}. Hence we simulate 
\begin{eqnarray}\label{BE}
E_i = \sqrt{Y_{1,i}^2+Y_{2,i}^2+Y_{3,i}^2} , \, i\in [0,N]
\end{eqnarray}
where $Y_{1,i},Y_{2,i}$ and $Y_{3,i}$ are three independent realisations of (\ref{BB}). $E_{[\tau N]}$ converges to a Brownian excursion $x_E(\tau)$.

\begin{verbatim}
//generation of Brownian Excursion
void BE ( int N, double *X, gsl_rng * r)
{
double X1[N];
double X2[N];
double X3[N];
  BB(N,X1,r);
  BB(N,X2,r);
  BB(N,X3,r);
    int i;
    for(i=1; i<N; i++)
    {
        X[i]=sqrt(X1[i]*X1[i]+X2[i]*X2[i]+X3[i]*X3[i]);
    }
}
\end{verbatim}

\subsection{Brownian meander}

For a Brownian meander $x_{Me}(\tau)$, a Brownian motion which begins at the origin and stays positive on $[0,1]$, one can show that the PDF of
its final position $x_F > 0$ at time $1$ is $p(x_F) = x_F e^{-x_F^2/2}$. One can then use the following representation of the meander ending at $x_F$: \cite{Wil70,Imh84} $\sqrt{[x_{B,1}(\tau)]^2 + [x_{B,2}(\tau)]^2+[x_{B,3}(\tau)+\tau \, x_F]^2}=x_{Me}(\tau)$ where $x_{{B},1},x_{{B},2}$ and $x_{{B},3}$ are three independent Brownian bridges and $x_F$ is a random variable drawn from $p(x_F) = x_F e^{-x_F^2/2}$. Hence the Brownian meander $x_{Me}(\tau)$ can be generated numerically as
\begin{eqnarray}\label{Bme}
M_i = \sqrt{Y_{1,i}^2+Y_{2,i}^2+\left(Y_{3,i}+f \frac{i}{N}\right)^2} , \, i\in [0,N]
\end{eqnarray}
where $Y_1,Y_2$ and $Y_3$ are three independent realizations of (\ref{BB}), where $f>0$ is a random variable, whose PDF is given by $p(f) = f e^{-f^2/2}$.
$M_{[tN]}$ converges to a Brownian meander $x_{Me}(\tau)$.

\begin{verbatim}
//generation of Brownian Meander
void BMe ( int N, double *X, gsl_rng * r)
{
  
double f=gsl_ran_rayleigh (r,1);
  
double X1[N];
double X2[N];
double X3[N];
  BB(N,X1,r);
  BB(N,X2,r);
  BB(N,X3,r);
  int i;
   for(i=0; i<N; i++)
            {
X[i]=sqrt((X1[i]+f*(double)i/(N-1))*(X1[i]+f*(double)i/(N-1))+X2[i]*X2[i]+X3[i]*X3[i]);
            }
}
\end{verbatim}

\end{appendix}

\end{document}